\documentclass[ALICE,manyauthors]{cernphprep}
\usepackage[comma,square,numbers,sort&compress]{natbib}

\usepackage{lineno}
\usepackage{xspace}
\usepackage{hyperref}
\usepackage[usenames,dvipsnames]{color}
\usepackage[T1]{fontenc}
\usepackage{orcidlink}

\begin{document}
%

\newcommand{\pp}           {pp\xspace}
\newcommand{\ppbar}        {\mbox{$\mathrm {p\overline{p}}$}\xspace}
\newcommand{\XeXe}         {\mbox{Xe--Xe}\xspace}
\newcommand{\PbPb}         {\mbox{Pb--Pb}\xspace}
\newcommand{\pA}           {\mbox{pA}\xspace}
\newcommand{\pPb}          {\mbox{p--Pb}\xspace}
\newcommand{\AuAu}         {\mbox{Au--Au}\xspace}
\newcommand{\dAu}          {\mbox{d--Au}\xspace}

\newcommand{\mee}         {\ensuremath{m_{\rm ee}}\xspace}
\newcommand{\ptee}         {\ensuremath{p_{\rm T,ee}}\xspace}
\newcommand{\ptll}         {\ensuremath{p_{\rm T,ll}}\xspace}
\newcommand{\pteesquare}    {\ensuremath{p_{\rm T,ee}^{\rm 2}}\xspace}
\newcommand{\meanpteesquare}       {$\langle p_{\rm T,ee}^{\rm 2}\rangle$\xspace}
\newcommand{\s}            {\ensuremath{\sqrt{s}}\xspace}
\newcommand{\snn}          {\ensuremath{\sqrt{s_{\mathrm{NN}}}}\xspace}
\newcommand{\pt}           {\ensuremath{p_{\rm T}}\xspace}
\newcommand{\pte}           {\ensuremath{p_{\rm T,e}}\xspace}
\newcommand{\etae}           {\ensuremath{\eta_{\rm e}}\xspace}
\newcommand{\meanpt}       {$\langle p_{\mathrm{T}}\rangle$\xspace}
\newcommand{\ycms}         {\ensuremath{y_{\rm CMS}}\xspace}
\newcommand{\ylab}         {\ensuremath{y_{\rm lab}}\xspace}
\newcommand{\etarange}[1]  {\mbox{$| \eta |\,<\,#1$}}
\newcommand{\etarangee}[1]  {\mbox{$| \eta_{\rm e} |<#1$}}
\newcommand{\yrange}[1]    {\mbox{$\left | y \right |~<~#1$}}
\newcommand{\dndy}         {\ensuremath{\mathrm{d}N_\mathrm{ch}/\mathrm{d}y}\xspace}
\newcommand{\dndeta}       {\ensuremath{\mathrm{d}N_\mathrm{ch}/\mathrm{d}\eta}\xspace}
\newcommand{\avdndeta}     {\ensuremath{\langle\dndeta\rangle}\xspace}
\newcommand{\dNdy}         {\ensuremath{\mathrm{d}N_\mathrm{ch}/\mathrm{d}y}\xspace}
\newcommand{\Npart}        {\ensuremath{N_\mathrm{part}}\xspace}
\newcommand{\Ncoll}        {\ensuremath{N_\mathrm{coll}}\xspace}
\newcommand{\dEdx}         {\ensuremath{\textrm{d}E/\textrm{d}x}\xspace}
\newcommand{\RpPb}         {\ensuremath{R_{\rm pPb}}\xspace}

\newcommand{\nineH}        {$\sqrt{s}=0.9$~Te\kern-.1emV\xspace}
\newcommand{\seven}        {$\sqrt{s}=7$~Te\kern-.1emV\xspace}
\newcommand{\twoH}         {$\sqrt{s}=0.2$~Te\kern-.1emV\xspace}
\newcommand{\twoHnn}         {$\sqrt{s_{\mathrm{NN}}}=0.2$~Te\kern-.1emV\xspace}
\newcommand{\twosevensix}  {$\sqrt{s}=2.76$~Te\kern-.1emV\xspace}
\newcommand{\five}         {$\sqrt{s}=5.02$~Te\kern-.1emV\xspace}
\newcommand{\twosevensixnn}{$\sqrt{s_{\mathrm{NN}}}=2.76$~Te\kern-.1emV\xspace}
\newcommand{\fivenn}       {$\sqrt{s_{\mathrm{NN}}}=5.02$~Te\kern-.1emV\xspace}
\newcommand{\LT}           {L{\'e}vy-Tsallis\xspace}
\newcommand{\GeVc}         {Ge\kern-.1emV$/c$\xspace}
\newcommand{\MeVc}         {Me\kern-.1emV$/c$\xspace}
\newcommand{\TeV}          {Te\kern-.1emV\xspace}
\newcommand{\GeV}          {Ge\kern-.1emV\xspace}
\newcommand{\MeV}          {Me\kern-.1emV\xspace}
\newcommand{\GeVmass}      {Ge\kern-.1emV$/c^2$\xspace}
\newcommand{\MeVmass}      {Me\kern-.2emV$/c^2$\xspace}
\newcommand{\lumi}         {\ensuremath{\mathcal{L}}\xspace}

\newcommand{\ITS}          {\rm{ITS}\xspace}
\newcommand{\TOF}          {\rm{TOF}\xspace}
\newcommand{\ZDC}          {\rm{ZDC}\xspace}
\newcommand{\ZDCs}         {\rm{ZDCs}\xspace}
\newcommand{\ZNA}          {\rm{ZNA}\xspace}
\newcommand{\ZNC}          {\rm{ZNC}\xspace}
\newcommand{\SPD}          {\rm{SPD}\xspace}
\newcommand{\SDD}          {\rm{SDD}\xspace}
\newcommand{\SSD}          {\rm{SSD}\xspace}
\newcommand{\TPC}          {\rm{TPC}\xspace}
\newcommand{\TRD}          {\rm{TRD}\xspace}
\newcommand{\VZERO}        {\rm{V0}\xspace}
\newcommand{\VZEROA}       {\rm{V0A}\xspace}
\newcommand{\VZEROC}       {\rm{V0C}\xspace}
\newcommand{\Vdecay} 	   {\ensuremath{V^{0}}\xspace}

\newcommand{\ee}           {\ensuremath{{\rm e}^{+}{\rm e}^{-}}\xspace} 
\newcommand{\pip}          {\ensuremath{\pi^{+}}\xspace}
\newcommand{\pim}          {\ensuremath{\pi^{-}}\xspace}
\newcommand{\kap}          {\ensuremath{\rm{K}^{+}}\xspace}
\newcommand{\kam}          {\ensuremath{\rm{K}^{-}}\xspace}
\newcommand{\pbar}         {\ensuremath{\rm\overline{p}}\xspace}
\newcommand{\kzero}        {\ensuremath{{\rm K}^{0}_{\rm{S}}}\xspace}
\newcommand{\lmb}          {\ensuremath{\Lambda}\xspace}
\newcommand{\almb}         {\ensuremath{\overline{\Lambda}}\xspace}
\newcommand{\Om}           {\ensuremath{\Omega^-}\xspace}
\newcommand{\Mo}           {\ensuremath{\overline{\Omega}^+}\xspace}
\newcommand{\X}            {\ensuremath{\Xi^-}\xspace}
\newcommand{\Ix}           {\ensuremath{\overline{\Xi}^+}\xspace}
\newcommand{\Xis}          {\ensuremath{\Xi^{\pm}}\xspace}
\newcommand{\Oms}          {\ensuremath{\Omega^{\pm}}\xspace}
\newcommand{\degree}       {\ensuremath{^{\rm o}}\xspace}

\begin{titlepage}
\PHyear{2023}       
\PHnumber{194}      
\PHdate{30 August}  

\title{Dielectron production in central Pb--Pb collisions at $\mathbf{\sqrt{\textit{s}_{NN}}=5.02}$~Te\kern-.1emV\xspace}
\ShortTitle{Dielectron production in central Pb--Pb collisions at \fivenn}   

\Collaboration{ALICE Collaboration\thanks{See Appendix~\ref{app:collab} for the list of collaboration members}}
\ShortAuthor{ALICE Collaboration} 

\begin{abstract}
The first measurement of the \ee pair production at midrapidity and low invariant mass in central \PbPb collisions at \fivenn at the LHC is presented. The yield of \ee pairs is compared with a cocktail of expected hadronic decay contributions in the invariant mass (\mee) and pair transverse momentum (\ptee) ranges $\mee < 3.5$~\GeVmass and $\ptee < 8$~\GeVc. For $0.18 < \mee < 0.5$~\GeVmass the ratio of data to the cocktail of hadronic contributions amounts to $1.40 \pm 0.11 \ ({\rm stat.}) \pm 0.23 \ ({\rm syst.}) \pm 0.16 \ ({\rm cocktail})$ and $1.42 \pm 0.11 \ ({\rm stat.}) \pm 0.23 \ ({\rm syst.}) ^{+0.24}_{-0.29} \ ({\rm cocktail})$, including or not including medium effects in the estimation of the heavy-flavor background, respectively. It is consistent with predictions from two different models for an additional contribution of thermal \ee pairs from the hadronic and partonic phases. In the intermediate-mass range ($1.2 < \mee < 2.6$~\GeVmass), the pair transverse impact parameter of the \ee pairs (DCA$_{\rm ee}$) is used for the first time in Pb--Pb collisions to separate displaced dielectrons from heavy-flavor hadron decays from a possible (thermal) contribution produced at the interaction point. The data are consistent with a suppression of \ee pairs from ${\rm c\overline{c}}$ and an additional prompt component. Finally, the first direct-photon measurement in the 10\% most central \PbPb collisions at \fivenn is reported via the study of virtual direct photons in the transverse momentum range  $1 < \pt < 5$~\GeVc. A model including prompt photons, as well as photons from the pre-equilibrium and fluid-dynamic phases, can reproduce the result, while being at the upper edge of the data uncertainties.
\end{abstract}
\end{titlepage}

\setcounter{page}{2} 


\section{Introduction}

Ultra-relativistic heavy-ion collisions at the Large Hadron Collider (LHC) allow the study of strongly interacting matter at high temperature and small net-baryon density~\cite{ALICE:2022wpn}. Under these conditions, the theory of the strong interaction, quantum chromodynamics (QCD), predicts a transition from ordinary matter made of hadrons to a quark--gluon plasma (QGP) in which quarks and gluons are deconfined~\cite{Bazavov:2011nk,Borsanyi:2013bia,HotQCD:2018pds,Borsanyi:2020fev}. In this QGP phase, chiral symmetry is also expected to be restored~\cite{Bilic:1997sh,Dominguez:2012bs}. Photons and dileptons, i.e. lepton--antilepton pairs originating from the internal conversion of virtual photons, are produced at all stages of the heavy-ion collision with negligible final-state interactions, unlike hadrons. Therefore, they carry undistorted information about the whole space--time evolution of the medium created in such collisions.

Direct photons are photons not originating from hadronic decays. They are emitted by various sources as discussed in Ref.~\cite{Gale:2009gc}. Prompt direct photons are produced in the initial hard parton--parton scatterings. Additional photons, still generated before the system reaches sufficient equilibrium to be described by relativistic viscous hydrodynamics, are called pre-equilibrium photons. Thermal photons are emitted from the deconfined QGP and hot hadronic matter and are characterized by the thermal distributions of partons and hadrons, respectively. Other direct photon production mechanisms are not excluded, like the interaction of hard-scattered partons from jets with the plasma~\cite{Turbide:2005fk,Qin:2009bk}. Each source of direct photons populates different transverse momentum (\pt) regions. Prompt direct photons follow a power-law spectrum and dominate at high \pt ($\pt \gtrsim 5$~\GeVc). At lower \pt ($\pt \lesssim 2$~\GeVc), thermal photons are expected to contribute significantly with an approximately exponential \pt spectrum containing information on the initial temperature and space--time evolution of the medium~\cite{Kapusta:1991qp,Turbide:2003si}. Pre-equilibrium photons may play a role in the intermediate \pt range~\cite{Gale:2021emg}.

The first measurement of a direct photon signal in relativistic heavy-ion collisions was reported by the WA98 collaboration~\cite{WA98:2000vxl} in central \PbPb collisions at the center-of-mass energy per nucleon pair $\snn = 17.3$~GeV. The spectrum is consistent with calculations for thermal photon radiation from a quark--gluon plasma and a hot hadronic phase, as well as with predictions including multiple soft scatterings of the incoming partons without the formation of a QGP~\cite{Turbide:2003si}. At the Relativistic Heavy Ion Collider (RHIC), the first \pt-differential yield of direct photons was measured by the PHENIX collaboration in \AuAu collisions at \twoHnn~\cite{PHENIX:2005yls}, followed by subsequent results at the same center-of-mass energy per nucleon pair~\cite{PHENIX:2008uif,PHENIX:2011oxq,PHENIX:2014nkk,PHENIX:2022rsx,PHENIX:2015igl} as well as at lower energies~\cite{PHENIX:2018for,PHENIX:2022qfp}. On the one hand, the measured direct-photon yields are well described by the expected yields of prompt direct photons at high \pt ($\pt \gtrsim 5$~\GeVc), showing that the high-\pt direct photons are predominantly from initial hard-scattering processes. The prompt component is estimated either from measurements in \pp and \dAu collisions at the same collision energy~\cite{PHENIX:2006duh,PHENIX:2012jgx,PHENIX:2012krx} or perturbative QCD calculations~\cite{Paquet:2015lta}, both scaled with the number of binary nucleon--nucleon collisions ($N_{\rm coll}$). On the other hand, a large excess of direct photons is observed at low \pt ($\pt < 2$~\GeVc) with respect to the expected prompt direct photon yields. The STAR collaboration also reported an enhancement of direct photons at low \pt in \AuAu collisions at \twoHnn~\cite{STAR:2016use}, which appears to be about a factor three smaller than the one measured by PHENIX. The large yield of direct photons is accompanied by a large azimuthal anisotropy (elliptic flow) with respect to the reaction plane~\cite{PHENIX:2011oxq,PHENIX:2015igl}, similar in magnitude to the one of charged hadrons, suggesting a large contribution from the late stages of the collision. A simultaneous description of the \pt-differential yields and elliptic flow results remains challenging for models~\cite{Chatterjee:2013naa,Gale:2021emg,Gale:2014dfa,Chatterjee:2017akg,Linnyk:2015tha}. At the LHC, direct photon yield~\cite{ALICE:2015xmh} and elliptic flow~\cite{ALICE:2018dti} measurements in \PbPb collisions at \twosevensixnn by the ALICE collaboration show similar trends as those from PHENIX within the experimental uncertainties. Recent results by the PHENIX collaboration, obtained with a factor ten larger dataset of \AuAu collisions at \twoHnn, reduce the tension between the measured direct-photon yield in central \AuAu collisions~\cite{PHENIX:2022rsx} and the predictions from a state-of-the-art model~\cite{Gale:2021emg} including prompt, pre-equilibrium, and thermal photons. The observed direct-photon yields in more peripheral \AuAu collisions remain nevertheless a factor two to three larger than the calculations. Effective temperatures ($T_{\rm eff}$) were extracted from the measured \pt spectra at low \pt in \AuAu collisions at \twoHnn~\cite{PHENIX:2022rsx} and in \PbPb collisions at \twosevensixnn~\cite{ALICE:2015xmh} by the PHENIX and ALICE collaborations, respectively. The measured $T_{\rm eff}$ values are significantly larger than the critical temperature for chiral symmetry restoration and color deconfinement. This does not prove that the radiation is emitted from a QGP, since radiation from the late stages of the collision with strong radial flow could lead to a significant blue shift. Detailed studies of the direct-photon yield as a function of the collision system size by PHENIX show a power-law dependence on the charged-particle multiplicity at midrapidity ($\frac{{\rm d}N_{\rm ch}}{{\rm d}\eta}|_{\rm \eta = 0}$) at low \pt. The measured power $\alpha$ seems to have a weak dependence on centrality or collision energy~\cite{PHENIX:2018for,PHENIX:2022qfp} and no apparent dependence on \pt, in particular in \AuAu collisions at \twoHnn~\cite{PHENIX:2022rsx}. However, model calculations predict different $\alpha$ for radiation from different phases~\cite{Shen:2013vja}.

In contrast to real photons, virtual photons, i.e. dileptons, carry a mass, the invariant mass of the lepton and antilepton pair ($m_{\rm l^{+}l^{-}}$). This provides an additional means to disentangle the different sources of electromagnetic radiation. For $m_{\rm l^{+}l^{-}} > 1.2$~\GeVmass, virtual direct photons are foreseen to originate from the partonic phase of the heavy-ion collision~\cite{Rapp:2014hha}. In the mass region $1.2 < m_{\rm l^{+}l^{-}} < 2.6$~\GeVmass thermal radiation from the QGP is expected to contribute significantly to the virtual direct-photon yield~\cite{Rapp:2013nxa}. The slope of their $m_{\rm l^{+}l^{-}}$ distribution is predicted to carry information about the early temperature in the medium without distortion due to blueshift effect~\cite{Rapp:2014hha}. Nevertheless, correlated background from semileptonic decays of open heavy-flavor hadrons has to be subtracted or rejected first. Moreover, initial hard parton--parton scatterings generate also direct dileptons via the Drell-Yan process. While this process contributes significantly to the measured dilepton spectra down to small invariant masses ($m_{\rm l^{+}l^{-}} \approx 1.5$~\GeVmass) at SPS energies, it is estimated to be negligible for masses up to about $3$~\GeVmass at the LHC. On the other hand, recent theoretical calculations for \PbPb collisions at \fivenn show that additional ${\rm l^{+}l^{-}}$ pairs, still produced before the system reaches local thermal equilibrium, may play a role for $m_{\rm l^{+}l^{-}} > 1.5$~\GeVmass~\cite{Coquet:2021lca,Coquet:2021gms}. At lower $m_{\rm l^{+}l^{-}}$ ($m_{\rm l^{+}l^{-}} < 1.2$~\GeVmass), the dilepton invariant mass can be used to study the decay of massive particles, such as the in-medium modified spectral shape of vector mesons. Dileptons produced in the medium in the vicinity of the transition temperatures are sensitive to effects related to chiral symmetry restoration~\cite{Pisarski:1995xu,Hohler:2013eba}. Due to its strong coupling to the $\pi^{+}\pi^{-}$ channel and its life time of only 1.3~fm$/c$, much smaller than that of the fireball at the LHC ($\approx 10$~fm$/c$), the $\rho$ meson is expected to be the most sensitive to such medium effects. Finally, in the zero mass limit, the fraction of virtual direct photons over inclusive virtual photons is expected to be identical to that of real photons~\cite{PHENIX:2009gyd}. Therefore the measured dilepton yield in the quasi-real virtual-photon region, where the pair transverse momenta \ptll is much larger than $m_{\rm l^{+}l^{-}}$, can be used to measure the \pt-differential yield of direct photons.

Precision measurements of virtual photons were performed at the CERN SPS by the CERES~\cite{CERESNA45:1997tgc,CERESNA45:2002gnc,CERES:2006wcq} and NA60~\cite{NA60:2008dcb,NA60:2006ymb,NA60:2008ctj,Specht:2010xu} collaborations in the dielectron and dimuon channels, respectively. An average temperature of about 200~MeV, significantly above the transition temperatures, was extracted from the mass spectrum in $1.2 < m_{\rm l^{+}l^{-}} < 2$~\GeVmass in 158A~GeV In--In collisions by the NA60 collaboration~\cite{Specht:2010xu}. The capability of the NA60 experiment to disentangle prompt muons, originating from the primary vertex, and non-prompt muons from displaced open-charm hadron decays, allowed them to explicitly attribute the observed dimuon signal in this mass range to a prompt source. Together with studies of the effective temperature $T_{\rm eff}$ from the dimuon \pt spectrum as a function of mass~\cite{NA60:2008ctj}, this temperature value suggested that dilepton radiation in the intermediate-mass region originates from the partonic phase~\cite{Renk:2006qr,Ruppert:2007cr}. For increasing \snn, the correlated background from heavy-flavor hadron decays becomes very large with respect to the QGP radiation, preventing  the PHENIX~\cite{PHENIX:2015vek}, STAR~\cite{STAR:2013pwb}, and ALICE~\cite{ALICE:2018ael} collaborations at RHIC and the LHC from extracting an unambiguous dilepton signal in the intermediate-mass range without the use of high-precision vertex detectors. At lower mass, the $\rho$ properties in the hot medium created in heavy-ion collisions were measured by the CERES and NA60 collaboration at SPS energies. The dilepton excess, obtained after subtraction of the expected contribution of other light-flavor meson decays happening at a later stage in the collision, was found to be well described by the calculations of a microscopic many-body model~\cite{Rapp:1999us,vanHees:2007th}. This model predicted a very strong broadening of the $\rho$ spectral function with essentially no mass shift, consistent with chiral symmetry restoration~\cite{Hohler:2013eba}. At the top RHIC energy, an enhancement of dileptons in the low-mass region was reported by both the STAR~\cite{STAR:2013pwb} and the PHENIX~\cite{PHENIX:2015vek} collaborations in central \AuAu collisions at \twoHnn. Models reproducing the SPS data and involving the broadening of the $\rho$ meson~\cite{Rapp:2000pe,Linnyk:2011vx} were shown to describe the RHIC data as well. The first low-mass dilepton results at the LHC, published by the ALICE collaboration in central \PbPb collisions at \twosevensixnn~\cite{ALICE:2018ael}, are in line with the expectations from these models, although no significant excess was observed within the limited precision of the data.

Measurements of direct photons and dileptons at LHC energies could shed light on the remaining discrepancy between those results and theory predictions. Heavy-ion collisions at the LHC provide nowadays the highest-temperature and longest-lived experimentally accessible QGP. Therefore, the thermal radiation yield is expected to be the largest with respect to measurements at other accelerator facilities. Moreover, the in-medium modifications of vector mesons can be probed under conditions that are most closely related to the regime accessible by lattice QCD (vanishing net-baryon density). However, the large combinatorial and heavy-flavor backgrounds pose a serious challenge.

In this article, the first dielectron and resulting direct-photon measurements in the 10\% most central \PbPb collisions at \fivenn are presented and compared to measurements at lower energies (RHIC and LHC) and with model predictions.
The uncertainties could be reduced compared to the previous publication in central \PbPb collisions at lower energy~\cite{ALICE:2018ael} due to the three times larger data sample. While direct photons have already been measured in central collisions at $\sqrt{s_{\rm NN}}=2.76$~TeV as part of a real-photon analysis~\cite{ALICE:2015xmh,ALICE:2018dti}, a direct-photon measurement is now available for the first time at the highest collision energy. Finally, the larger data sample allows for the first studies in Pb--Pb collisions of a topological separation of prompt dielectrons and dielectrons originating from displaced correlated heavy-flavor hadron decays.
The article is organized as follows. Section~\ref{Detector} contains a brief description of the ALICE apparatus and the used data sample. Section~\ref{analysis} illustrates the data analysis techniques. Section~\ref{hadroniccocktail} explains how the expected dielectron yields from known hadronic sources are calculated. The results are presented and discussed in Sec.~\ref{inclusiveproduction} for the inclusive dielectron production, Sec.~\ref{DCAeeresults} for the topological separation of \ee sources, and Sec.~\ref{directphoton} for the direct-photon measurement. Finally, they are summarized in Sec.~\ref{summary}.

\section{Detector and data sample}
\label{Detector}

A detailed description of the ALICE apparatus and its performance can be found in Refs.~\cite{ALICE:2008ngc,ALICE:2014sbx}. The main detectors used for the track reconstruction and electron identification at midrapidity ($|\eta| < 0.9$) are the Inner Tracking System (ITS)~\cite{ALICE:2010tia}, the Time Projection Chamber (TPC)~\cite{Alme:2010ke}, and the Time-Of-Flight (TOF) detector~\cite{Akindinov:2013tea}. They are located  inside a large solenoidal magnet, providing a uniform magnetic field of 0.5~T parallel to the LHC beam direction. Charged-particle trajectories are reconstructed from their signals in the ITS and the TPC. The ITS consists of a six-layer silicon detector with the innermost layer installed at a radius of 3.9~cm from the beam axis. It is used for tracking, for the reconstruction of the main vertex of the collision, and for the measurement of the distance-of-closest-approach (DCA) of the track to this primary vertex. Moreover, the four outer layers, made of silicon drift detectors and silicon strip detectors, provide charged-particle identification (PID) via the measurement of their specific energy loss (${\rm d}E/{\rm d}x$) in the detector material. The TPC, a $500$~cm long gas filled cylinder providing up to $159$ three-dimensional space points per track, is the main tracking detector. The measured ${\rm d}E/{\rm d}x$ of the charged particles in the gas allows electron identification over a large momentum range (up to a momentum of 10~\GeVc). The TOF detector extends at intermediate momenta ($0.4 < p < 1.3$~\GeVc) the PID capabilities of the TPC and ITS via the measurement of the time-of-flight of charged particles from the interaction point to the detector.

The data used in this article were collected by ALICE in 2018 during \PbPb runs at \fivenn. A minimum bias interaction trigger was provided by the V0 detector~\cite{ALICE:2013axi}. Simultaneous signals in the two scintillator arrays covering the pseudorapidity intervals $-3.7 < \eta < -1.7$ (V0C) and $2.8 < \eta < 5.1$ (V0A) were required. In addition, events were selected online based on the signal amplitude in these detectors in order to enrich the sample of central \PbPb collisions. Further selection criteria were applied later offline. Events due to the interaction of the beams with residual gas in the beam pipe were rejected using the V0 and the Zero Degree Calorimeter (ZDC)~\cite{Arnaldi:1999zz} timing information. Pileup collisions occurring during the TPC readout time were removed using the correlation between the number of tracks reconstructed in the TPC and in the ITS, which grants a faster readout. Only events with a primary vertex reconstructed within $\pm 10$\,cm from the center of the detector along the beam axis were considered. Finally, the 10\% most central \PbPb collisions were defined in terms of percentiles of the hadronic \PbPb cross section using the sum of the V0 signal amplitudes as described in detail in Ref.~\cite{ALICE:2018tvk}. The corresponding average value of the nuclear overlap function, $\langle T_{\rm AA} \rangle$, is $(23.26 \pm 0.17)~{\rm mb^{-1}}$~\cite{ALICE:2018tvk}. The total number of selected events ($N_{\rm events}$) is about $65\times10^{6}$ corresponding to an integrated luminosity of ${\rm 85~\mu b^{-1}}$~\cite{ALICE:2022xir}.

\section{Data analysis}
\label{analysis}

\subsection{Dielectron raw yields}

\subsubsection{Electron selection}

Electron candidates are selected in the transverse momentum range $0.2 < \pte < 10$~\GeVc and at midrapidity ($|\etae| < 0.8$). For the analysis using the DCA information to separate dielectrons from different sources, the tracks must have a $\pt > 0.4$~\GeVc to assure a sufficient separation between prompt \ee pairs originating from the primary vertex and non-prompt ones arising from displaced open heavy-flavor hadron decays. The DCA resolution worsens at low \pt. For tracks reconstructed in the ITS and TPC with a transverse momentum smaller than $0.35$~\GeVc the DCA resolution in the plane perpendicular to the beam axis is larger than $150$~ $\mu$m~\cite{ALICE:2023kzv} and thus comparable to the decay length of the D$^{\rm 0}$ and D$^{\pm}_{\rm s}$ mesons. The same tracking selection criteria as those described in Ref.~\cite{ALICE:2022hvk} are applied. In particular, a hit in the first ITS layer is required to be attached to the reconstructed track and a maximum of one ITS cluster, not placed in the first ITS layer, may be shared with any other track candidate. These requirements reduce the amount of electron tracks originating from photon conversion in the detector material by $98.8$\%, keeping $93.7$\% of the signal electrons.

The electron identification is based on the complementary information provided by the ITS, TPC, and TOF. The detector PID signals, $n(\sigma_{i}^{\rm DET})$, are expressed in terms of the deviation between the measured and expected value of the specific ionization energy loss in the ITS (TPC) or time-of-flight in the TOF for a given particle hypothesis $i$ and momentum, normalized to the respective detector resolution. Electrons are identified in a similar way as described in Ref.~\cite{ALICE:2022hvk}. To increase the electron purity, an asymmetric electron selection criterium is applied in the TPC, i.e. $-2 < n(\sigma_{\rm e}^{\rm TPC}) < 3$. Kaons, protons and pions are rejected with the requirements $|n(\sigma_{\rm K}^{\rm TPC})| > 3$, $|n(\sigma_{\rm p}^{\rm TPC})| > 3$ and $n(\sigma_{\rm \pi}^{\rm TPC}) > 3.5$, respectively. To increase the PID efficiency ($\epsilon_{\rm PID}$) and avoid a strong momentum dependence of $\epsilon_{\rm PID}$, the ITS and TOF are used to recover electrons with an energy loss in the TPC in the range where the charged kaon and proton bands cross the one of electrons. It exploits the fact that kaons (protons) and electrons are still separated in the energy loss measurements in the ITS or the measured time-of-flight in the TOF, at momenta where they have a very similar energy loss in the TPC. Tracks which fulfill only the TPC electron selection and pion rejection but have an associated TOF signal with $-3 < n(\sigma_{\rm e}^{\rm TOF}) < 3$ or an associated ITS signal with $n(\sigma_{\rm e}^{\rm ITS}) < 2$ not consistent with the kaon ($|n(\sigma_{\rm K}^{\rm ITS})| < 2$) or proton ($|n(\sigma_{\rm p}^{\rm ITS})| < 2$) hypothesis are accepted. The final hadron contamination in the single-electron candidate sample is estimated to be less than $3.3$\% integrated over \pt.

\subsubsection{Electron pairing and combinatorial background subtraction}

A statistical approach is applied to extract the yield of correlated signal pairs ($S$). To this end, all electron and positron candidates from the same event are combined into opposite-sign pairs ($N^{\rm same}_{\rm +-}$), characterised by their invariant mass \mee, pair transverse momentum \ptee and pair DCA$_{\rm ee}$, similarly to the approach followed by the NA60 collaboration\,\cite{Shahoyan:2009zz}. The latter is calculated from the single-electron DCAs as
\begin{equation}
{\rm DCA_{ee}} = \sqrt{\frac{({\rm DCA}_{xy,1}/\sigma_{xy,1})^{2}+({\rm DCA}_{xy,2}/\sigma_{xy,2})^{2}}{2}},
\label{DCAdefinition}
\end{equation}
where DCA$_{xy,i}$ is the DCA of the electron $i$ in the transverse plane and $\sigma_{xy,i}$ is its resolution estimated from the covariance matrix of the track reconstruction parameters obtained with the Kalman filter technique, similarly as in Ref.~\cite{ALICE:2018fvj}.

\begin{figure}[tbh]
    \begin{center}
    \includegraphics[width = 0.495\textwidth]{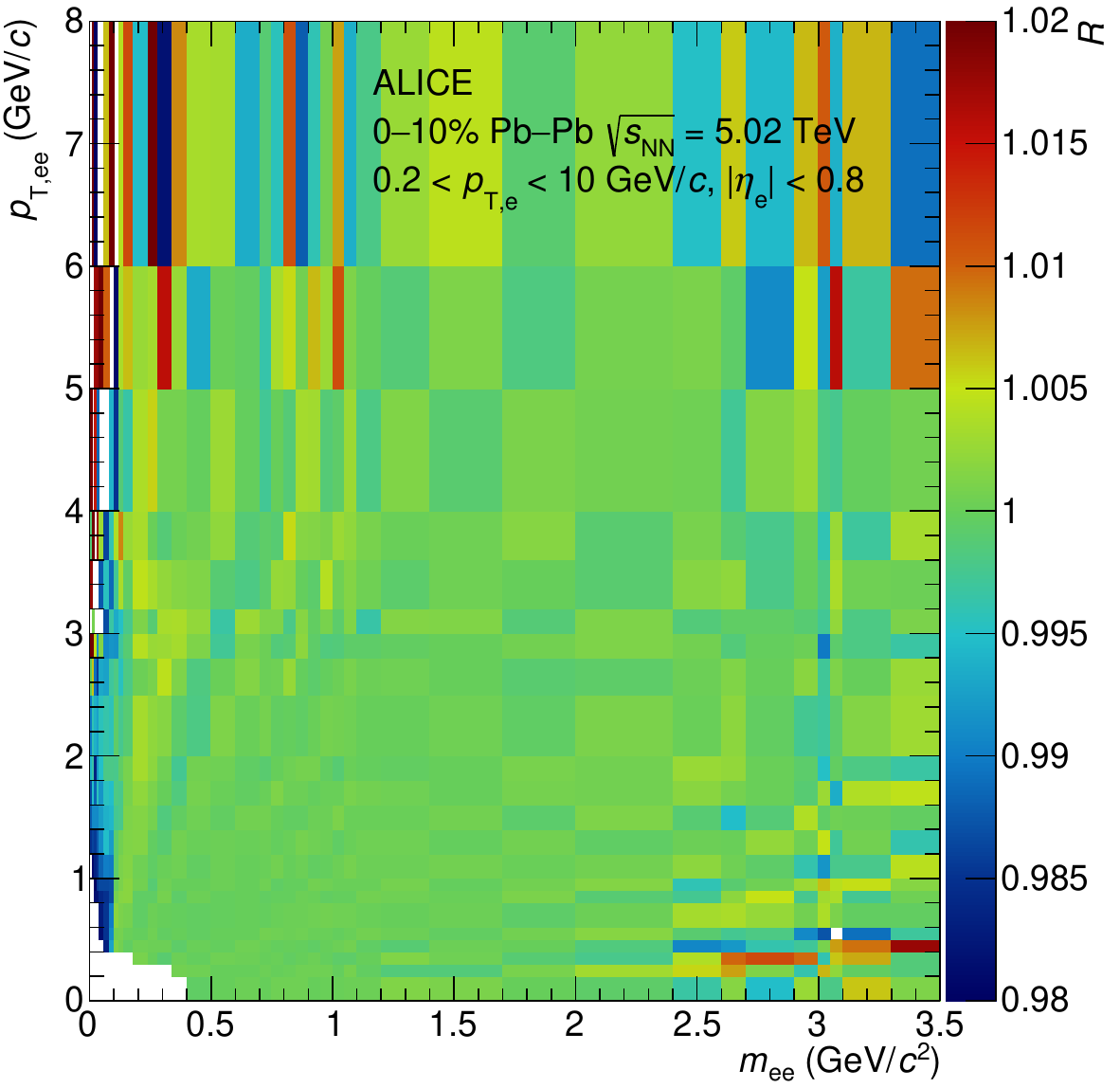}
    \includegraphics[width = 0.495\textwidth]{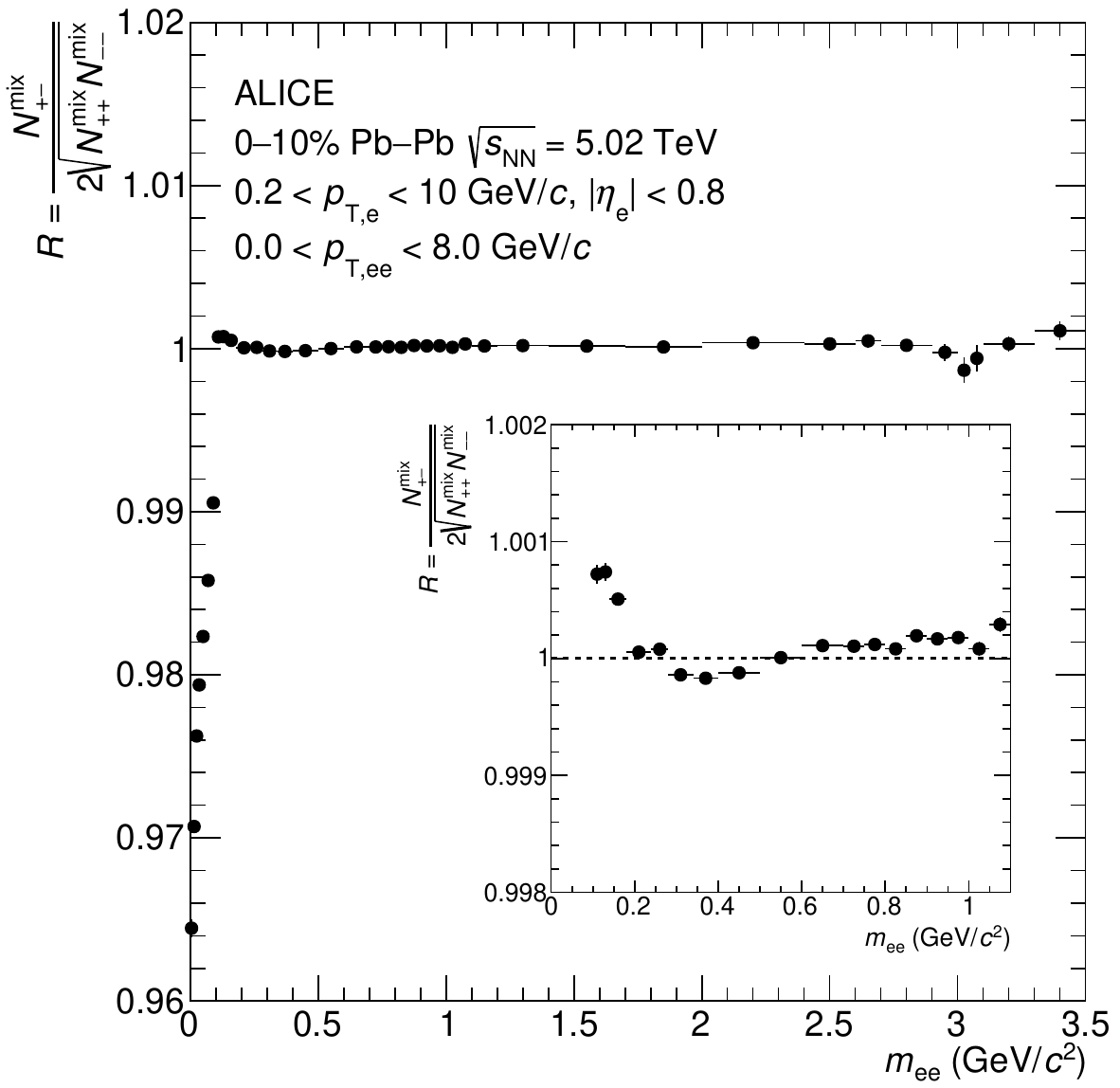}
    \end{center}
    \caption{Relative acceptance correction factor $R$ as a function of \mee and \ptee (left panel) and as a function of \mee  integrated over \ptee (right panel) in the 10\% most central \PbPb collisions at \fivenn. Statistical uncertainties are represented by vertical bars.}
    \label{fig:Rfactor}
\end{figure}

The combinatorial background ($B$) is estimated from the distribution of same-sign pairs from the same event~\cite{ALICE:2018fvj,ALICE:2018ael}. This approach has the advantage of subtracting at the same time correlated and uncorrelated background arising from charge-symmetric processes such as jet fragmentation or conversions of decay photons from the same mother particle that are present in $B$~\cite{PHENIX:2009gyd}. The geometric mean of same-sign pairs from the same event $\sqrt{N^{\rm same}_{\rm ++}N^{\rm same}_{\rm --}}$ needs to be corrected for the different acceptance for opposite-sign and same-sign pairs.  The correction factor $R$ is calculated as $R=N^{\rm mix}_{\rm +-}/(2\sqrt{N^{\rm mix}_{\rm ++}N^{\rm mix}_{\rm --}})$ using uncorrelated opposite- ($N^{\rm mix}_{\rm +-}$) and same-sign ($N^{\rm mix}_{\rm ++}$ and $N^{\rm mix}_{\rm --}$) pairs from mixed events with similar global properties according to the $z$-position of the reconstructed primary vertex (ten intervals between -10 and 10\,cm), the centrality of the collision (two intervals between 0 and 10\% centrality), the event-plane angle estimated with the TPC detector (four intervals), and the polarity of the magnetic field in the central barrel. Thanks to the full coverage of the ALICE central barrel in azimuth such acceptance differences are small. The correction factor $R$ is presented as a function of \mee and \ptee in the left panel of Fig.~\ref{fig:Rfactor}. The deviation of $R$ from unity is driven by the sector structure of the TPC readout plane. The observed characteristic pattern arises from modulations of the relative azimuthal acceptance as a function of the opening angle of the pairs, and its dependence on the electron transverse momenta. In the right panel of Fig.~\ref{fig:Rfactor} the correction factor is shown as a function of \mee integrated over \ptee. Around $\mee = 0.5$\,\GeVmass, $|R-1|$ reaches values of about 0.0002, which is more than one order of magnitude smaller than what is observed in STAR~\cite{STAR:2015tnn}, demonstrating the good tracking and electron identification capabilities of the ALICE central barrel. The acceptance correction is applied differentially in \mee, \ptee, and DCA$_{\rm ee}$ to obtain the final combinatorial background $B = 2R\sqrt{N^{\rm same}_{\rm ++}N^{\rm same}_{\rm --}}$.

\begin{figure}[tbh]
    \begin{center}
    \includegraphics[width = 0.495\textwidth]{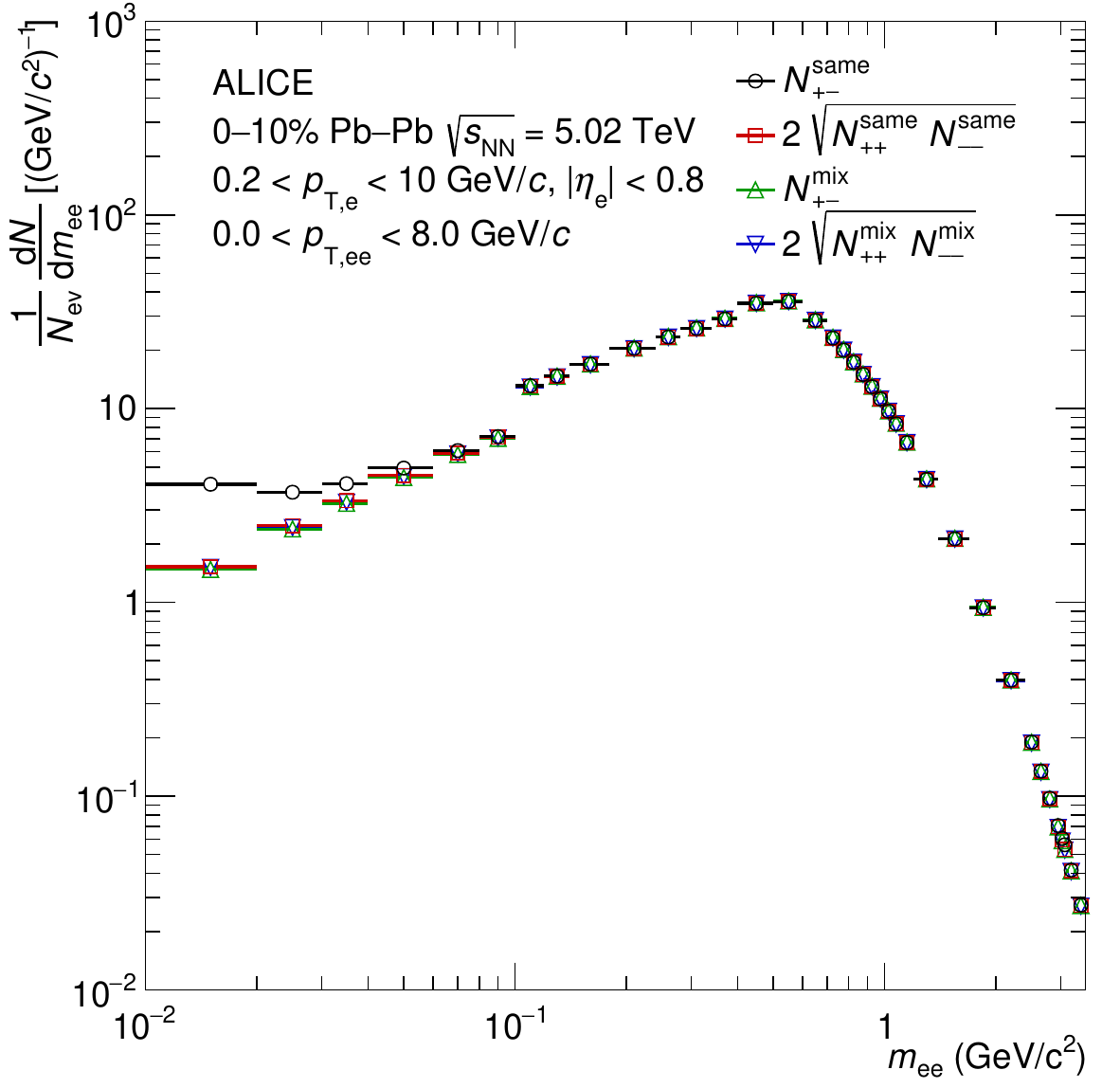}
    \includegraphics[width = 0.495\textwidth]{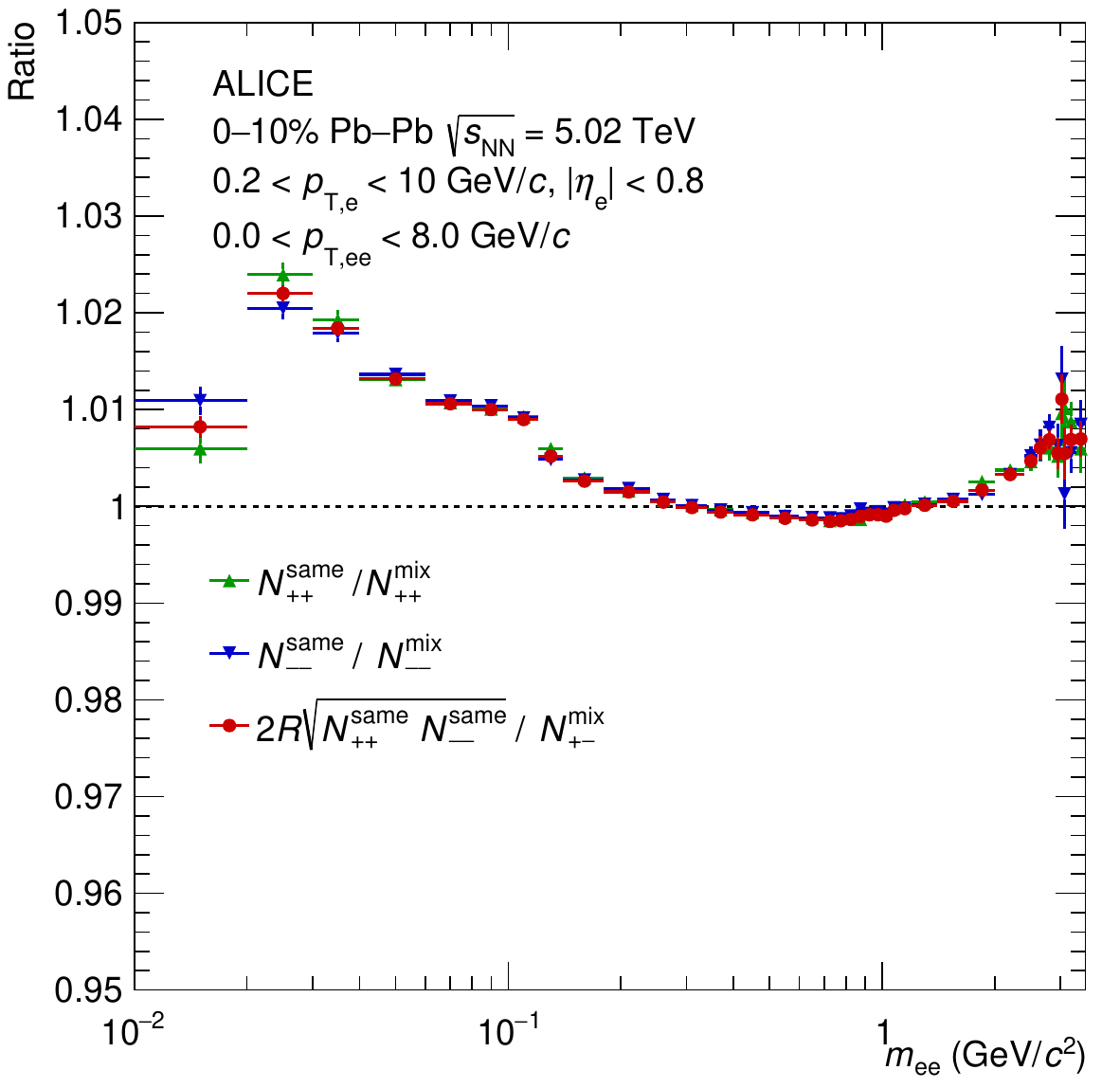}
    \end{center}
    \caption{Left: raw mass distributions of same-sign and opposite-sign pairs from same events and mixed events in the 10\% most central \PbPb collisions at \fivenn. Right panel: corresponding ratios of same-event and mixed-event same-sign distributions, as well as estimations of the full and uncorrelated combinatorial backgrounds. Statistical uncertainties are represented by vertical bars.}
    \label{fig:mixedevents}
\end{figure}

Alternatively, the distribution of opposite-sign pairs from mixed events provides the shape of the uncorrelated combinatorial background. Following the approach explained in Ref.~\cite{PHENIX:2009gyd}, the normalization factor can be estimated by comparing the same-sign pair spectra from same events to the ones from mixed events. In kinematic regions where the contributions of correlated pairs is negligible in the same-sign sample, the distributions are expected to match. The choice of such a kinematic region is susceptible to systematic biases. In the following, the same-sign distributions from mixed events are normalized to the integrals of the corresponding spectra from same events in the mass range 0-3.5\,\GeVmass. In the left panel of Fig.~\ref{fig:mixedevents} the normalized raw mass distributions of mixed-event opposite-sign and same-sign pairs are shown together with the same-event raw yields. The ratios of the same-sign distributions, depicted on the right panel of Fig.~\ref{fig:mixedevents}, are not flat as a function of \mee, pointing to correlated pairs in same events. This makes the choice of a normalization region particularly difficult. Finally, the ratio of the full combinatorial background estimated with same-sign pairs from same events ($2R\sqrt{N^{\rm same}_{\rm ++}N^{\rm same}_{\rm --}}$) and the uncorrelated combinatorial background evaluated with opposite-sign pairs from mixed events ($N_{\rm \pm}^{\rm mix}$) is plotted in the right panel of Fig.~\ref{fig:mixedevents} as a function of \mee. At low mass ($m_{\rm ee} < 0.5$\,\GeVmass), the difference from unity is expected from the contributions of pairs originating from the decays of correlated mesons produced in the same jet and from the contributions of cross-pairs. The latter are created when there are more than one $\rm{e^{+}e^{-}}$ pairs in the final state of a single meson decay, e.g. in a $\pi^{0}$-Dalitz decay followed by a photon conversion $\pi^{0} \to \rm{e^{+}e^{-}\gamma} \to e^{+}e^{-}e^{+}e^{-}$. In the high mass region, the ratio of the full background to the uncorrelated background is predicted to increase due to electrons generated in the same jet or in back-to-back jets. At LHC energies, same-sign pairs originating from $\rm{b\bar{b}}$ pairs produced in one hard-scattering are also expected to play a role. Therefore, mixed events are not directly used to calculate the combinatorial background in this analysis, but only employed to compute the acceptance difference between opposite-sign and same-sign pairs.

\subsubsection{Signal extraction}

\begin{figure}[tbh]
    \begin{center}  
     \includegraphics[width = 0.495\textwidth]{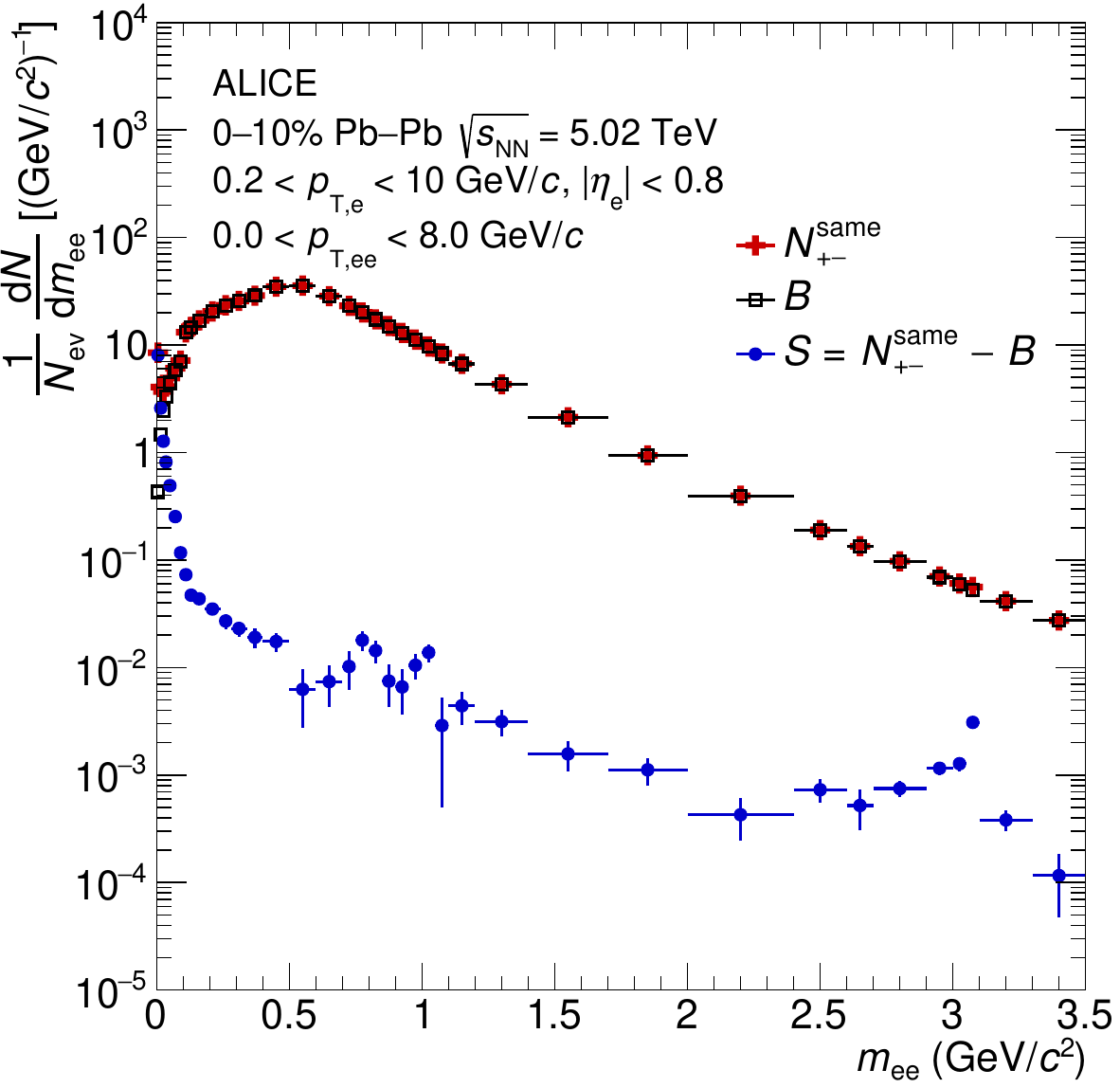}
    \includegraphics[width = 0.495\textwidth]{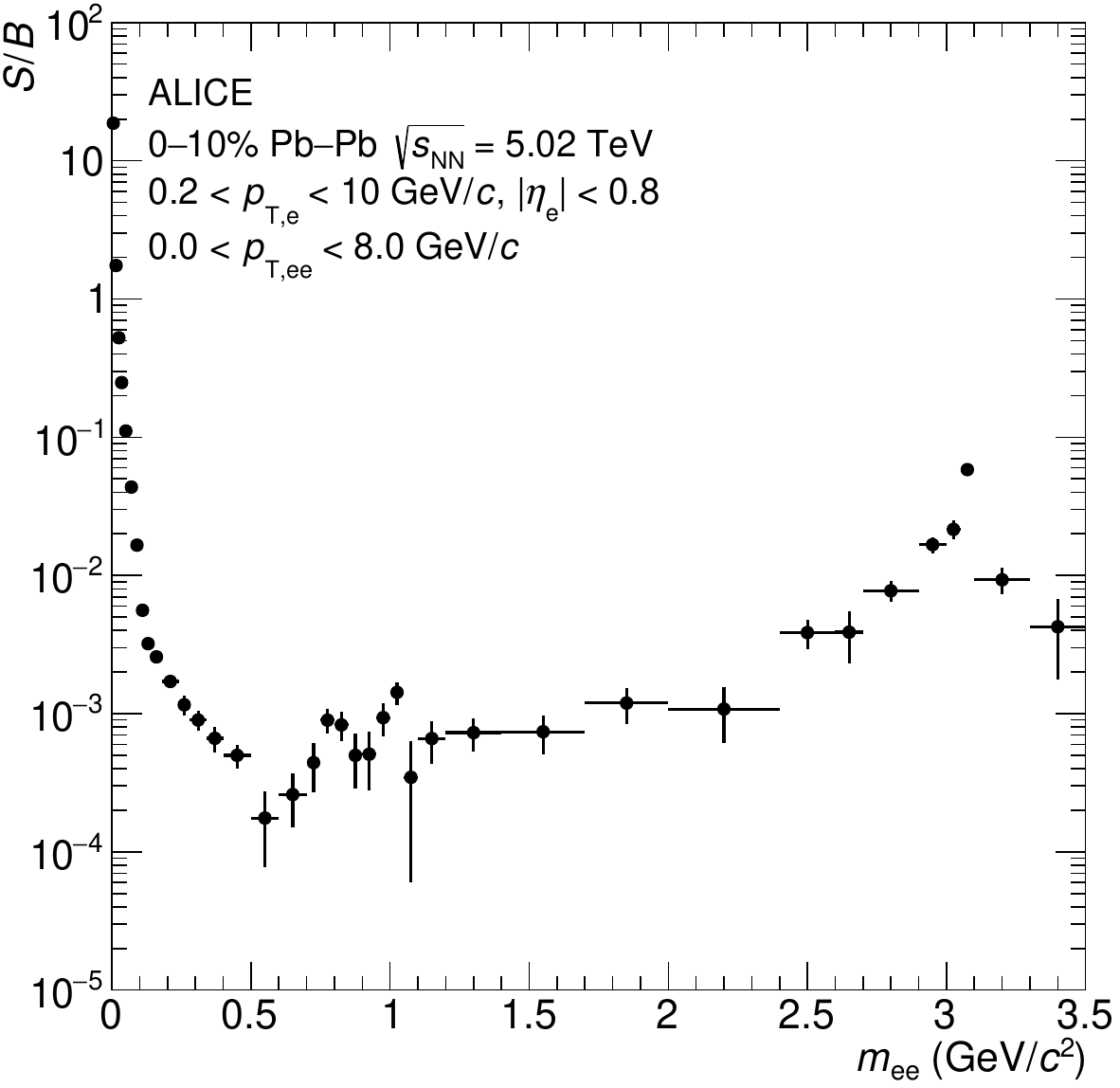}
    \end{center}
    \caption{Raw yield ($S$) overlaid with the opposite-sign pair distribution and the combinatorial background (left panel), as well as signal-over-background ratio (right panel), in the 10\% most central \PbPb collisions at \mbox{\fivenn}.}
    \label{fig:rawyield}
\end{figure}

The raw dielectron yield is obtained as $S=N^{\rm same}_{\rm +-}-2R\sqrt{N^{\rm same}_{\rm ++}N^{\rm same}_{\rm --}}$. Additional photon conversion rejection is achieved at low \mee ($\mee < 0.1$~\GeVmass) by removing pairs based on their characteristic orientation relative to the magnetic field quantified by the so-called $\varphi_{\rm V}$ angle defined in Ref.~\cite{ALICE:2018ael}. The remaining contribution of dielectrons from photon conversions is found to be negligible, below 1\%. The opposite-sign pair spectrum, the combinatorial background and the extracted raw dielectron yield are shown on the left panel of Fig.~\ref{fig:rawyield} as a function of \mee in the 10\% most central \PbPb collisions, together with the $S/B$ ratio on the right panel.

\subsection{Yield corrections}

The corrected yield of \ee pairs in the ALICE acceptance ($0.2$ or  $0.4 < \pte < 10$~\GeVc and $|\etae| < 0.8$) before integration over one or the other variables is expressed as
\begin{equation}
\frac{{\rm d^{3}}N_{\rm \ee}}{{\rm d}\mee{\rm d}\ptee{\rm dDCA_{\rm ee}}} = \frac{1}{\Delta\ptee}\frac{1}{\Delta\mee}\frac{1}{\Delta {\rm DCA_{ee}}}\frac{S(\mee,\ptee,{\rm DCA_{\rm ee}})}{\epsilon^{\rm ee}_{\rm rec}(\mee,\ptee)\times N_{\rm events}},
\end{equation}

where $\Delta \ptee$, $\Delta \mee$ and $\Delta \rm DCA_{ee}$ are the width of the \ptee, \mee and DCA$_{\rm ee}$ intervals, respectively. The pair efficiency $\epsilon^{\rm ee}_{\rm rec}(\mee,\ptee)$, including all tracking and PID selection criteria, is calculated using detailed MC simulations. A realistic detector response is modelled using GEANT 3~\cite{Brun:1119728} with the same detector configurations as in data and the DCA detector resolution is corrected a posteriori to reproduce the one measured in the data. To estimate $\epsilon^{\rm ee}_{\rm rec}(\mee,\ptee)$ with a high statistical precision, samples of dielectron sources are embedded into hadronic collisions simulated with the HIJING event generator~\cite{Wang:1991hta}. Light-flavor hadrons ($\pi^{0}$,$\eta$,$\eta^{\prime}$,$\rho$,$\omega$, and $\phi$) and $J/\psi$ mesons, forced to decay into dielectrons with a phenomenological generator~\cite{ALICE:2018fvj} and PHOTOS~\cite{Golonka:2005pn}, respectively, are used together with an enriched sample of heavy-flavor hadron sources with enforced semileptonic decay channels generated with the Perugia 2011 tune~\cite{Skands:2010ak} of PYTHIA 6.4~\cite{Sjostrand:2006za}. The pair efficiency is computed as explained in Ref.~\cite{ALICE:2022hvk}. Since the reconstruction efficiency for single electron tracks does not show any significant dependence on the electron DCA, for which loose selection criteria are applied, $\epsilon^{\rm ee}_{\rm rec}(\mee,\ptee)$ is applied to the data as a function of \mee and \ptee. The average reconstruction efficiency of a signal \ee pair ranges from 15\% to 25\%.

\subsection{Systematic uncertainties of measured dielectron spectra}

The systematic uncertainties on the measured \ptee- and \mee-differential yields originate from tracking, electron identification and purity, and background subtraction. They are evaluated by simultaneous variations of the tracking and PID selection criteria, as described in Ref.~\cite{ALICE:2018fvj}, implying relative changes of the pair efficiencies and $S/B$ ratio by up to 30\% and 40\%, respectively. In particular, modifying the requirements on the ITS, TPC and TOF PID signals allows for probing possible biases due to differences in the detector responses in data and MC and remaining hadron contamination in the electron sample. The systematic uncertainty is calculated as the root mean square of the variation of the data points. It is found to be larger at low \ptee, up to about 10\% for $\ptee < 2$~\GeVc where the $S/B$ is smaller.

\begin{figure}[tbh]
    \begin{center}
    \includegraphics[width = 0.495\textwidth]    {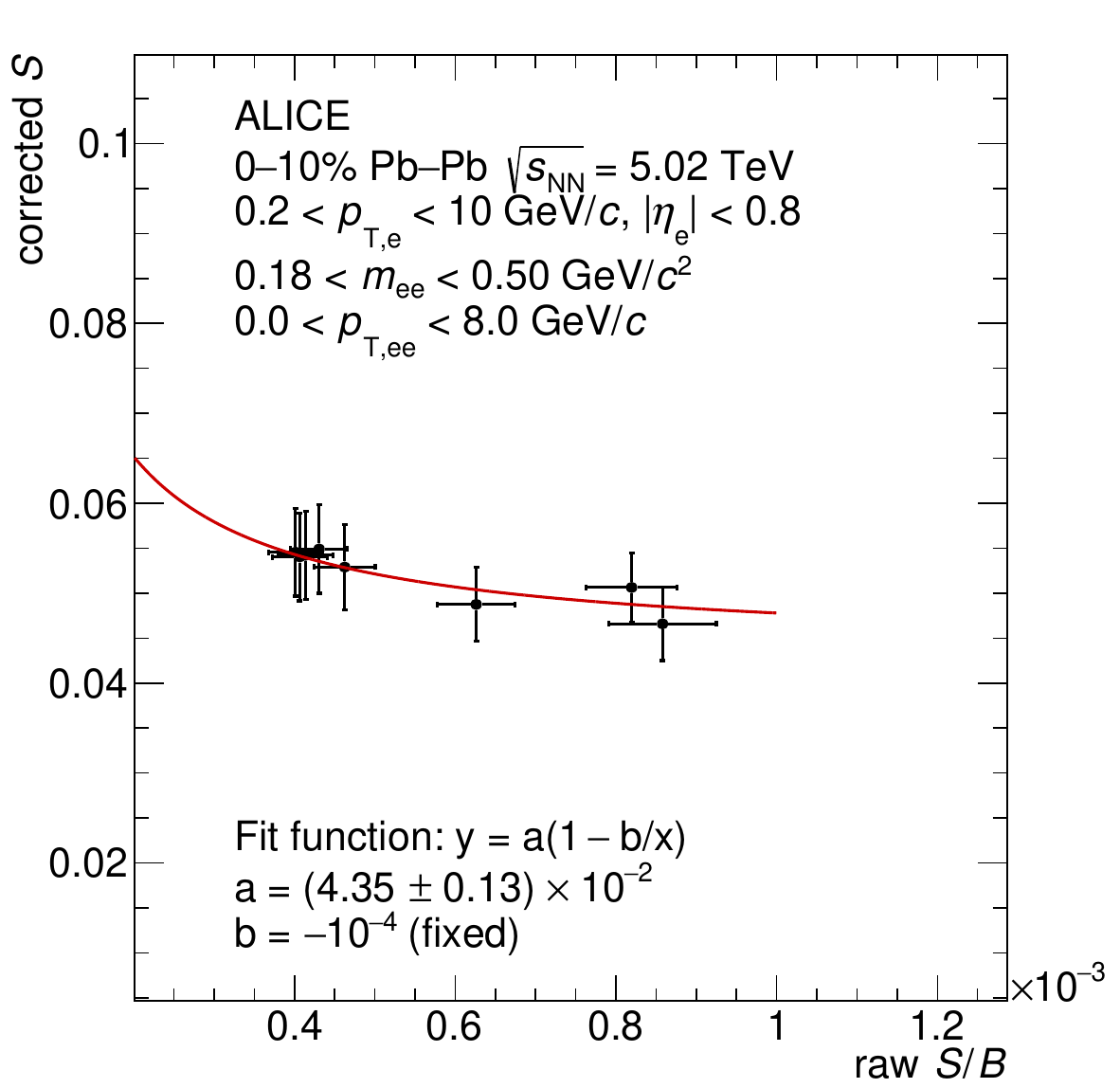}
    \includegraphics[width = 0.495\textwidth]{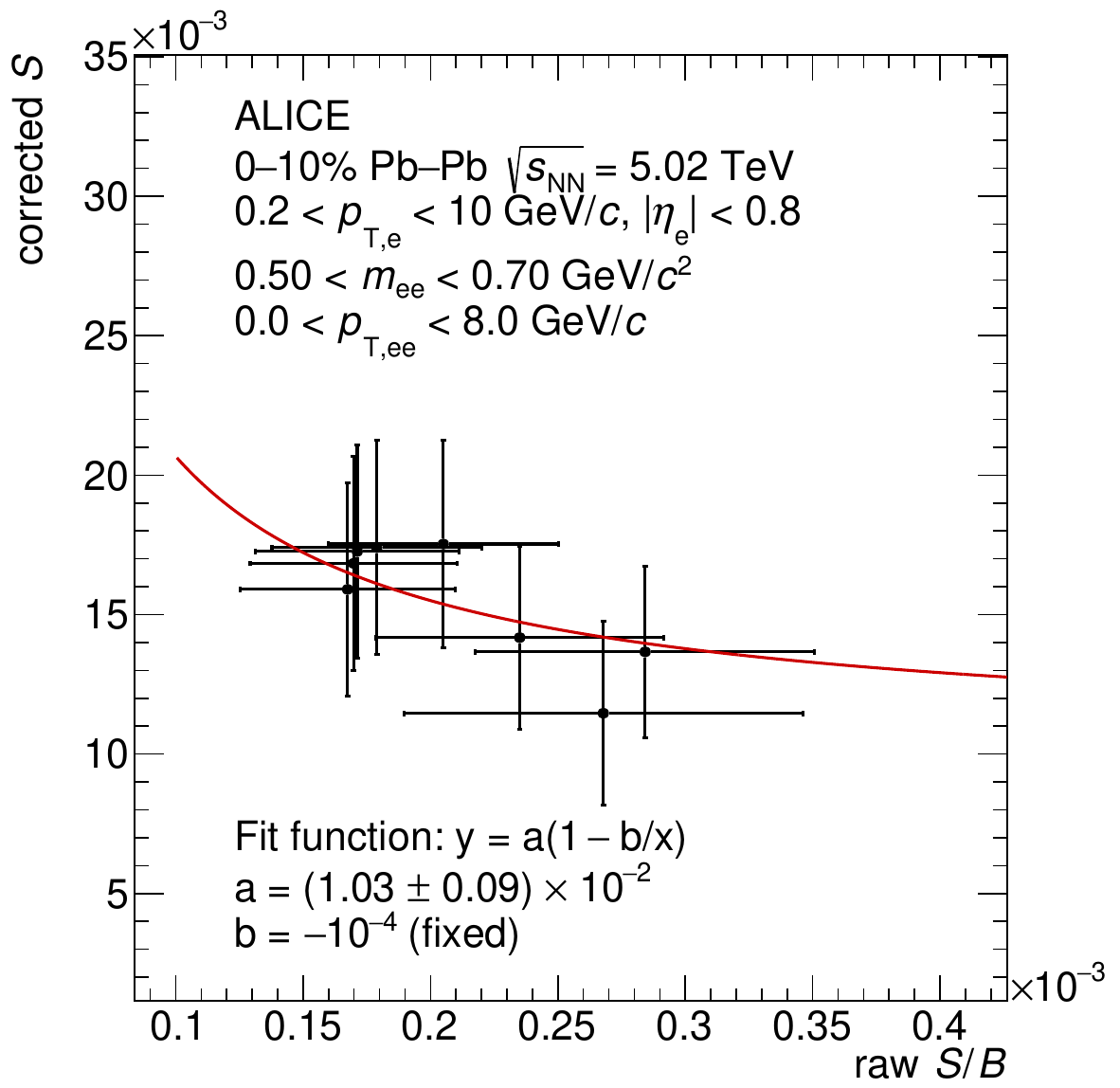}
    \end{center}
    \caption{Corrected dielectron yields as a function of $S/B$ in two different mass intervals, i.e. 0.18-0.5\,\GeVmass (left panel) and 0.5-0.7\,\GeVmass (right panel), for $0.2 < p_{\rm T,e} < 10$\,\GeVc and $|\eta_{\rm e}| < 0.8$ in the 10\% most central Pb--Pb collisions at \fivenn. The red curves show the expected dependence assuming a bias of the estimated background of the order of $\pm 10^{-4}$ ($b = -10^{-4}$ in Eq.~\ref{sysb}).}
    \label{fig:sysbackgrounda}
\end{figure}

\begin{figure}[tbh]
    \begin{center}
    \includegraphics[width = 0.53\textwidth]{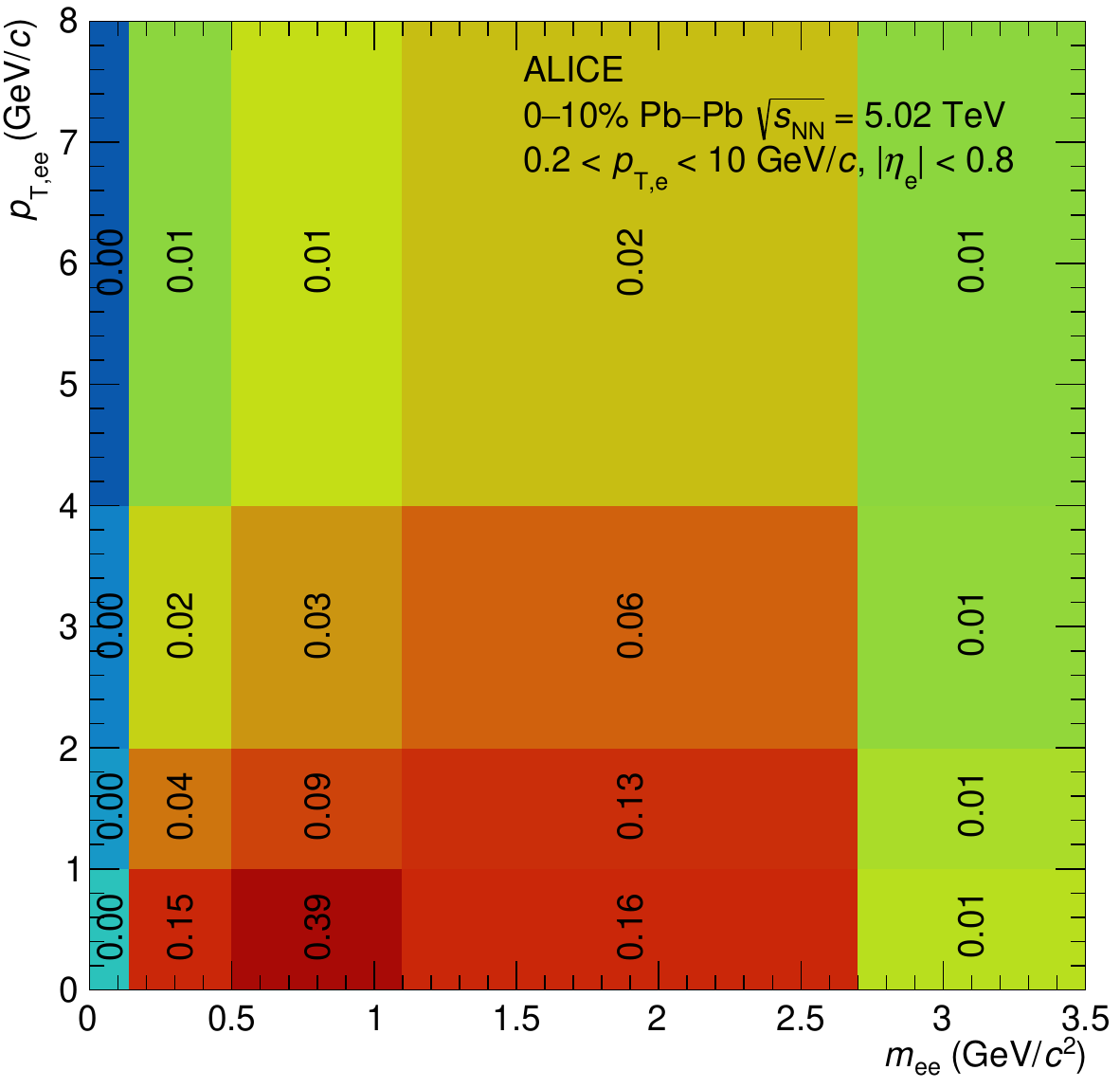}
    \end{center}
    \caption{Relative systematic uncertainty from the background uncertainty of $10^{-4}$ on the corrected dielectron yield in bins of \mee and \ptee for $0.2 < p_{\rm T,e} < 10$\,\GeVc and $|\eta_{\rm e}| < 0.8$ in the 10\% most central Pb--Pb collisions at \fivenn.}
    \label{fig:sysbackgroundb}
\end{figure}

To take into account a possible bias in the estimation of the combinatorial background, an additional uncertainty is considered. The dependence of the corrected yields, obtained by varying the selection criteria, on the $S/B$ ratios is shown for two different mass intervals in Fig.~\ref{fig:sysbackgrounda}. The data indicate a possible decreasing trend with $S/B$. In a simple approximation, a small bias $b$ of the background is assumed, i.e. 

\begin{equation}
  B’ = (1+b)\times B,
\end{equation}

leading to a bias of the extracted signal $S’$:

\begin{equation}
    S’ = S + B - B’
\end{equation}

\begin{equation}
S’ = S\times (1 - b \times B/S) 
\label{sysb}
\end{equation}

where $S$ and $B$ are the true signal and the true background, whereas $S’$ and $B'$ are the extracted signal and the estimated background. The data are compatible with a bias $b$ of the order of $10^{-4}$, as can be seen with the red curves plotted in Fig.~\ref{fig:sysbackgrounda}. This corresponds to about 50\% of the maximum acceptance correction applied to the same-sign pair distribution $\sqrt{N^{\rm same}_{\rm ++}N^{\rm same}_{\rm --}}$ used to estimate $B$, in the region where the $S/B$ is small ($\mee > 0.2$\,\GeVmass). Such uncertainty leads to the \mee and \ptee dependent systematic uncertainty shown in Fig.~\ref{fig:sysbackgroundb} for $0.2 < p_{\rm T,e} < 10$\,\GeVc obtained by $10^{-4} \times B/S$. It is driven by the $S/B$ ratio and ranging from 0 to 39\%. For the DCA$_{\rm ee}$-differential analysis, for which the electron candidates must fulfill $p_{\rm T} > 0.4$\,\GeVc, the systematic uncertainty is smaller due to the larger $S/B$.

Similar to Ref.~\cite{ALICE:2022hvk}, additional uncertainties related to the conversion rejection criterion ($\varphi_{\rm V}$), the criteria on the number of shared ITS clusters, and the requirement of a hit in the first ITS layer, as well as to the TPC--ITS and TPC--TOF matching efficiencies, are added in quadrature. The first two sources of systematic uncertainties are estimated by varying the selection criteria, whereas the others are determined for single electrons and propagated to the \ee pairs using a MC method. The resulting systematic uncertainties are listed in Table~\ref{systable}.

\begin{figure}[tbh]
    \begin{center}
    \includegraphics[width = 0.495\textwidth]{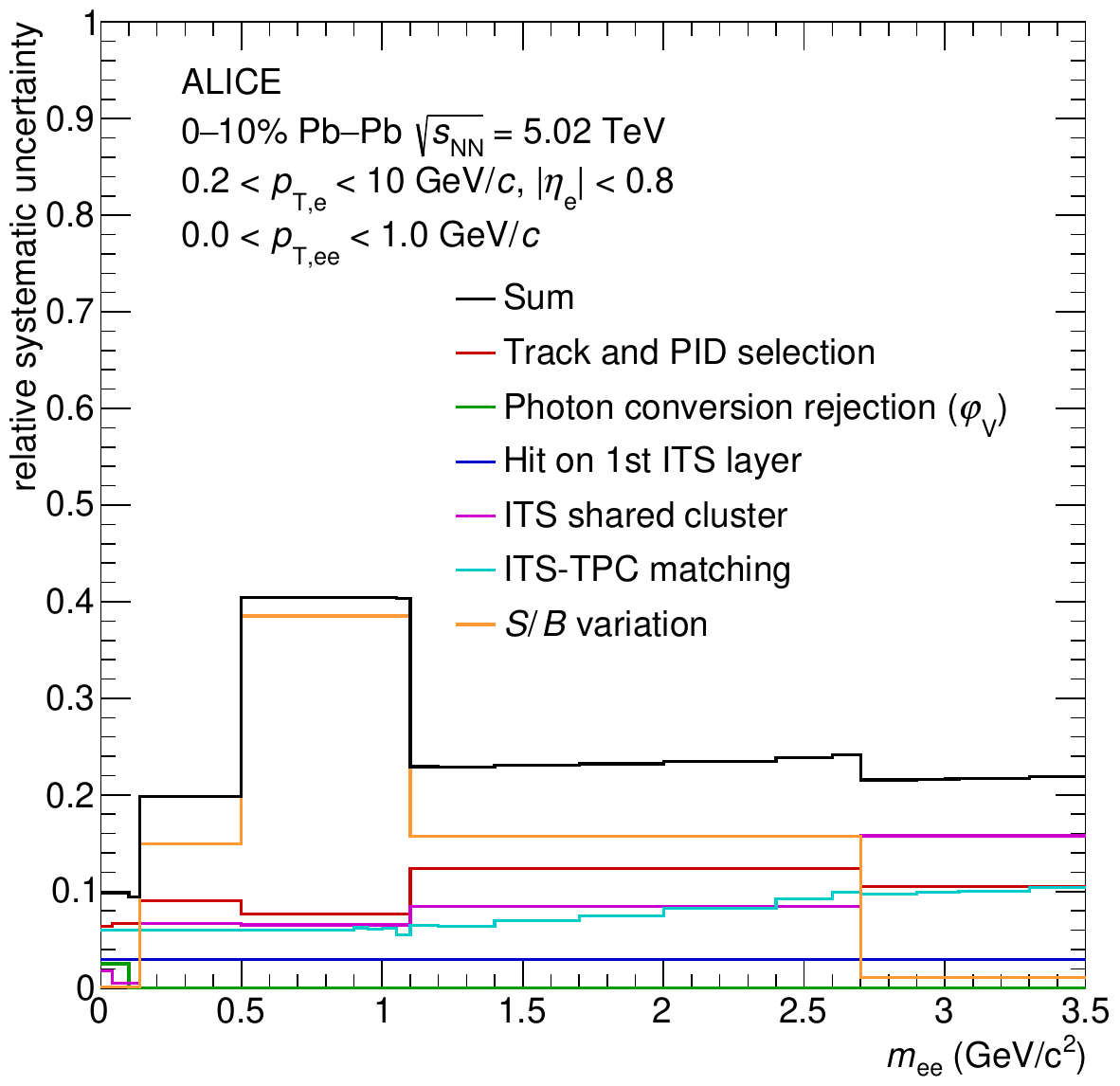}
    \includegraphics[width = 0.495\textwidth]{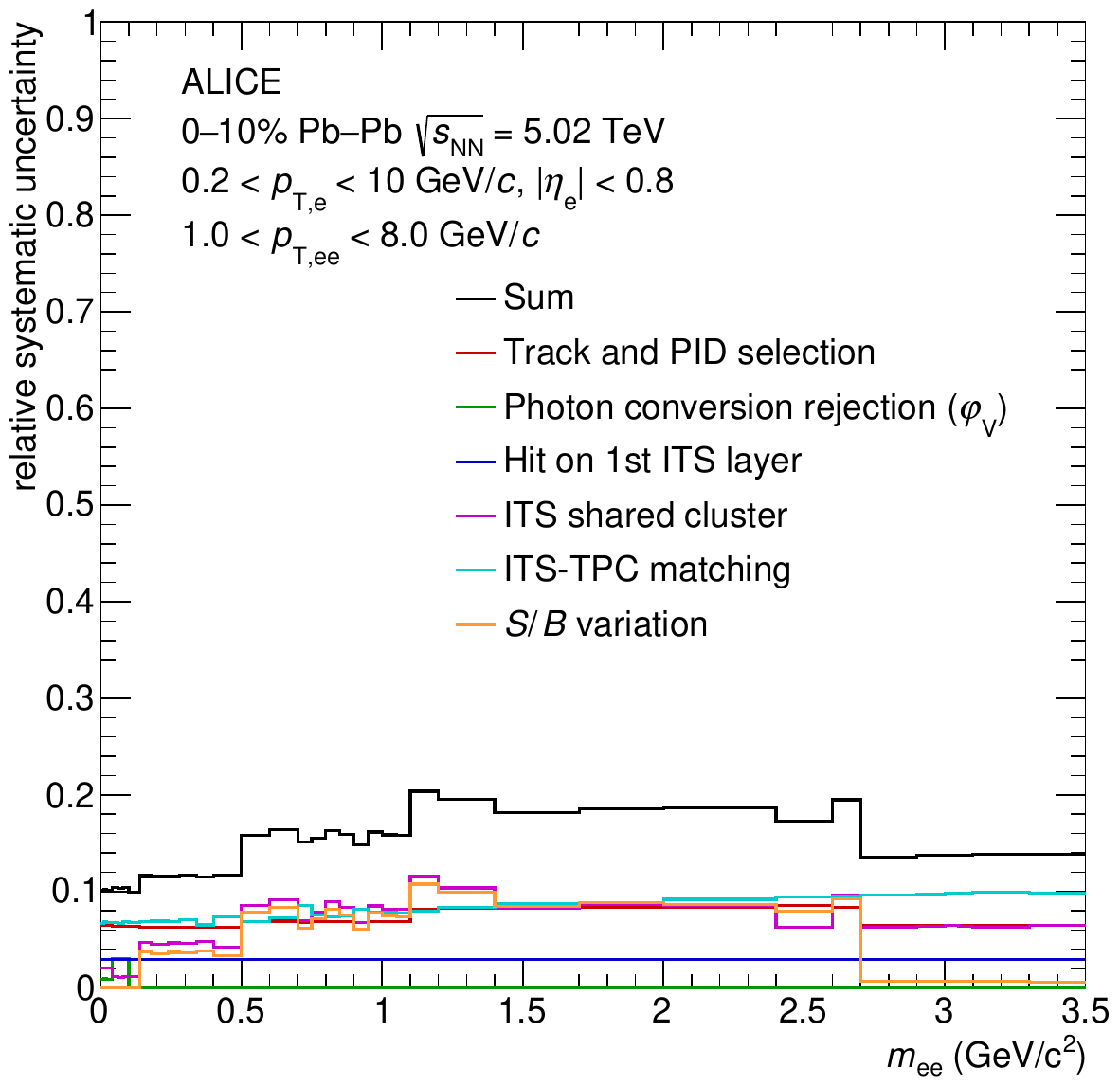}
    \end{center}
    \caption{Summary of the systematic uncertainties of the measured dielectron yields for $0.2 < p_{\rm T,e} < 10$\,\GeVc and $|\eta_{\rm e}| < 0.8$ as a function of \mee in two different \ptee intervals 0-1\,\GeVc (left panel) and 1-8\,\GeVc (right panel) for the 10\% most central Pb--Pb collisions at \fivenn.}
    \label{fig:sysbackgroundc}
\end{figure}

Finally, an additional source of systematic uncertainty is considered for the DCA$_{\rm ee}$-differential analysis. To take into account possible systematic effects related to the correlation between \ptee and DCA$_{\rm ee}$, the latter is studied in each mass region under interest. The difference of the pair efficiency at the maximum and minimum mean \ptee, seen as a function of DCA$_{\rm ee}$, in a given mass region is used to assign a systematic uncertainty~\cite{ALICE:2018fvj}. It is found to be of the order of 3\%. The total systematic uncertainty is given in Table~\ref{systable} for the \ptee and \mee analysis, as well as the DCA$_{\rm ee}$ analysis. Moreover, the dependence of the different sources of systematic uncertainties on \mee is shown in two different \ptee intervals 0-1\,\GeVc and 1-8\,\GeVc in Fig.~\ref{fig:sysbackgroundc} for the \ptee and \mee analysis with $0.2 < p_{\rm T,e} < 10$\,\GeVc and $|\eta_{\rm e}| < 0.8$.

\begin{table}[ht!]
\begin{center}
\centering
\caption{Summary of the total systematic uncertainties of the measured dielectron yields in the 10\% most central \PbPb collisions at \fivenn. The values presented as a range correspond to the smallest and largest observed systematic uncertainties.}
\begin{tabular}{lll}
 \hline 
  Analysis    & (\mee,\ptee)  & DCA$_{\rm ee}$  \\
/sources  &   &    \\
\hline
Tracking and PID selection  & 5--11\% & 4--11\% \\\hline
Combinatorial background  & 0--39\% & 1--10\% \\\hline
Conversion rejection  & 2--4\% & N/A \\\hline  
Number of shared ITS clusters   & 4--12\% & 4--12\% \\\hline
Hit in the first ITS layer  & 3\% & 3\% \\\hline
TPC--ITS matching  & 8--12\% & 8--12\% \\\hline
TPC--TOF matching & 0--3\% & 0--3\% \\\hline
DCA$_{\rm ee}$--\ptee correlation & N/A & 3\% \\\hline
Total & 15--40\% & 17--22\% \\
\hline 
\end{tabular}
\label{systable}
\end{center}
\end{table}

\section{Estimation of hadronic sources}
\label{hadroniccocktail}

The dielectron measurement is compared with the sum of expected contributions from light- and heavy-flavor hadron decays within the kinematic range under study, i.e.~the hadronic cocktail. They are estimated with simulations including the angular, momentum and DCA resolution of the detector for the given data-taking period, as well as bremsstrahlung effects, which are not corrected for in the data analysis~\cite{ALICE:2017tau,ALICE:2018fvj}.

\subsection{Dielectron yield from hadronic decays as a function of \texorpdfstring{\mee}{mee} and \texorpdfstring{\ptee}{ptee}}

\subsubsection{Decays of light-flavor hadrons and \texorpdfstring{$J/\psi$}{Jpsi} mesons}
\label{cocktailinvariantmasslight}

\begin{figure}[tbh]
    \begin{center}  
     \includegraphics[width = 0.495\textwidth]{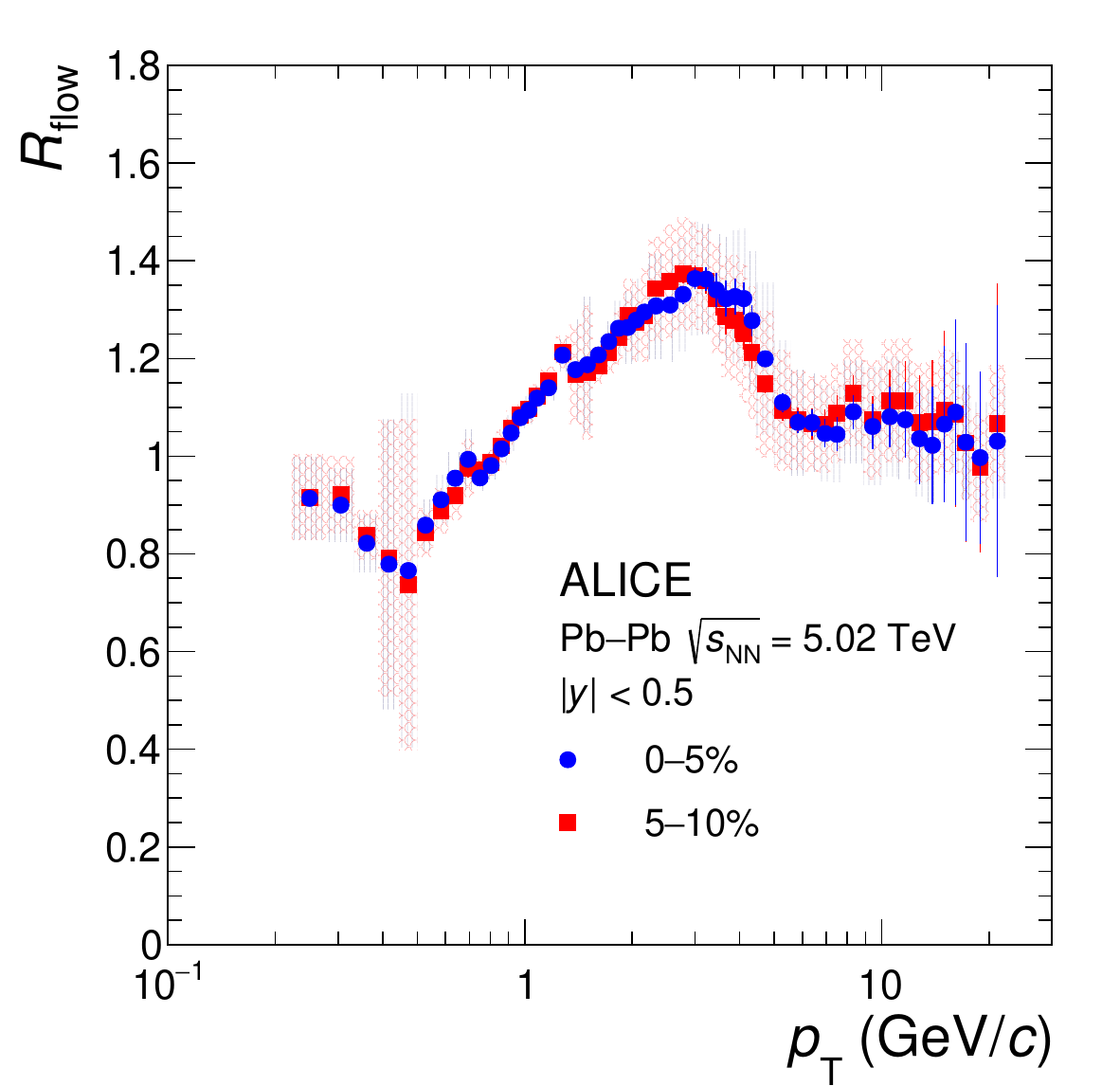}
    \includegraphics[width = 0.495\textwidth]{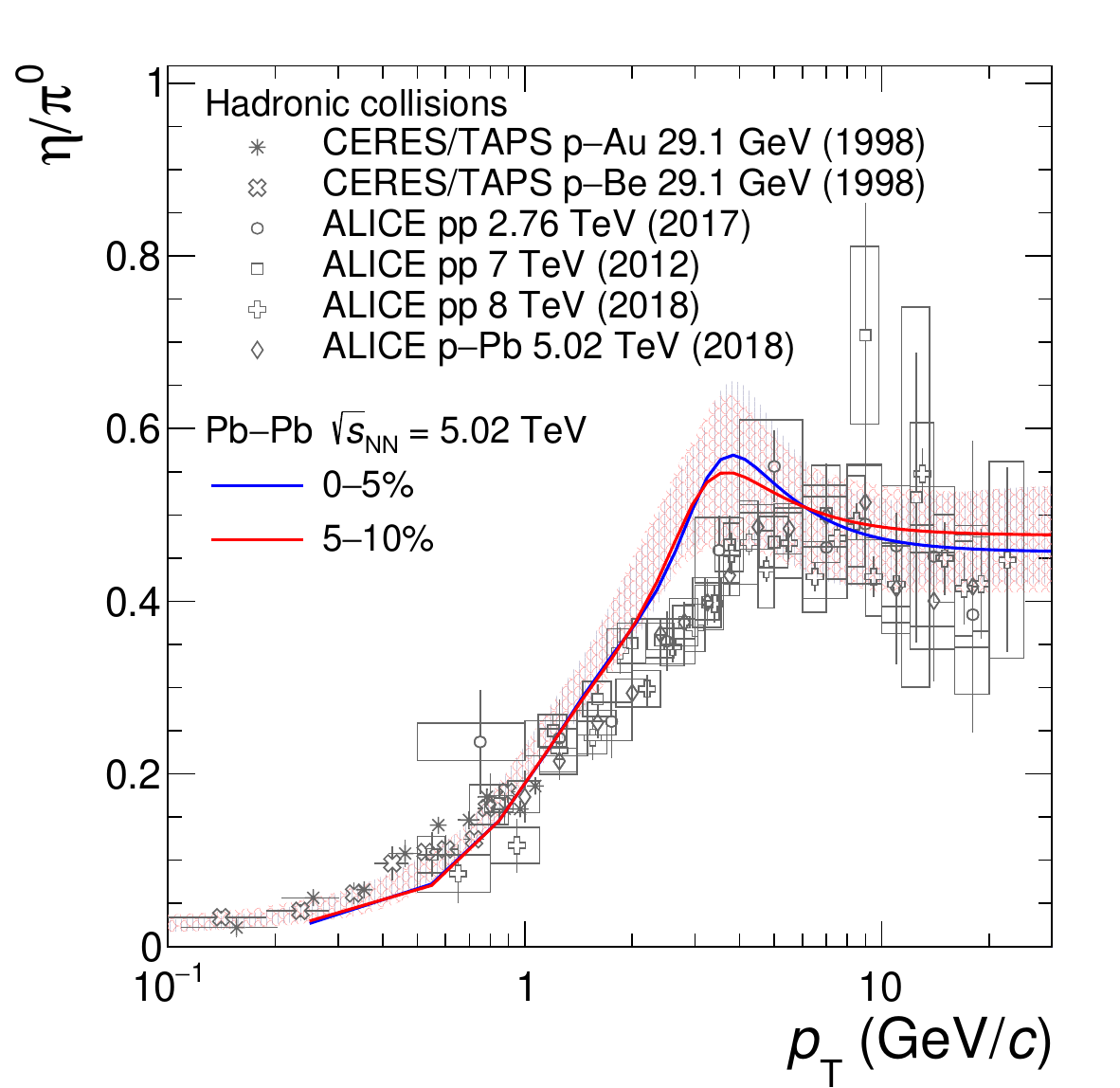}
    \end{center}
    \caption{Left panel: double ratio of the measured ${\rm K^{\pm}/\pi^{\pm}}$ \pt spectra in central \PbPb collisions and in pp collisions at \fivenn~\cite{ALICE:2019hno} dilated as a function of \pt by the ratio of the $\eta$ and $\rm K^{\pm}$ meson mass (see text). Right panel: measured~\cite{Agakichiev:1998ign,ALICE:2012wos,ALICE:2017ryd,ALICE:2018vhm,ALICE:2017nce} and parameterized $\frac{\eta}{\pi^{0}}$ ratios in hadronic collisions and central \PbPb collisions, respectively.}
    \label{fig:etapi0}
\end{figure}

The expected dielectron yield from the decays of known light-flavor ($\pi^{0}$, $\eta$, $\eta^{\prime}$, $\omega$, $\rho$, $\phi$) and $J/\psi$ hadrons is calculated as described in Ref.~\cite{ALICE:2022hvk}. Parameterizations of the \pt-differential yields of the different hadrons are taken as input to a fast Monte Carlo simulation, that performs the decays and allows retrieving the corresponding expected dielectron yields~\cite{ALICE:2018fvj}. Since the $\pi^{\pm}$ measurement extends down to $\pt = 0.1$~\GeVc and exhibits small uncertainties~\cite{ALICE:2019hno}, 
charged pions are used to approximate neutral pions. 
An additional 10\% systematic uncertainty is considered to take into account differences due to isospin-violating decays as described in Refs.~\cite{ALICE:2020umb,ALICE:2022hvk}. The \pt spectrum of $\eta$ is computed using the same approach as in Ref.~\cite{Ren:2021pzi}. First, the $\eta/\pi^{0}$ ratios measured as a function of \pt in hadronic collisions by ALICE~\cite{ALICE:2012wos,ALICE:2017ryd,ALICE:2018vhm,ALICE:2017nce} and CERES/TAPS~\cite{Agakichiev:1998ign} at different energies are parameterized to obtain a pp reference. Second, radial flow effects in \PbPb collisions are evaluated with the double ratio of the ${\rm K^{\pm}/\pi^{\pm}}$ \pt spectra measured at midrapidity ($|y| < 0.5$) in the 10\% most central \PbPb collisions and in pp collisions~\cite{ALICE:2019hno}. Due to their similar masses, ${\rm K^{\pm}}$ and $\eta$ are expected to be affected by radial flow in a similar way. The corresponding $\eta/\pi^{0}$ ratio in central \PbPb collisions is computed as
\begin{equation}
\left(\frac{\eta}{\pi^{0}}\right)_{\rm \scriptsize PbPb} = \left( \frac{\eta}{\pi^{0}} \right)_{\rm \pp} \times R_{\rm flow} = \left( \frac{\eta}{\pi^{0}} \right)_{\rm \pp} \times \frac{ \left( \frac{{\rm K^{\pm}}}{\pi^{\pm}} \right)_{\rm \scriptsize PbPb}^{*}} {\left( \frac{{\rm K^{\pm}}}{\pi^{\pm}} \right)_{\rm \pp}^{*}},
\end{equation}
where $^{*}$ means that the value is not taken at $p_{\rm T}$ but $m_{\eta}/m_{\rm K^{\pm}} \times p_{\rm T}$ to take into account the small difference in mass between $\eta$ and $\rm K^{\pm}$ of about 10\%. In the left panel of Fig.~\ref{fig:etapi0}, the correction factor $R_{\rm flow}$ is presented for the 0--5\% and 5--10\% centrality intervals based on the $\rm K^{\pm}$ and $\pi^{\pm}$ measurements~\cite{ALICE:2019hno}, considering the systematic uncertainties of the published ${\rm K^{\pm}/\pi^{\pm}}$ ratios as uncorrelated between the \PbPb and pp colliding systems. For $\pt > 9$~\GeVc $R_{\rm flow}$ is consistent with unity and an universal constant behavior in hadronic and heavy-ion collisions is assumed. The resulting $\eta/\pi^{0}$ ratio in central \PbPb collisions is parameterized as a function of $\pt$ for the two centrality classes separately, as shown in the right panel of Fig.~\ref{fig:etapi0} together with the measured $\eta/\pi^{0}$ ratios in hadronic collisions~\cite{Agakichiev:1998ign,ALICE:2012wos,ALICE:2017ryd,ALICE:2018vhm,ALICE:2017nce}. A similar trend is observed in central \PbPb collisions at \twosevensixnn, where the $\eta/\pi^{0}$ ratio was measured by the ALICE collaboration~\cite{ALICE:2018mdl} at midrapidity. The corresponding calculated dielectron yields in both centrality classes are combined together to obtain the expected \ee yield in the 10\% most central \PbPb collisions assuming fully correlated systematic uncertainties of the $\eta/\pi^{0}$ ratios. The $\eta^{\prime}$, $\omega$, and $\rho$ \pt distributions are generated assuming $m_{\rm T}$-scaling, implying that the spectra of all light mesons as a function of $m_{\rm T} = \sqrt{m^{\rm 2}+p_{\rm T}^{\rm 2}}$ follow a universal form and differ only by a normalization factor. Finally, the measured $\phi$~\cite{ALICE:2019xyr} and $J/\psi$~\cite{ALICE:2023hou} \pt-differential yields in central \PbPb collisions are fitted and in case of the $\phi$ extrapolated down to low $\pt$ ($\pt < 0.4$~\GeVc) using $m_{\rm T}$-scaling.

\subsubsection{Open-charm and open-beauty hadron decays}
\label{cocktailinvariantmassheavyflavour}

\begin{figure}[tbh]
    \begin{center}  
     \includegraphics[width = 0.495\textwidth]{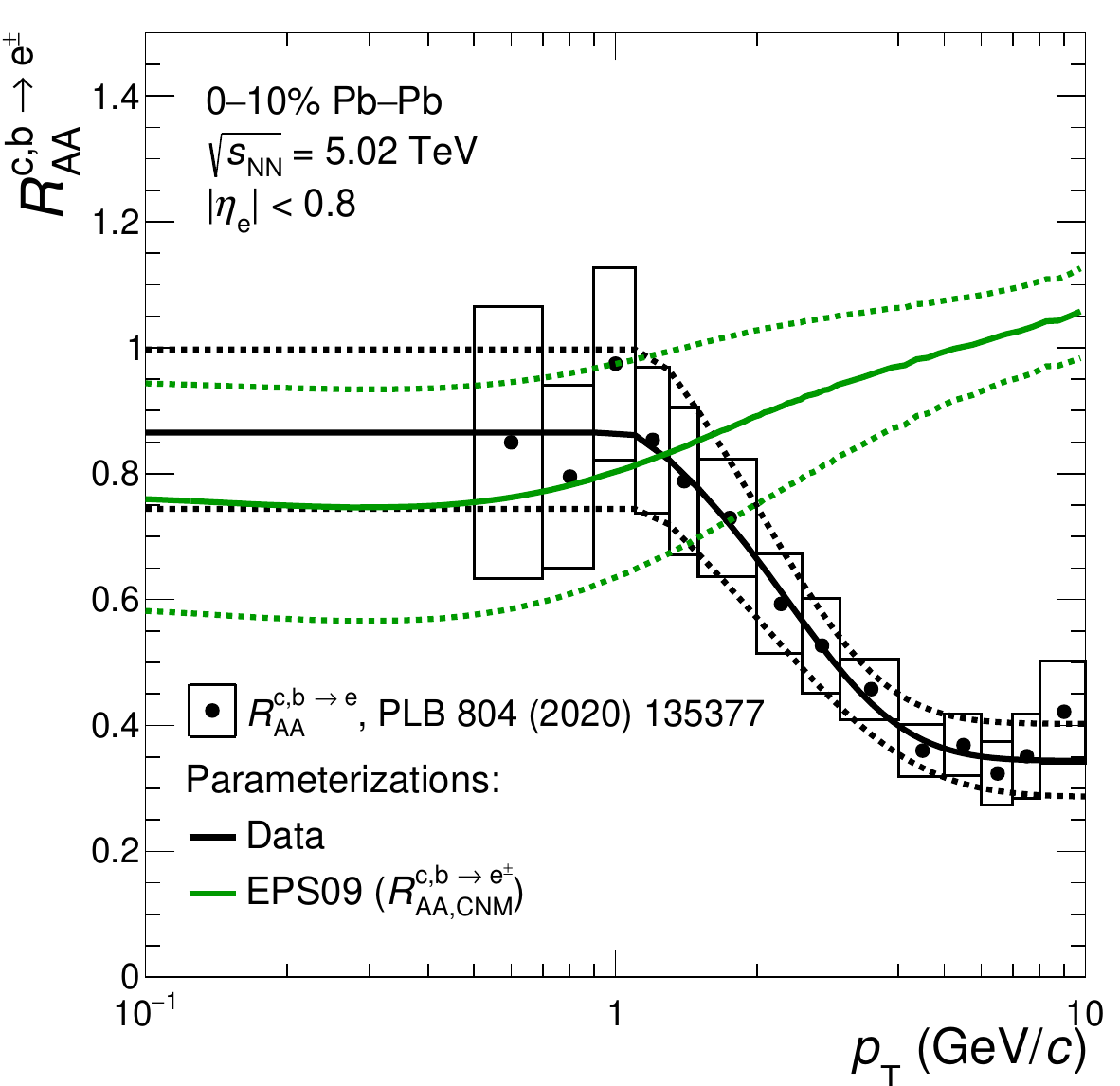}
    \includegraphics[width = 0.495\textwidth]{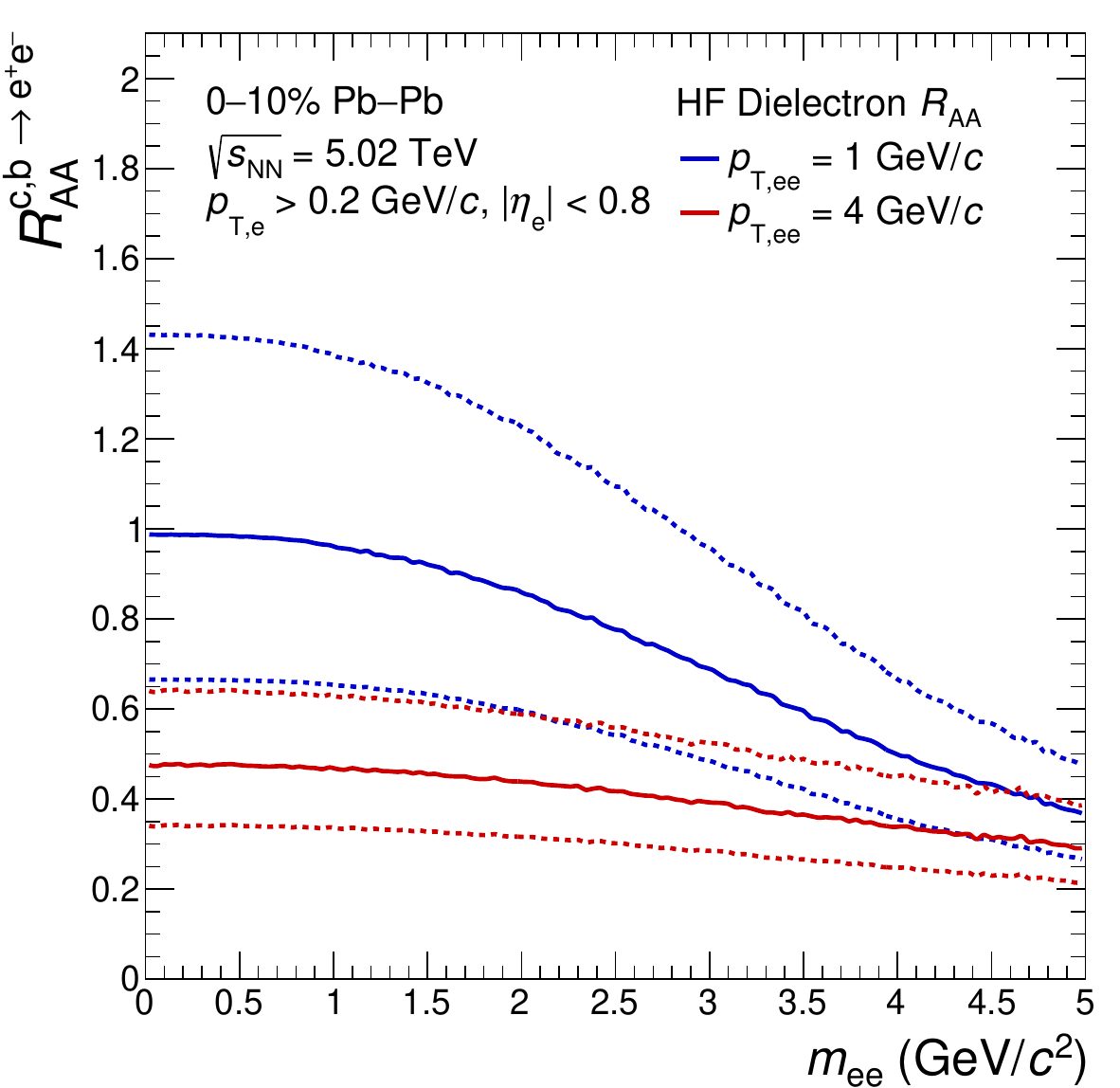}
    \end{center}
    \caption{Left panel: measured nuclear modification factor of single electrons from heavy-flavor hadron decays~\cite{ALICE:2019nuy}, as well as its parameterization and the estimated nuclear modification factor caused by pure cold-nuclear matter effects with EPS09 nPDFs\,\cite{Eskola:2009uj}, in the 10\% most central \PbPb collisions at \fivenn. Right panel: resulting nuclear modification factor of \ee pairs from correlated open heavy-flavor hadron decays computed with a toy MC (see text) for two different \ptee.}
    \label{fig:hferaacocktail}
\end{figure}

Two different calculations for the contributions of correlated semileptonic heavy-flavor (HF) hadron decays are considered. One is based on the dielectron measurements in pp collisions at the same center-of-mass energy per nucleon pair~\cite{ALICE:2020mfy} scaled with the average number of binary nucleon--nucleon collisions $\langle N_{\rm coll} \rangle$ to obtain the $N_{\rm coll}$-scaled HF contributions. The resulting full hadronic cocktail is called in the following Cocktail1. For this purpose, the expected \ee yields of correlated semileptonic heavy-flavor hadron decays, estimated with the POWHEG event generator~\cite{powheg1,powheg2,powheg3,powheg4} using PYTHIA 6~\cite{Sjostrand:2006za} as parton shower, are first fitted to the measured \ee spectra in pp collisions to extract ${\rm d\sigma_{c\overline{c}}}/{\rm d}y|_{{y=0}}$ and ${\rm d\sigma_{b\overline{b}}}/{\rm d}y|_{{y=0}}$, as explained in Ref.~\cite{ALICE:2020mfy}. The calculations are then scaled with the average nuclear overlap function in the 0--10\% centrality interval. The uncertainties originating from the branching ratio of the semileptonic decays of the open heavy-flavor hadrons and the fragmentation functions of charm and beauty quarks are omitted assuming that these do not change from \pp to central \PbPb collisions. This approach ignores shadowing and suppression effects due to the interactions of charm and beauty quarks with other partons in the medium, as well as any modification of the hadronization processes. Measurements of the nuclear modification factor $R^{{\rm c,b\to e^{\pm}}}_{\rm AA}$ of single electrons from charm and beauty hadron decays, however, do show significant modifications of the \pt spectra in \PbPb collisions compared to pp collisions~\cite{ALICE:2019nuy}. These modifications are due to initial cold nuclear matter (CNM) effects as well as final hot nuclear matter (HNM) effects like partonic energy loss and collective flow. Both affect differently the \ee pairs. Whereas the former alters the initial production of heavy-quark pairs and per consequence of dielectrons, the latter acts on the ${\rm e^{+}}$ and ${\rm e^{-}}$ mostly independently with the exception of flow.

In a second approach, a simple model is considered to include the single heavy-flavor decay electron measurements in the cocktail inputs. First, CNM and HNM effects are disentangled assuming
\begin{equation}
R^{{\rm c,b\to e^{\pm}}}_{\rm AA} = R^{{\rm c,b\to e^{\pm}}}_{\rm AA, CNM} \times R^{{\rm c,b\to e^{\pm}}}_{\rm AA, HNM},
\end{equation}
where $R^{{\rm c,b\to e^{\pm}}}_{\rm AA}$ is the parameterized single-electron measurement and $R^{{\rm c,b\to e^{\pm}}}_{\rm AA, CNM}$ is computed using the EPS09 nuclear parton distribution function (nPDF)~\cite{Eskola:2009uj}. Both are shown as a function of \pt on the left panel of Fig.~\ref{fig:hferaacocktail}, together with the measured nuclear modification factor of heavy-flavor decay electrons in the 10\% most central \PbPb collisions at \fivenn. The modifications are then propagated to \ee pairs using Monte Carlo simulations and a weighting procedure. The HNM effects are assumed to fully factorize between the ${\rm e^{+}}$ and ${\rm e^{-}}$:
\begin{equation}
R_{\rm AA, HNM}^{\rm c,b\to e^{+}e^{-}} = R^{{\rm c,b\to e^{\pm}}}_{\rm AA, HNM}(p_{\rm T,e^{+}}) \times R^{{\rm c,b\to e^{\pm}}}_{\rm AA, HNM} (p_{\rm T,e^{-}}),
\end{equation}
where $R_{\rm AA, HNM}^{\rm c,b\to e^{+}e^{-}}$ is the dielectron nuclear modification factor due to HNM effects. Similarly, the CNM effects are estimated on the \ee pair with the dielectron nuclear modification factor due to CNM effects ($R_{\rm AA, CNM}^{\rm c,b\to e^{+}e^{-}}$) by taking the mean of the single electron $R^{{\rm c,b\to e^{\pm}}}_{\rm AA, CNM}$
\begin{equation}
R_{\rm AA, CNM}^{\rm c,b\to e^{+}e^{-}} = \frac{R^{{\rm c,b\to e^{\pm}}}_{\rm AA, CNM}(p_{\rm T,e^{+}}) + R^{{\rm c,b\to e^{\pm}}}_{\rm AA, CNM} (p_{\rm T,e^{-}})}{2}.
\end{equation}
The resulting dielectron nuclear modification factor ($R_{\rm AA}^{\rm c,b\to e^{+}e^{-}}$),
\begin{equation}
R^{{\rm c,b\to e^{+}e^{-}}}_{\rm AA} = R^{{\rm c,b\to e^{+}e^{-}}}_{\rm AA, CNM} \times R^{{\rm c,b\to e^{+}e^{-}}}_{\rm AA, HNM},
\end{equation}
is reported on the right panel of Fig.~\ref{fig:hferaacocktail} for two different \ptee values as a function of \mee. Dielectrons with a large \mee and/or \ptee originate on average from heavy-flavor hadrons with higher \pt than \ee pairs at smaller \mee or \ptee and are therefore more affected by HNM effects. In this approach, CNM and HNM effects seem to cancel out within large uncertainties at low \mee and \ptee ($R^{{\rm c,b\to e^{+}e^{-}}}_{\rm AA, CNM} \times R^{{\rm c,b\to e^{+}e^{-}}}_{\rm AA, HNM} \approx 1$). The dielectron $R_{\rm AA}^{\rm c,b\to e^{+}e^{-}}$ is then applied to the contributions of correlated semileptonic open-charm and open-beauty hadron decays estimated from \pp collisions at the same \snn, as explained above, to obtain the $R^{{\rm c,b\to e^{\pm}}}_{\rm AA}$-modified HF contributions. The resulting full hadronic cocktail is called Cocktail2.

\subsubsection{Systematic uncertainties}
\label{syscocktail}

The systematic uncertainties of the hadronic cocktail originate from the following sources: the $\pi^{\pm}$, $\phi$, and $J/\psi$  parameterizations as a function of \pt, the $\eta$/$\pi^{\rm 0}$ ratio, the $m_{\rm T}$-scaling parameters used for $\rho$, $\omega$, $\eta^{\prime}$ and $\phi$, the branching ratios of the different light-flavor hadron decay channels, and the heavy-flavor cross sections in \pp collisions. In case of Cocktail1, the nuclear overlap function is also considered as a source of uncertainty, whereas for Cocktail2 the uncertainties of the measured $R^{{\rm c,b\to e^{\pm}}}_{\rm AA}$ and the EPS09 nPDF are propagated to the final expected dielectron yield resulting in two uncorrelated uncertainties of similar magnitude depending on \mee and \ptee. The different contributions added in quadrature amount to a total systematic uncertainty between 10\% and 20\% depending on \mee and \ptee for Cocktail1. Cocktail2 has significantly larger uncertainties, where the heavy-flavor contributions are dominant, i.e. up to about 40\% in the intermediate-mass range ($1.2 < \mee < 2.6$~\GeVmass).

\subsection{\texorpdfstring{DCA$_{\rm ee}$}{DCAee} template distributions}

The DCA$_{\rm ee}$ distributions of the different dielectron sources are determined by the convolution of the DCA detector resolution, the decay length of the mother particles, and the decay kinematics. Dielectrons produced at the primary vertex are expected to have smaller DCA$_{\rm ee}$ than \ee pairs originating from the decays of correlated open-charm and open-beauty hadrons with finite decay lengths ($c\tau_{\rm D} \approx 150$~$\mu{\rm m}$ and $c\tau_{\rm B} \approx 470$~$\mu{\rm m}$). A full GEANT 3~\cite{Brun:1119728} simulation of the ALICE central barrel, tuned to describe the performance of each detector subsystem, in particular the ITS, is utilized. The same analysis selection criteria as in data are applied to extract the DCA$_{\rm ee}$ spectra of \ee pairs originating from the decays of prompt and non-prompt $J/\psi$ mesons, as well as open-beauty and open-charm hadrons. Contrary to open-beauty hadrons, the various open-charm hadrons have significantly different decay lengths (e.g.~$59.9$~$\mu$m for $\Lambda^{\pm}_{\rm c}$ and  $311.8$~$\mu$m for ${\rm D^{\pm}}$ mesons~\cite{ParticleDataGroup:2022pth}). Therefore their relative yields are particularly relevant. The measured production ratios of prompt open-charm hadrons in pp~\cite{ALICE:2021dhb,ALICE:2021rxa} and \PbPb collisions~\cite{ALICE:2021bib}, together with their semileptonic decay branching ratios~\cite{ParticleDataGroup:2022pth}, are used to obtain the ${\rm c\overline{c}}$ DCA$_{\rm ee}$ distribution calculated as the combination of semi-leptonic decays of the different open-charm hadrons. The ${\rm D^{\pm}/D^{0}}$ ratio is found to be similar in pp and \PbPb collisions, whereas possible enhancements of the $\rm D^{\pm}_{s}/D^{0}$ and $\Lambda_{\rm c}/{\rm D^{0}}$ ratios in \PbPb compared to pp collisions are taken into account. Finally, the fraction of non-prompt $J/\psi$, $f_{\rm B}$, as a function of \pt measured by ALICE~\cite{ALICE:2023hou} in central \PbPb collisions is taken as input to build the inclusive DCA$_{\rm ee}$ template of the $J/\psi$. Each contribution is normalized to its expected yield from the hadronic cocktail in the same \mee and \ptee range after the same fiducial selection criteria ($|\eta_{\rm e}| < 0.8$ and $\pte > 0.4$~\GeVc). 

Three sources of systematic uncertainties additional to those mentioned in Section~\ref{syscocktail}, depending on DCA$_{\rm ee}$, are considered and added in quadrature. First, the DCA single-track resolution is not perfectly reproduced in the simulation. The DCA templates obtained with MC simulations using a post-correction of the DCA resolution in order to reproduce the one in data are compared to those estimated without. As a conservative estimate, half of the difference between the two spectra is attributed to the systematic uncertainty. This leads to a systematic uncertainty between 10 and 25\% varying with the DCA$_{\rm ee}$ value for prompt \ee pairs, while for non-prompt dielectrons (${\rm c\overline{c}}$ and $\rm b\overline{b}$) a maximum of 10\% systematic uncertainty is observed at small DCA$_{\rm ee}$, i.e. for DCA$_{\rm ee}$ values smaller than one. Second, the uncertainties of the charm and beauty-hadron production ratios, of their semileptonic decay branching ratios, and of $f_{\rm B}$ are propagated to the corresponding DCA$_{\rm ee}$ distributions and amount to about 10\%. Third, the uncertainty related to the \pt spectra of the open-charm and open-beauty mother hadrons is found to be of the order of 5\% at large DCA$_{\rm ee}$.

\section{Inclusive dielectron production}
\label{inclusiveproduction}

\subsection{Dielectron invariant-mass spectra}
\label{invariantmass}

\begin{figure}[tbh]
    \begin{center}  
     \includegraphics[width = 0.495\textwidth]{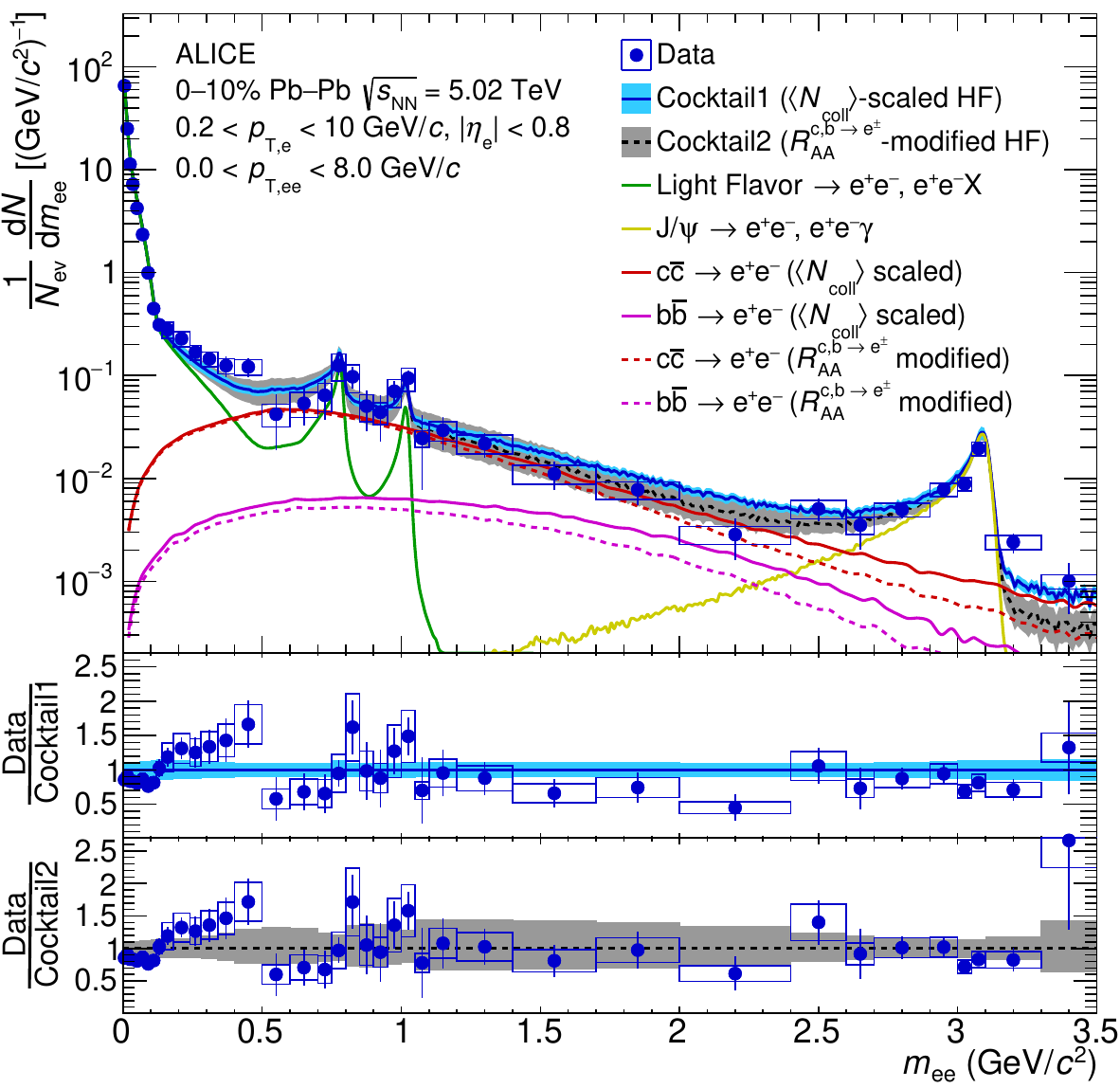}
    \includegraphics[width = 0.495\textwidth]{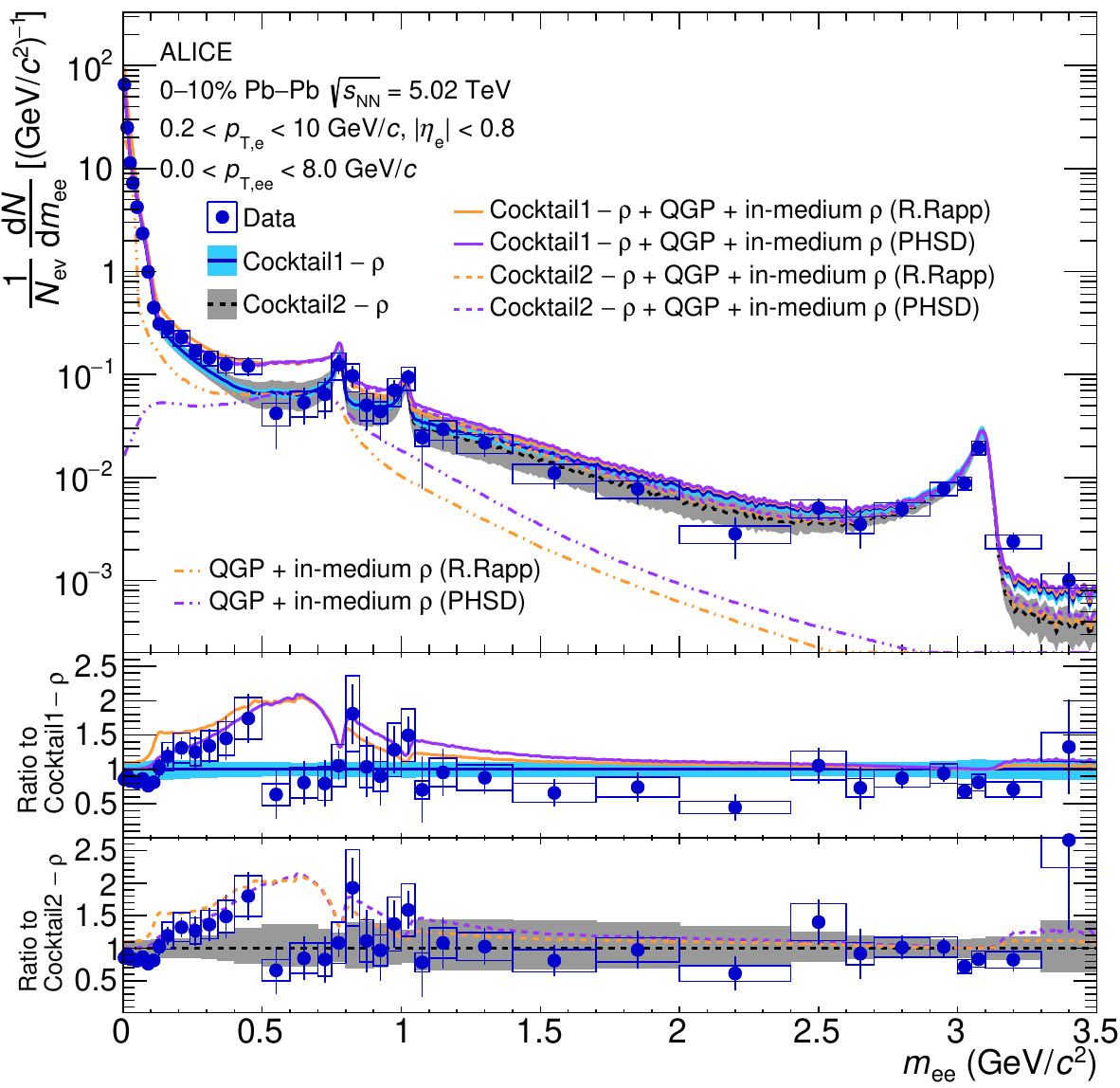}
    \end{center}
    \caption{Upper panels: dielectron $m_{\rm ee}$-differential yields in the 10\% most central Pb--Pb collisions at \mbox{\fivenn,} compared with the expected e$^{+}$e$^{-}$ contributions from known hadronic decays, including two different estimations for dielectrons from correlated heavy-flavor hadron decays (left panel), and two predictions for thermal radiation from the medium~\cite{Rapp:2013nxa,Song:2018xca} (right panel). Bottom panels: ratios data to cocktails, together with the expected ratios from the models. The error bars and boxes represent the statistical and systematic uncertainties of the data, respectively, whereas the bands show the uncertainties of the hadronic cocktails.}
    \label{fig:invariantmass}
\end{figure}

The yield of \ee pairs in the ALICE acceptance ($|\eta_{\rm e}| < 0.8$ and $0.2 < \pte < 10$~\GeVc) is shown as a function of \mee in the 10\% most central \PbPb collisions at \fivenn in Fig.~\ref{fig:invariantmass}. In the left panel, the data are compared with the expected contributions from known hadronic sources, i.e. Cocktail2 shown with dashed lines and Cocktail1 displayed as full lines, both explained in Section~\ref{cocktailinvariantmassheavyflavour}. In the $\pi^{0}$ ($\mee < 0.135$~\GeVmass) and $J/\psi$ ($2.7 < \mee < 3.3$~\GeVmass) mass regions, where dielectrons from $\pi^{0}$-Dalitz and $J/\psi$ decays are expected to dominate the yield of \ee pairs, respectively, the data are well reproduced by the hadronic cocktails.   

In the bottom and middle left panels of Fig.~\ref{fig:invariantmass}, the ratio of the data and the different hadronic cocktails are presented. At low \mee ($0.18 < \mee < 0.5$~\GeVmass), the ratios are systematically above one, although consistent with unity within $1.3 \sigma$ ($1.1 \sigma$) for Cocktail1 (Cocktail2). Integrated in the \mee range $0.18 < \mee < 0.5$~\GeVmass they are found to be $R = 1.40 \pm 0.11 \ {\rm (stat.)} \pm 0.23 \ {\rm(syst.)} \pm 0.16 \ {\rm(cocktail)}$ for Cocktail1 and $R = 1.42 \pm 0.11 \ {\rm (stat.)} \pm 0.23 \ {\rm(syst.)} ^{+0.24}_{-0.29} \ {\rm(cocktail)}$ for Cocktail2, where the systematic uncertainty related to the cocktail is reported separately to those originating from the data. The hint for an excess does not depend significantly on the method used to estimate the heavy-flavor contribution. In both cocktails, the contribution from $\rho$ mesons is estimated neglecting any medium effect and amounts to about 18\% of the total yield of known hadronic sources at $m_{\rm \rho}$. However, a significant contribution of \ee pairs originating from $\rho$ mesons produced thermally in the medium is expected at low \mee. Due to its short lifetime compared to the one of the hot fireball and its strong coupling to the $\pi^{+}\pi^{-}$ channel, the $\rho$ meson is likely to be regenerated in the hot hadronic phase with a medium-modified spectral function broader than in vacuum. In the intermediate-mass range (IMR), i.e.~$1.2 < \mee < 2.6$~\GeVmass, Cocktail2 including medium effects in the estimation of the heavy-flavor background can better describe the data, systematically below the expectations from Cocktail1. Nevertheless, the former cocktail suffers from large uncertainties.

In the right panel of Fig.~\ref{fig:invariantmass}, the data are compared with Cocktail1 and Cocktail2 excluding the contribution from the $\rho$ meson. In addition, expectations from two theoretical models for thermal dielectrons from the partonic and hadronic phase, referred to as QGP and in-medium $\rho$, are shown. As for the hadronic cocktail, detector resolution effects were applied to the predictions~\cite{ALICE:2017tau,ALICE:2022hvk}. In both calculations, $\rho$ mesons produced during the full evolution of the system are considered, in particular those generated in the hot hadronic phase which are a relevant source of thermal radiation for $\mee < 0.8$~\GeVmass.
The model from R. Rapp~\cite{Rapp:2013nxa} is an expanding fireball model, where the thermal emission rate of dielectrons from the hadronic phase is calculated based on a hadronic many-body theory with in-medium modified $\rho$, whereas a lattice-QCD inspired approach is used for the equation of state in the QGP. The Parton-Hadron-String Dynamics (PHSD) model is a transport model~\cite{Song:2018xca} with in-medium modified electromagnetic spectral functions of low-mass vector mesons. The predictions were added to the hadronic cocktails to obtain the total expected yield of dielectrons. In the bottom and middle right panels of Fig.~\ref{fig:invariantmass}, the ratio between the data and the different hadronic cocktails excluding the $\rho$ meson contribution are shown together with the ratio of the total expected dielectron yield to the hadronic cocktails. Both models can reproduce the data for $0 < \mee < 0.5$~\GeVmass, but tend to overestimate them for $0.5 < \mee < 0.7$~\GeVmass by $2.7 \sigma$ and $4.0 \sigma$ including or not including medium effects in the estimation of the heavy-flavor background. In the IMR, the QGP is the main source of predicted thermal radiation. The expected thermal contribution to the total dielectron yield is small, i.e. around 15\% integrated over this mass range, which is below the sensitivity of the data and much smaller than the uncertainties of the heavy-flavor medium-modified cocktail (Cocktail2).

\begin{figure}[tbh]
    \begin{center}  
     \includegraphics[width = 0.495\textwidth]{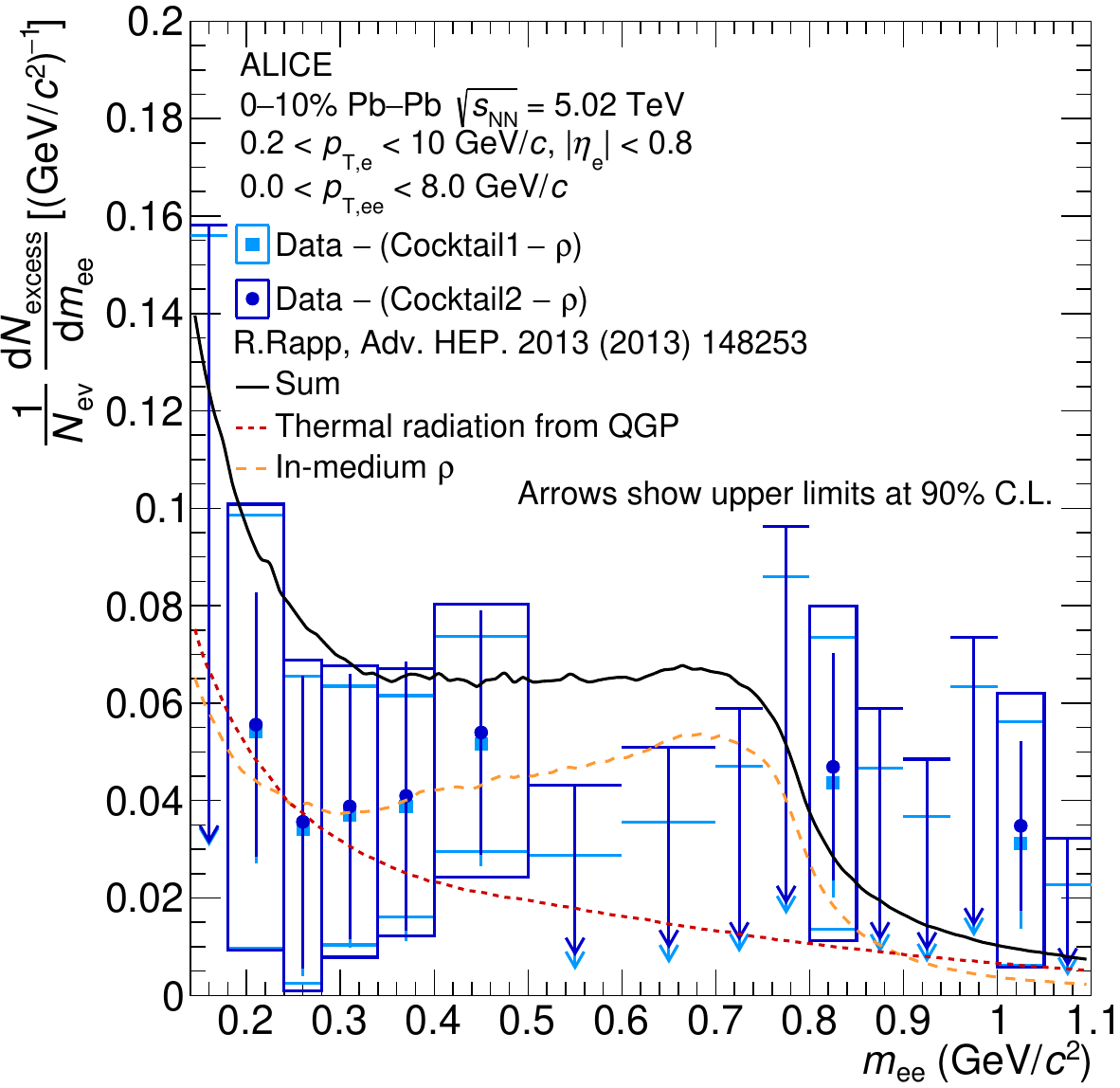}
    \includegraphics[width = 0.495\textwidth]{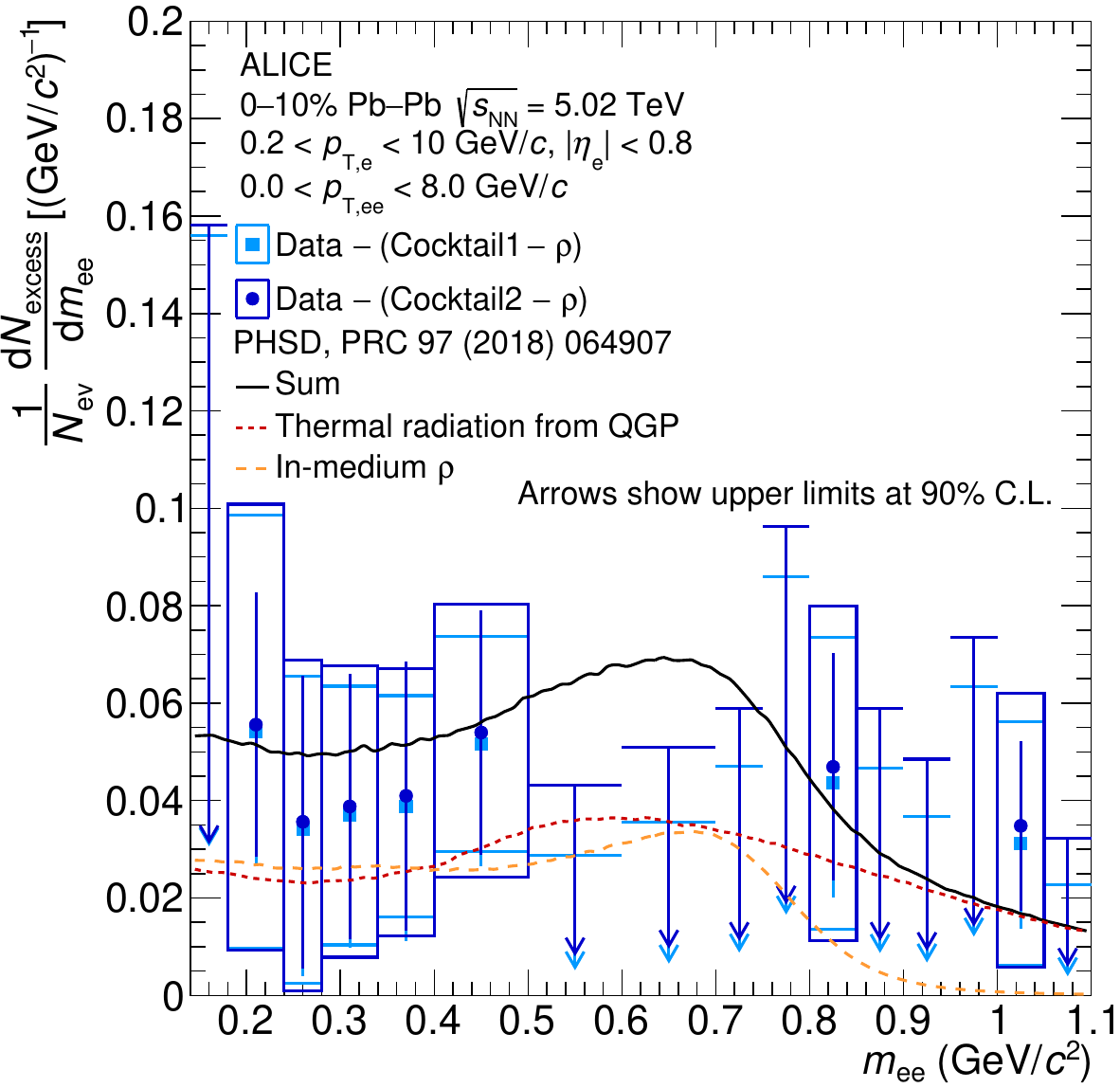}
    \end{center}
    \caption{Excess yield of dielectrons in the 10\% most central \PbPb collisions at \fivenn with respect to the expected \ee contributions from known hadronic sources, including (Cocktail2) or not including (Cocktail1) medium effects for the heavy-flavor contributions, and compared with predictions from the model of R. Rapp~\cite{Rapp:2013nxa} (left) and from the PHSD transport approach~\cite{Song:2018xca} (right).}
    \label{fig:excess}
\end{figure}

To better investigate the hint for an excess of \ee pairs at low \mee, the expected dielectron yield from known hadronic sources is subtracted from the data to obtain a so-called excess spectrum. The excess spectra computed with the two different cocktails (Cocktail1 $- \ \rho$ and Cocktail2 $- \ \rho$) are shown in Fig.~\ref{fig:excess} for $0.14 < \mee < 1.1$~\GeVmass. Upper limits at 90\% C.L. using the Feldman and Cousins method\,\cite{Feldman:1997qc} are reported for the values which are found to be statistically consistent with zero within one standard deviation. In the \mee range $0.18 < \mee < 0.5$~\GeVmass, where the significance of the excess reaches $1.34 \sigma$ and $1.49 \sigma$ using Cocktail2 $- \ \rho$ and Cocktail1 $- \ \rho$, respectively, the result is not very sensitive to the different implementations of the heavy-flavor background. The excess spectra are compared with the same models~\cite{Rapp:2013nxa,Song:2018xca} as above. The computed contributions from $\rho$ mesons produced thermally in the hot hadronic matter and thermal radiation from the QGP are shown separately. In the case of the PHSD model (right panel of Fig.~\ref{fig:excess}), the dielectron production in the QGP is calculated by implementing the off-shell cross sections of ${\rm q\overline{q}}\to {\rm e^{+}e^{-}}$, ${\rm q\overline{q}}\to {\rm e^{+}e^{-}g}$ and ${\rm qg}\to {\rm q e^{+}e^{-}}$ (${\rm \overline{q}g}\to {\rm \overline{q} e^{+}e^{-}}$) reactions from the  dynamical quasiparticle model\,\cite{Linnyk:2011ee} into the PHSD transport approach. The yield of thermal radiation from the hadronic phase (in-medium $\rho$) predicted by PHSD is smaller than the one estimated by the model of R. Rapp (right panel of Fig.~\ref{fig:excess})~\cite{Rapp:2013nxa}. Both calculations are nevertheless compatible with the data within their statistical and systematic uncertainties.

\subsection{Dielectron \texorpdfstring{\ptee}{ptee} spectra}
\label{pteespectra}

\begin{figure}[tbh]
    \begin{center}
    \includegraphics[width = 0.495\textwidth]{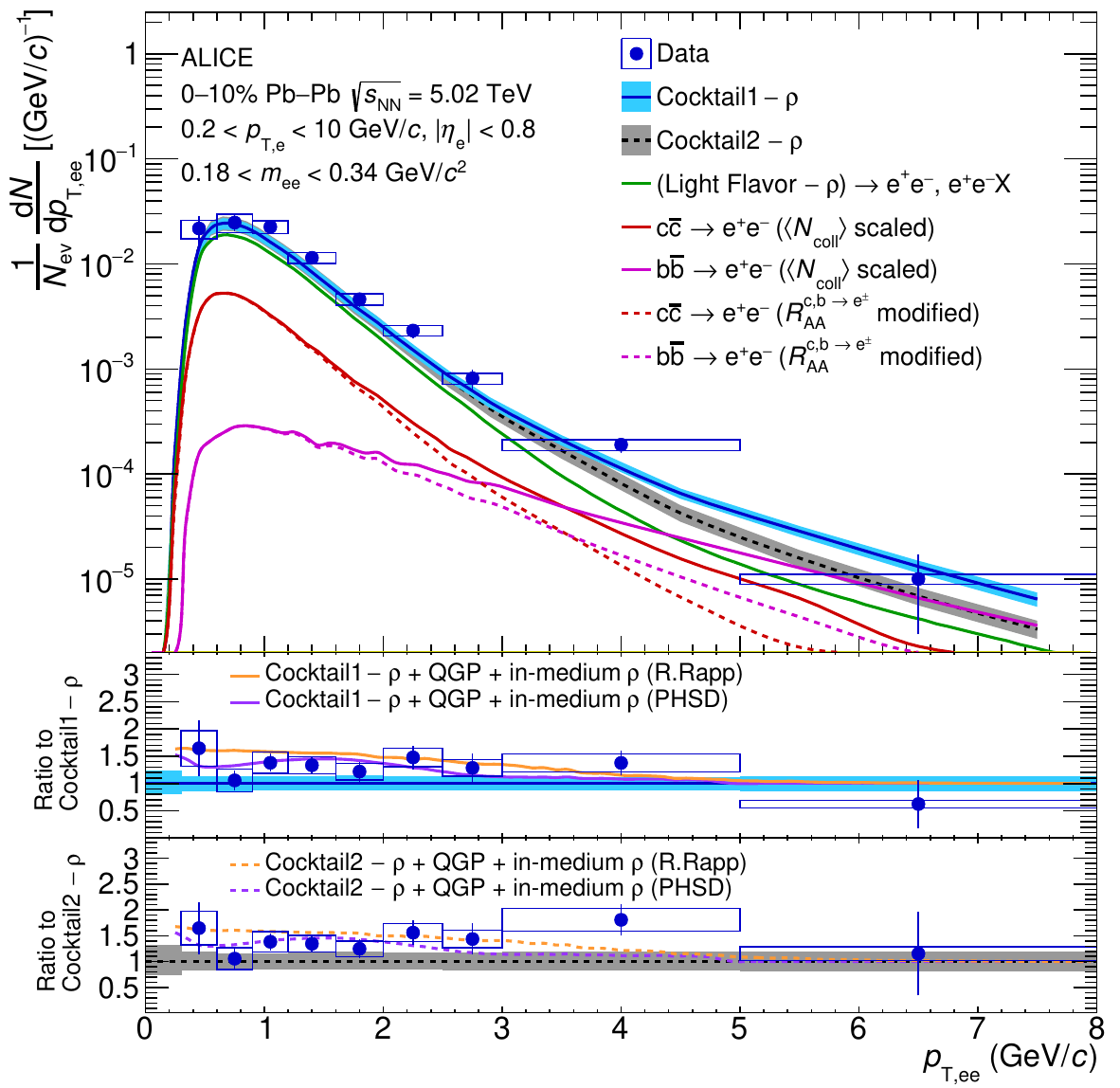}
    \includegraphics[width = 0.495\textwidth]{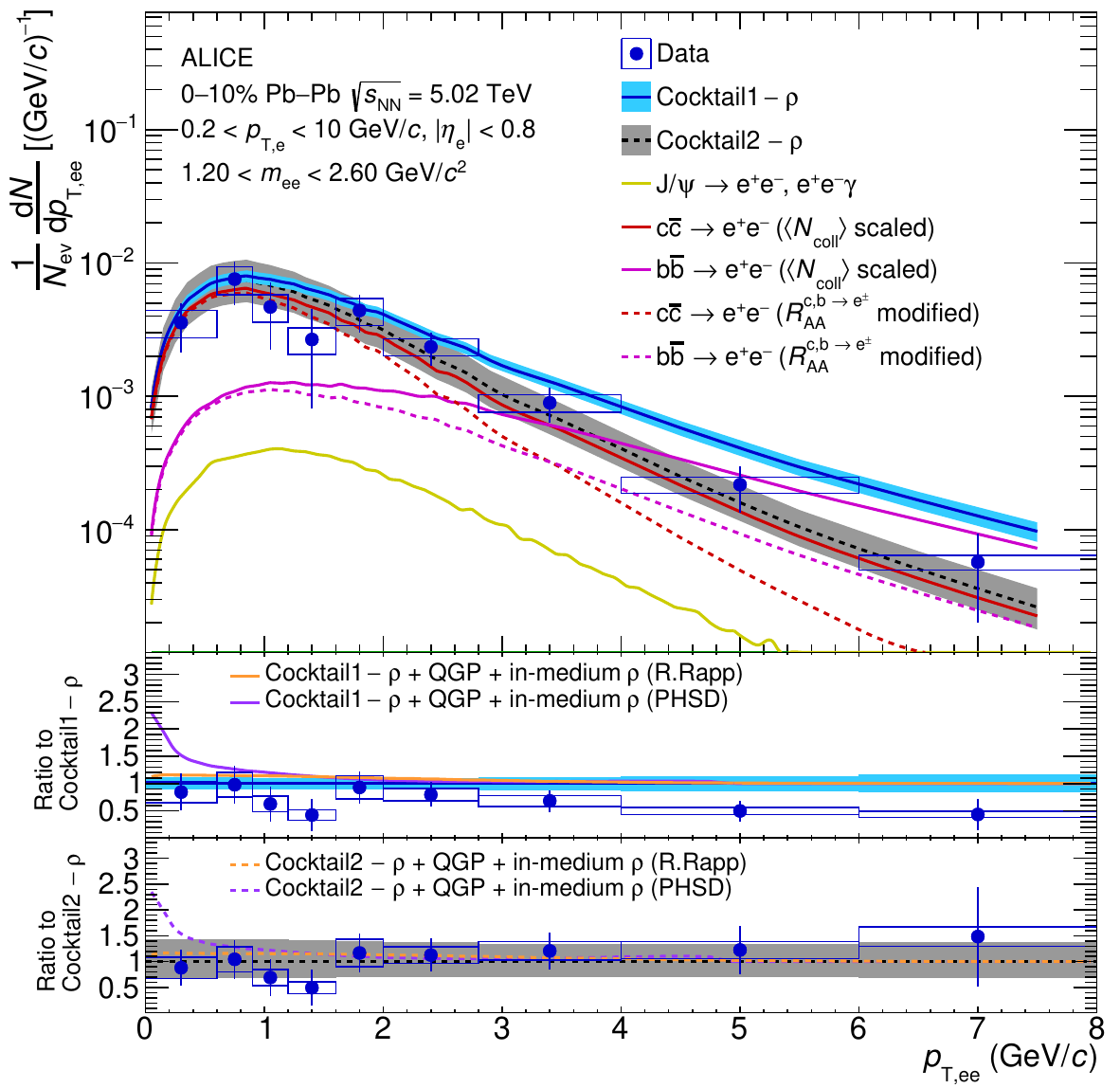}
    \end{center}
    \caption{Upper panels: dielectron \ptee-differential yields in two different \mee ranges, $0.18 < \mee < 0.34$~\GeVmass (left) and $1.2 < \mee < 2.6$~\GeVmass (right), in the 10\% most central Pb--Pb collisions at \fivenn, compared with the expected e$^{+}$e$^{-}$ contributions from known hadronic decays, including two different estimations for dielectrons from correlated heavy-flavor hadron decays (see text). Bottom panels: ratios data to cocktails, excluding the contribution from vacuum $\rho$, together with two predictions for thermal radiation from the medium~\cite{Rapp:2013nxa,Song:2018xca}. The error bars and boxes represent the statistical and systematic uncertainties of the data, respectively, whereas the bands show the uncertainties of the hadronic cocktails.}
    \label{fig:ptee}
\end{figure}

The yield of \ee pairs in the \mee range where a hint for an excess is observed and the heavy-flavor contribution is expected to be still relatively small ($0.18 < \mee < 0.34$~\GeVmass) is shown as a function of \ptee in the left panel of Fig.~\ref{fig:ptee}. The data are compared with the two different cocktails discussed in Sec.~\ref{hadroniccocktail} excluding the $\rho$ contribution. In this \mee range, $\eta$ Dalitz decays ($\eta \to {\rm \gamma e^{+}e^{-}}$) are expected to contribute significantly at low \ptee, whereas correlated open-beauty hadron decays play a role at high \ptee. The data-to-cocktail ratios, shown on the bottom and middle left panels of Fig.~\ref{fig:ptee}, are consistent with unity, while being systematically above but still within $1.3 \sigma$ ($1.2 \sigma$) in the \ptee range $0.9 < \ptee < 2$~\GeVc for Cocktail1 (Cocktail2) $- \ \rho$. The ratios are compared with the expected inclusive dielectron yield-to-cocktail ratio estimated with the two different calculations for thermal radiation from the partonic and hadronic phase~\cite{Rapp:2013nxa,Song:2018xca} presented in Sec.~\ref{invariantmass}. Both models predict thermal contributions relevant only at low \ptee ($\ptee < 4$~\GeVc) and are compatible with the data within uncertainties.

The IMR is dominated by correlated \ee pairs from semileptonic decays of charm and beauty hadrons. The \ptee-differential yield of \ee pairs measured in this \mee region is shown in comparison with the two different hadronic cocktails, as well as predictions for thermal radiation~\cite{Rapp:2013nxa,Song:2018xca} in the data-to-cocktail ratios, in the right panel of Fig.~\ref{fig:ptee}. The contribution of \ee pairs from ${\rm c\overline{c}}$ is expected to be the dominant dielectron source for $\ptee < 2$~\GeVc, whereas most of the \ee pairs originate from ${\rm b\overline{b}}$ for $\ptee > 4$~\GeVc. The data are systematically below Cocktail1 not incorporating any heavy-flavor medium modifications, particularly at high \ptee where the difference reaches $2 \sigma$ in the last \ptee interval ($6 < \ptee < 8$~\GeVc). This is in agreement with a heavy-flavor suppression, increasing with \ptee,  with respect to \pp collisions at the same center-of-mass energy per nucleon pair. The trend is reproduced by Cocktail2 based on the measured single heavy-flavor decay electron $R_{\rm AA}$~\cite{ALICE:2019nuy}, although this cocktail has large uncertainties. The contribution of thermal radiation from the QGP is predicted to be at most 15\% of the inclusive dielectrons for $\ptee < 2$~\GeV/c, as can be seen on the bottom and middle right panels of Fig.~\ref{fig:ptee}. It is much smaller than the uncertainties of Cocktail2. Therefore a different approach is mandatory to address the production of \ee pairs in the partonic phase.

\section{Topological separation of \texorpdfstring{\ee}{ee} sources}
\label{DCAeeresults}

Dielectron measurements in the IMR provide a handle to disentangle the contributions of hadronic and QGP thermal radiation. However, dielectrons from correlated heavy-flavor hadron decays dominate the \ee pair production in this phase space and make the extraction of the thermal QGP signal very challenging. The pair ${\rm {DCA_{ee}}}$ variable offers experimental means to disentangle displaced dielectrons produced in the decays of heavy-flavor hadrons from a prompt thermal contribution.

\subsection{\texorpdfstring{$J/\psi$}{Jpsi} invariant-mass region}

\begin{figure}[tbh]
    \begin{center}
    \includegraphics[width = 0.6\textwidth]{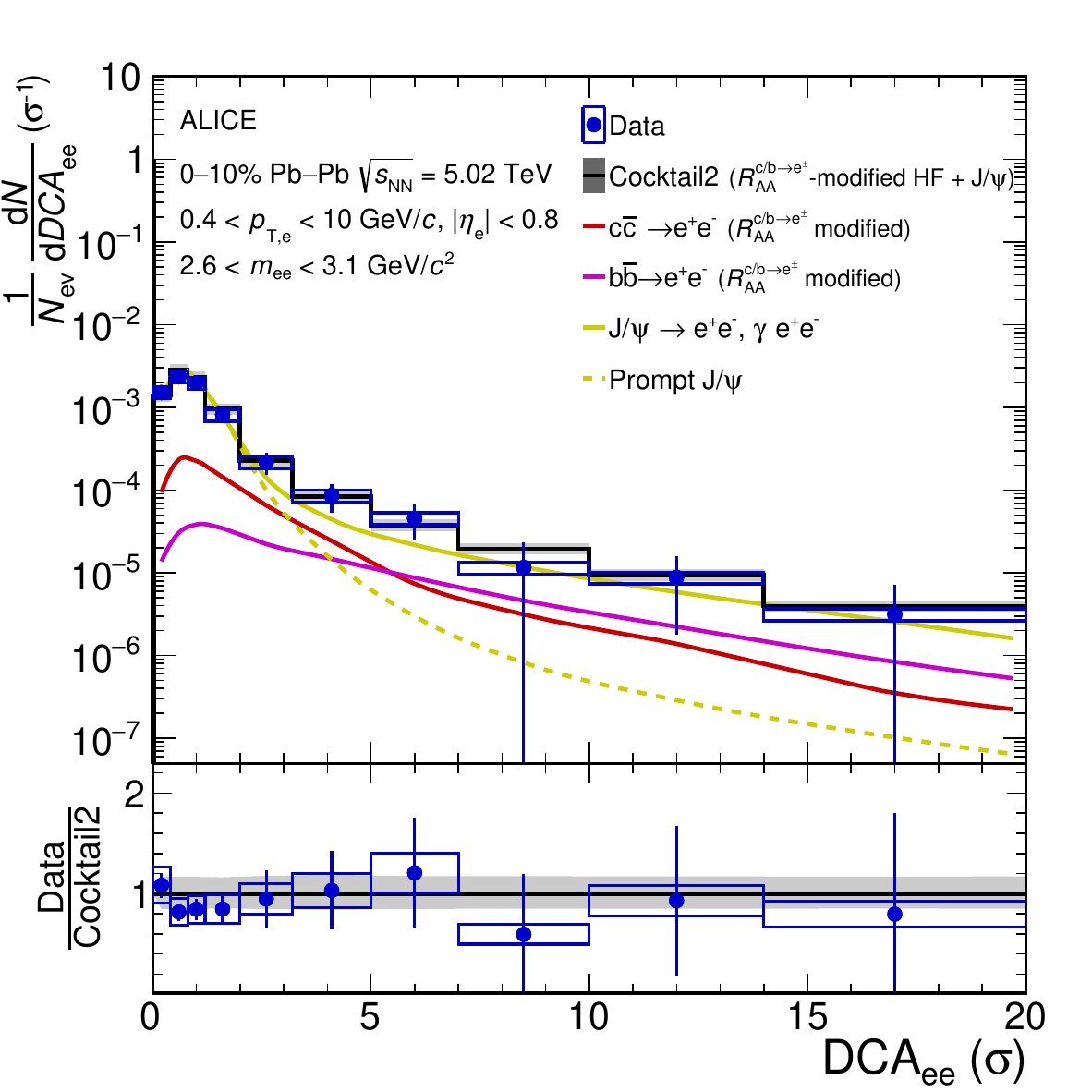}
    \end{center}
    \caption{Upper panel: inclusive \ee yield in the 10\% most central \PbPb collisions at \fivenn as a function of DCA$_{\rm ee}$ in the mass range $2.6 < m_{\rm ee} < 3.1$~\GeVmass, compared with a cocktail of expected sources, including medium effects for the heavy-flavor contributions. Bottom panel: corresponding data to cocktail ratio. Statistical and systematic uncertainties on the data are shown as vertical bars and boxes, respectively. The total uncertainty of the cocktail is represented as a grey band.}
    \label{fig:dcajpsi}
\end{figure}

The contributions from prompt and non-prompt $J/\psi$ are well constrained by independent ALICE measurements which combine the inclusive yield with the fraction of $J/\psi$ originating from beauty hadron decays~\cite{ALICE:2023hou}. These components are dominant for $2.6 < \mee < 3.1$~\GeVmass, making this mass range well suited to test the understanding of the detector DCA$_{\rm ee}$ resolution. The measured DCA$_{\rm ee}$-differential yield in this mass range is compared with the different expected hadronic components, i.e. dielectrons from prompt and non-prompt $J/\psi$ decays, as well as from open heavy-flavor hadron decays in Fig.~\ref{fig:dcajpsi}. The $R^{{\rm c,b\to e^{\pm}}}_{\rm AA}$-modified heavy-flavor calculations (Cocktail2) are used to estimate the heavy-flavor yields, which are found to be a subleading contribution at all DCA$_{\rm ee}$. The data are described by the sum of the DCA$_{\rm ee}$ templates, validating the description of the DCA$_{\rm ee}$ resolution in the simulations. 

\subsection{Intermediate-mass range and QGP thermal radiation}

\begin{figure}[tbh]
    \begin{center}
    \includegraphics[width = 0.6\textwidth]{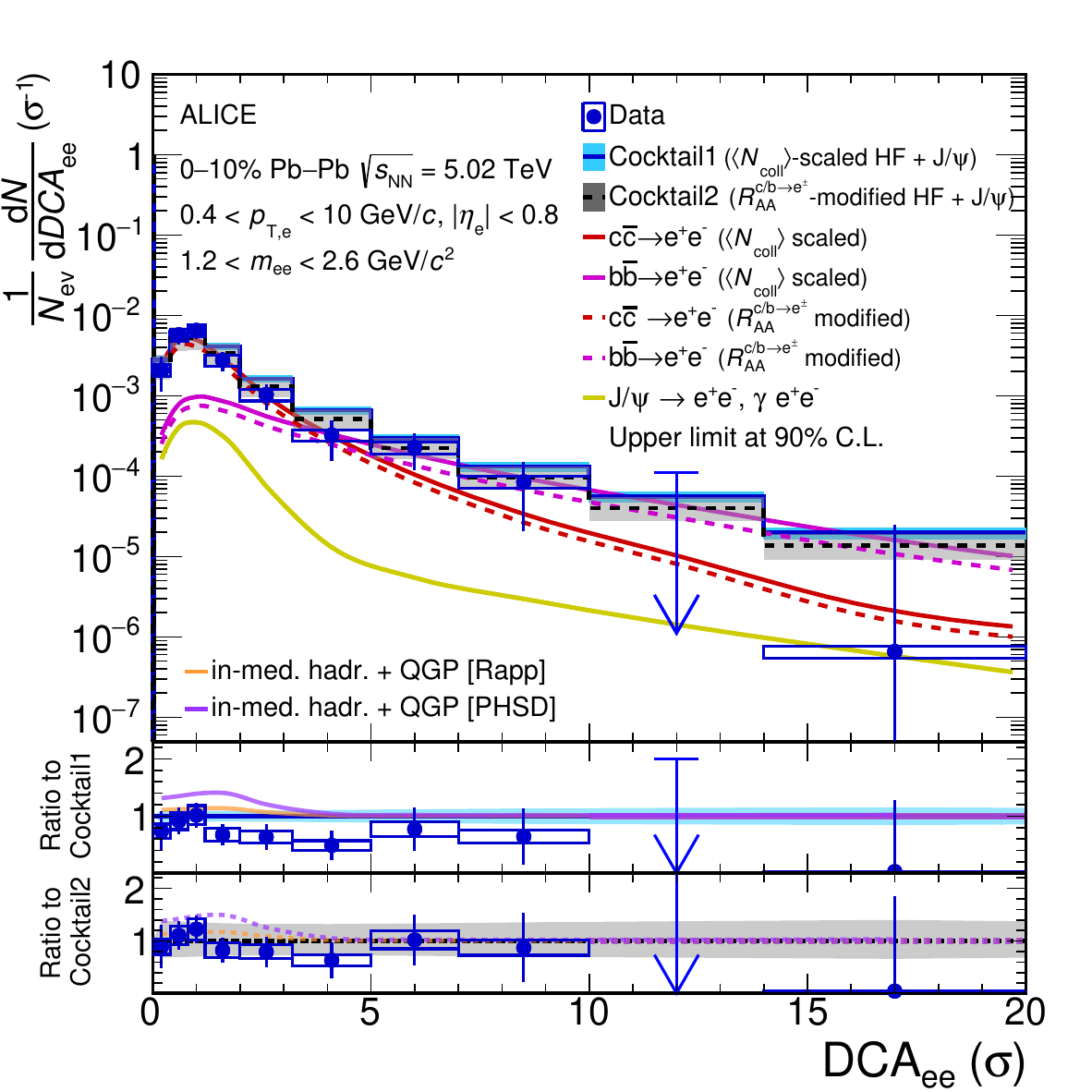}
    \end{center}
    \caption{Upper panel: inclusive \ee yield in the 10\% most central \PbPb collisions at \fivenn as a function of DCA$_{\rm ee}$ in the mass range $1.2 < m_{\rm ee} < 2.6$~\GeVmass, compared with the expected e$^{+}$e$^{-}$ contributions from known hadronic decays, including two different estimations for dielectrons from correlated heavy-flavor hadron decays (see text). Bottom panels: ratios data to cocktail, together with two predictions for thermal radiation from the medium~\cite{Rapp:2013nxa,Song:2018xca}. Statistical and systematic uncertainties on the data are shown as vertical bars and boxes, respectively. The total uncertainty of the cocktails are represented as band.}
    \label{fig:dcainv}
\end{figure}

The DCA$_{\rm ee}$-differential yield measured in the mass range $1.2 < \mee < 2.6$~\GeVmass is shown in Fig.~\ref{fig:dcainv} together with the expected contributions of known hadronic decays, including (Cocktail2) or not including (Cocktail1) medium effects for dielectrons from heavy-flavor hadron decays. The $J/\psi$ contribution in this mass range originates principally from electrons affected by bremsstrahlung in the detector material and reconstructed with a smaller \pt compared to their true one, shifting the measured dielectron invariant mass towards smaller values. The upper boundary of the \mee range (2.6~\GeVmass) was chosen such that the $J/\psi$ contribution is one order of magnitude smaller than the measured dielectron yield. On the contrary, the ${\rm b\overline{b}}\to {\rm ee}$ contribution, with the large decay length of open-beauty hadrons ($c\tau_{\rm B} \approx$ 470\,$\mu{\rm m}$), dominates the spectrum at high DCA$_{\rm ee}$, while the ${\rm c\overline{c}}\to {\rm ee}$ component defines the spectrum at low DCA$_{\rm ee}$ values ($c\tau_{\rm D} \approx$ 150\,$\mu{\rm m}$). Overall, the data are below Cocktail1, indicating a heavy-flavor suppression in \PbPb collisions compared to pp collisions. A better agreement is observed with Cocktail2. The data-to-cocktail ratios are compared with the ratio of the total expected dielectron yield and the hadronic cocktail, using predictions for thermal radiation from the two different models~\cite{Rapp:2013nxa,Song:2018xca} discussed in Sec.~\ref{invariantmass}. For this purpose, the DCA$_{\rm ee}$ template for prompt \ee pairs is normalized to the expected thermal radiation yield from the models. The contribution of thermal radiation is expected to be between 10 and 40\% around DCA$_{\rm ee}$ values of 1.5$\sigma$ according to these calculations and the DCA$_{\rm ee}$ detector resolution. It is consistent with the data using some medium modifications (Cocktail2) to describe the heavy-flavor background.

\begin{figure}[tbh]
    \begin{center}  
     \includegraphics[width = 0.49\textwidth]{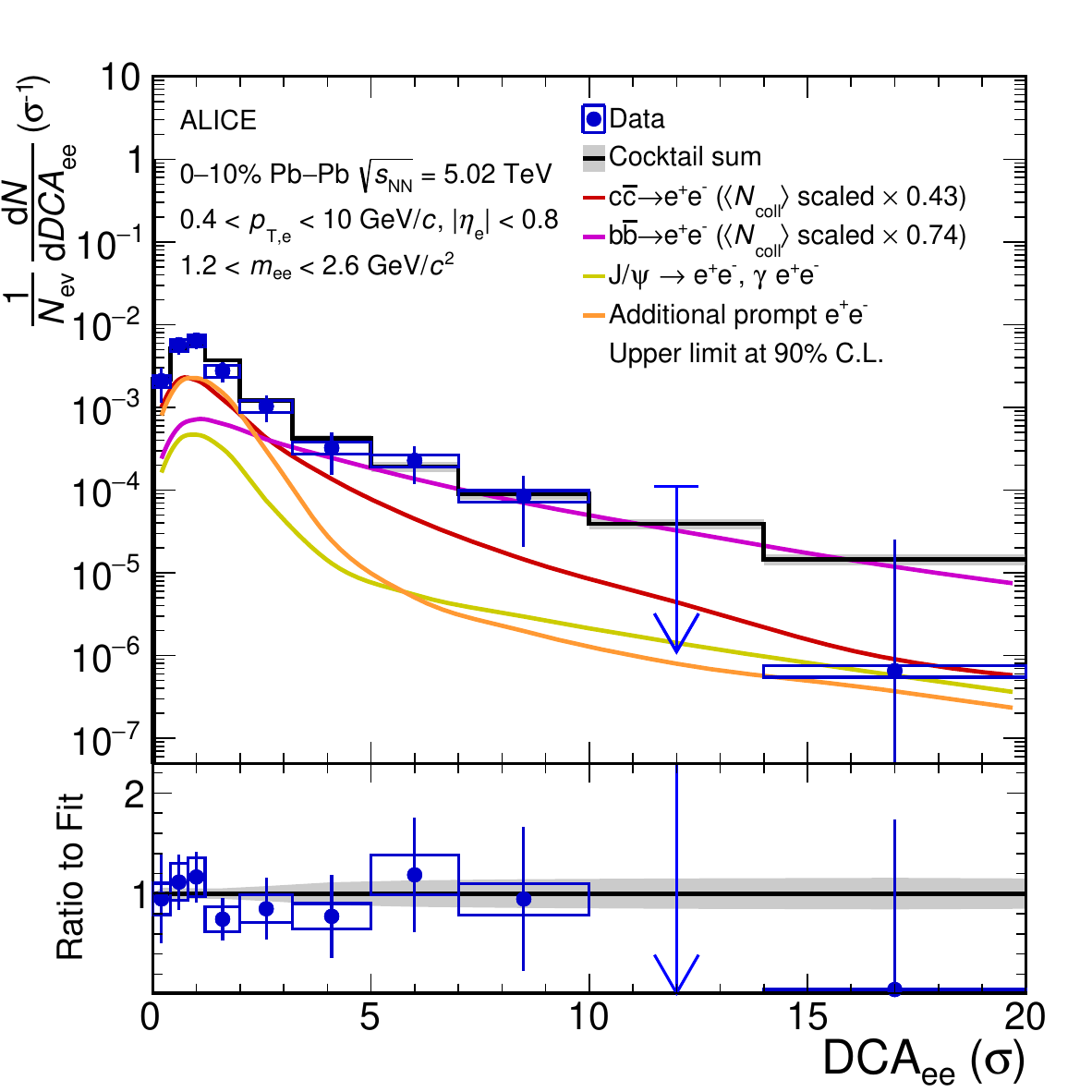}
    \includegraphics[width = 0.5\textwidth]{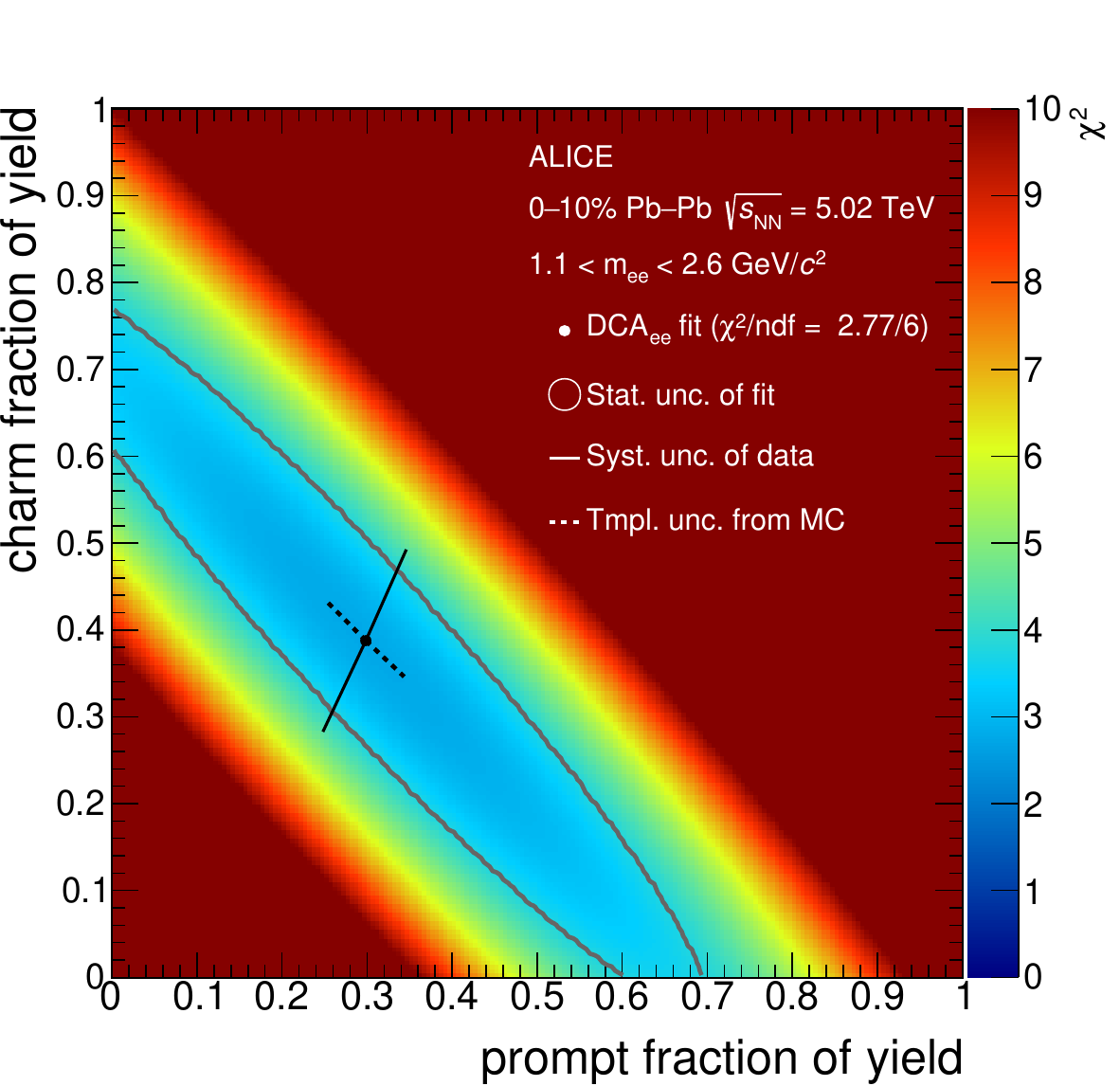}    
    \end{center}
    \caption{Left panel: fit of the inclusive \ee yield in the 10\% most central \PbPb collisions at \fivenn as a function of DCA$_{\rm ee}$ in the mass range $1.2 < m_{\rm ee} < 2.6$~\GeVmass. Right panel: corresponding $\chi^{2}$ as a function of the two fit parameters, i.e. the ratio of the yield of \ee pairs originating from a prompt source or from open-charm hadron decays to the measured \ee yield. See text for more details.}
    \label{fig:fitandchi2dca}
\end{figure}

The description of the data can be improved by fitting the DCA$_{\rm ee}$ templates of the different expected contributions to the data as shown in Fig.~\ref{fig:fitandchi2dca} (left) and explained below. The $J/\psi$ DCA$_{\rm ee}$ template $\left (f^{\rm fixed}_{\rm DCA}(J/\psi \rm \to e^{+}e^{-})\right )$ is given by the cocktail. The beauty contribution $\left (f^{\rm fixed}_{\rm DCA}(\rm b\overline{b} \to e^{+}e^{-})\right )$ is fixed to reproduce the data at high \ptee ($\ptee > 3$~\GeVc). In this \ptee region, the beauty contribution dominates the spectra, while other components (charm, $J/\psi$, and thermal) are expected to contribute less than 26\% to the dielectron yield. This approach leads to a suppression by a factor $0.74 \pm 0.24 \ {\rm (stat.)} \pm 0.12 \ {\rm (syst.)}$ for beauty compared to $N_{\rm coll}$-scaling. Finally, the data points are fitted with a four-component function    
\begin{equation}
f_{\rm DCA} = a \times f_{\rm DCA}({\rm prompt}) + b \times f_{\rm DCA}({\rm c\overline{c} \to e^{+}e^{-}}) + f^{\rm fixed}_{{\rm DCA}}({\rm b\overline{b} \to e^{+}e^{-}}) + f^{{\rm fixed}}_{{\rm DCA}}({\rm J/\psi \to e^{+}e^{-}}),
\end{equation}
where $a$ and $b$ are the two fit parameters, whereas $f_{\rm DCA}(\rm prompt)$ and $f_{\rm DCA}(\rm c\overline{c} \to e^{+}e^{-})$ are the thermal and open charm contributions, whose yields are given by the model of R. Rapp~\cite{Rapp:2013nxa} and Cocktail1 based on $N_{\rm coll}$-scaling, respectively.
The data are found to be consistent with a charm suppression by a factor $0.43 \pm 0.4 \ {\rm (stat.)} \pm 0.12 \ {\rm (syst.)}$ with respect to $N_{\rm coll}$-scaling and an additional prompt component that is a factor $3.17 \pm 3.81 \ {\rm (stat.)} \pm 0.5 \ {\rm (syst.)}$ larger than the predictions from R. Rapp~\cite{Rapp:2013nxa} and $1.15 \pm 1.38 \ {\rm (stat.)} \pm 0.18 \ {\rm (syst.)}$ times the expected yield from the PHSD model~\cite{Song:2018xca}. The $\chi^{2}$ of the DCA$_{\rm ee}$ fit is reported as a function of the ratio of the yield of \ee pairs originating from a prompt source and from correlated open-charm hadron decays to the total measured dielectron yield on the right-hand panel of Fig.~\ref{fig:fitandchi2dca}. The full line shows the systematic uncertainties originating from those of the data, determined by moving the data points coherently upward and downward by their systematic uncertainties. The smallest source of uncertainty is related to the systematic uncertainties of the DCA$_{\rm ee}$ templates, represented as a dashed line. The fit procedure is repeated after changing the DCA$_{\rm ee}$ resolution in the simulations used to build the templates, as well as after varying the semileptonic decay branching ratios and fragmentation functions of the charm hadrons according to their uncertainties. The former source of systematic uncertainty dominates the latter one. Finally, the statistical uncertainties of the fit, directly related to those of the data, are displayed by an ellipse. They are by far the biggest uncertainty and limit the interpretation of the results, since the data are consistent with no as well as a 65\% prompt contribution in the measured dielectron yield. The larger expected data sample of \PbPb collisions during the LHC Run 3 and 4 periods, up to a factor 100, will allow a significant reduction of the statistical uncertainties~\cite{CERN-LHCC-2013-020,CERN-LHCC-2015-002,ALICETPC:2020ann,newReadoutALICE,newCompFrameworkALICE}. At the same time, the better single-track DCA resolution of the detector, by a factor 3 and 6 in the transverse plane and in the direction of the beam axis, respectively, for Run 3~\cite{Musa:1475244} and even more for Run 4\,\cite{ALICE:2018fuj}, will further improve the separation of prompt and non-prompt dielectron sources~\cite{Citron:2018lsq}. According to the present pioneer analysis of DCA$_{\rm ee}$, a significant extraction of QGP radiation in the IMR should be then possible in Run 3 and 4~\cite{ALICE:2023udb}.

\section{Direct-photon production}
\label{directphoton}

\subsection{Inclusive to decay photon ratio}

\begin{figure}[tbh]
    \begin{center}  
     \includegraphics[width = 0.495\textwidth]{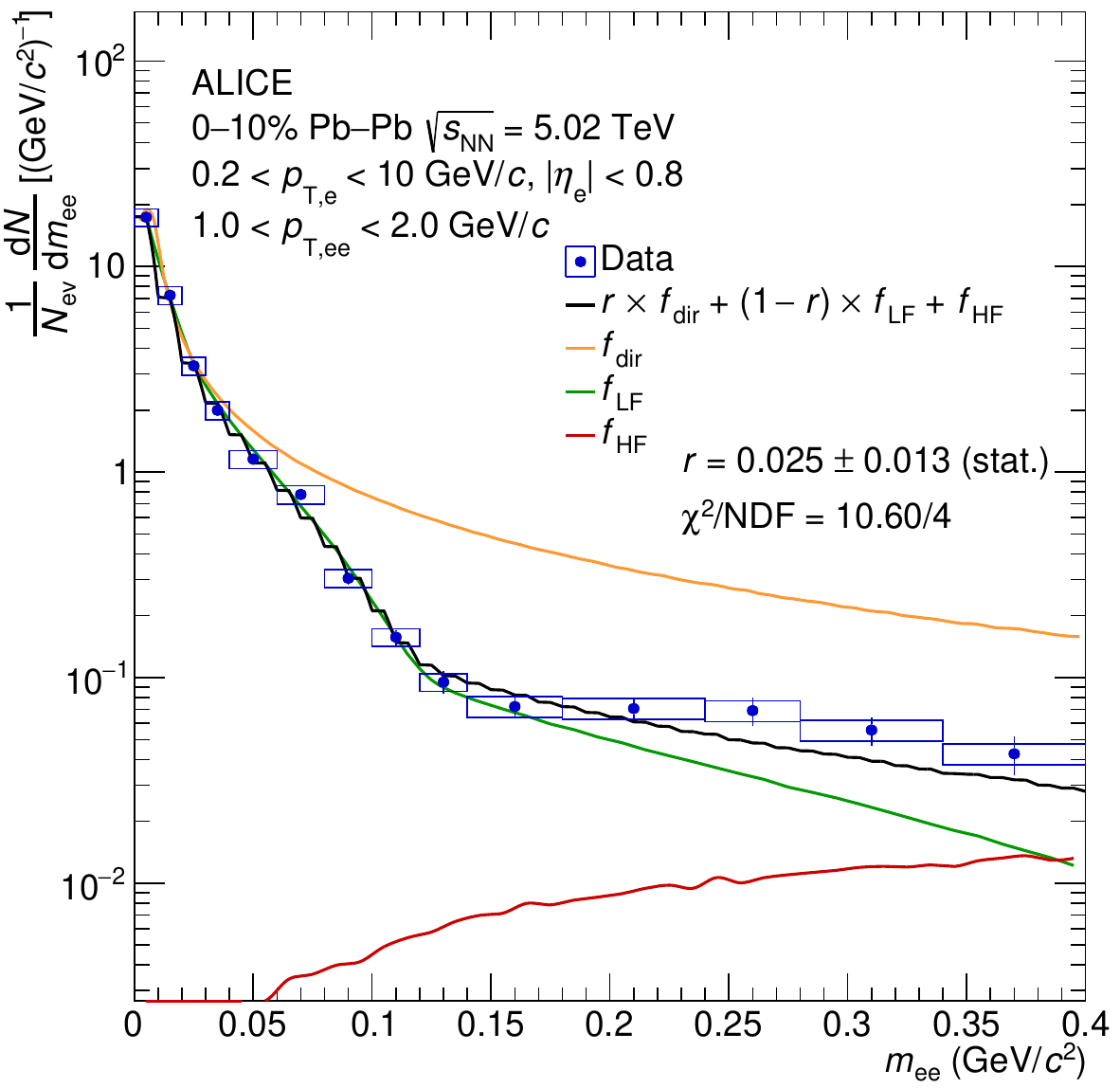}
    \includegraphics[width = 0.495\textwidth]{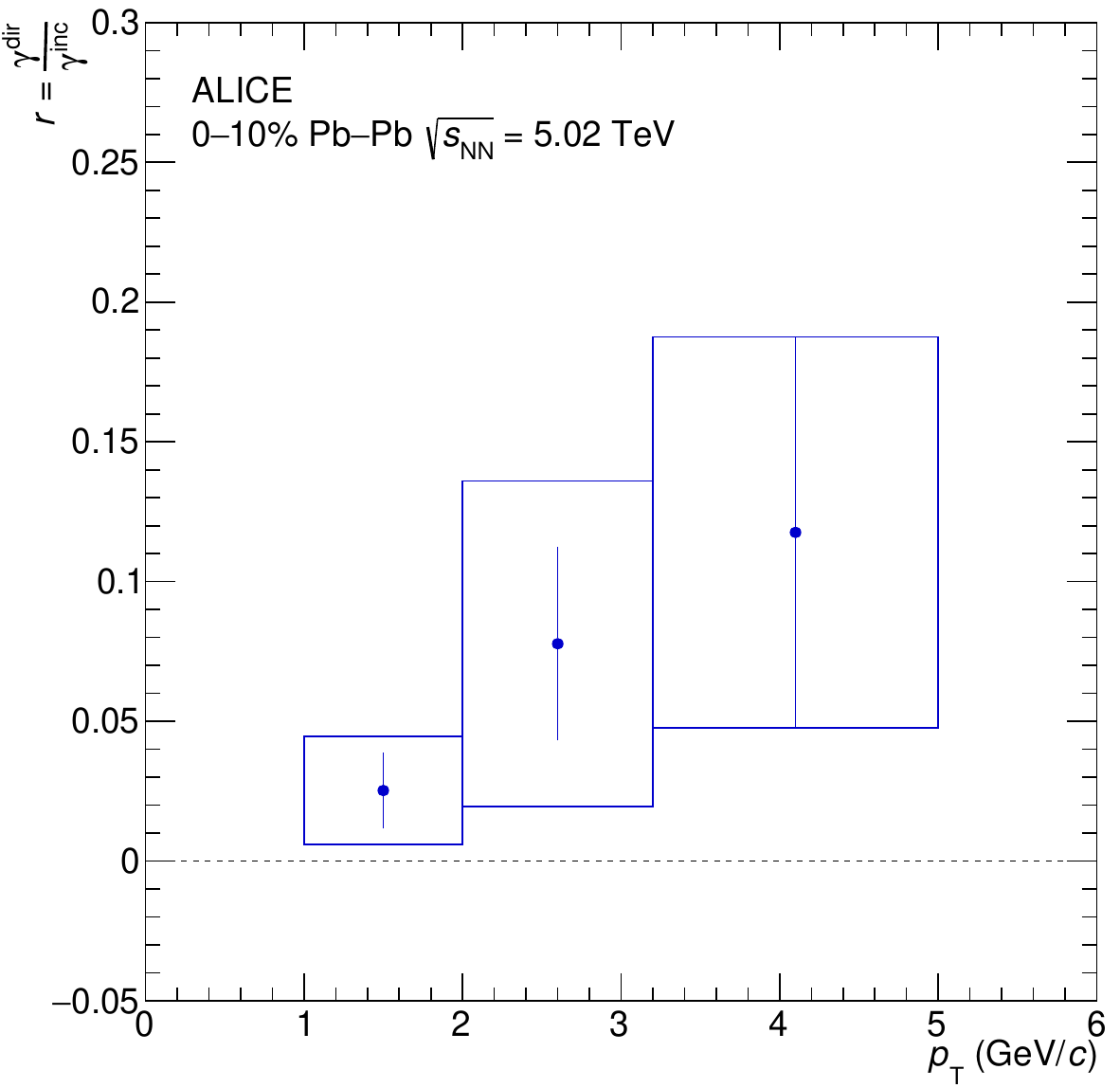}    
    \end{center}
    \caption{Left panel: fit of the dielectron yield with the three-component function defined by Eq.~\ref{equationdir} in the 10\% most central \PbPb collisions at \fivenn as a function of \mee in the \ptee range $1 < \ptee < 2$~\GeVc. Right panel: ratio of direct to inclusive photon yields extracted from the dielectron spectra as a function of \pt 
    in central \PbPb collisions at \fivenn. Statistical and systematic uncertainties on the data are shown separately as vertical bars and boxes, respectively.}
    \label{fig:fitandRgamma}
\end{figure}

In the quasi-real virtual-photon region where the \ptee of the dilepton pair is much larger than its mass ($p^{2}_{\rm T,ee} \gg m^{2}_{\rm ee}$), the fraction of virtual direct photons over inclusive photons can be extracted from the measured dielectron yield, as performed for the first time by the PHENIX Collaboration\,\cite{PHENIX:2008uif,PHENIX:2009gyd}. To this end, the \mee spectra in different \ptee intervals are fitted in the range $0.12 < \mee < 0.34$~\GeVmass with a three-component function
\begin{equation}
f{\rm (}\mee{\rm )} = r f_{\rm dir}{\rm (}\mee{\rm )} + {\rm ( 1 -}r{\rm )} f_{\rm LF}{\rm (}\mee{\rm )} + f_{\rm HF}{\rm (}\mee{\rm )}.
\label{equationdir}
\end{equation}
In the above equation, $f_{\rm dir}{\rm (}\mee{\rm )}$ is the expected invariant-mass distribution of virtual direct photons, described by the Kroll--Wada equation\,\cite{Kroll:1955zu}, and $f_{\rm LF}{\rm (}\mee{\rm )}$ and $f_{\rm HF}{\rm (}\mee{\rm )}$ are the mass distributions of the light-flavor contributions and the heavy-flavor components of the hadronic cocktail, respectively. Cocktail2 including the $\rho$ contribution and some heavy-flavor medium modifications as estimated in Section~\ref{cocktailinvariantmasslight} and ~\ref{cocktailinvariantmassheavyflavour}, respectively, is used for this purpose. Both $f_{\rm LF}{\rm (} \mee {\rm ) }$ and $f_{\rm dir}{\rm (} \mee {\rm ) }$ are independently normalized to the data in the mass range $0 < \mee < 0.03$~\GeVmass, whereas $f_{\rm HF}{\rm (}\mee{\rm )}$ is normalized accordingly to the hadronic cocktail. The only fit parameter $r$ represents the fraction of virtual direct photons at vanishing mass, directly related to the one of real direct photons:
\begin{equation}
r = \left(\frac{\gamma^{*\,\rm dir}}{\gamma^{*\,\rm inc}}\right)_{\mee\to 0} = \left(\frac{\gamma^{\rm dir}}{\gamma^{\rm inc}}\right ).
\end{equation}
This approach has the advantage that the background of photons originating from hadronic decays, dominated by $\pi^{0}$ decays, can be suppressed by selecting \ee pairs with sufficiently large \mee ($\mee > m_{\rm \pi^{0}}$). However, the internal conversion probability is small ($\approx 10^{-2}$) and the dielectron yield decreases rapidly as a function of \mee ($\propto 1/\mee$). The fit of the measured \mee-differential dielectron yield in the \ptee range $1 < \ptee < 2$~\GeVc is shown in the left panel of Fig.~\ref{fig:fitandRgamma}, together with its individual contributions.

Systematic uncertainties arise from the data, the hadronic cocktail components and the choice of the normalization and fit ranges. They are estimated in a similar way as in Refs.~\cite{ALICE:2018fvj,ALICE:2018gev,ALICE:2018ael}. On the one hand, the uncertainties from the data are evaluated by shifting all data points coherently to their upper and lower systematic uncertainties and by repeating the fit procedure. On the other hand, the different contributions of the cocktail are shifted separately according to their respective uncertainties. The limited knowledge of the $\eta/\pi^{0}$ ratio dominates the systematic uncertainties for the first \ptee interval, whereas the choice of the fit range becomes relevant with increasing \ptee. The latter is estimated by varying the lower bound of the fit, keeping the upper one fixed to assure the $p^{2}_{\rm T,ee} \gg m^{2}_{\rm ee}$ condition and avoid a significant contribution from $\rho$ hadron decays. The total systematic uncertainties on $r$ are obtained by summing in quadrature all individual contributions and are ranging from 78\% to 60\% from low to high \pt.

The ratio of direct to inclusive photons ($r$) measured in the 10\% most central \PbPb collisions is shown as a function of \pt ($=\ptee$) in the right panel of Fig.~\ref{fig:fitandRgamma}. 
The data differ from zero suggesting a source of photons that is not originating from hadronic decays.
The significance of a direct photon signal is ranging from $1.05$ to $1.2 \sigma$ from the low to high \pt interval for virtual photons.

\subsection{Direct-photon \texorpdfstring{\pt}{pt} spectrum}

\begin{figure}[tbh]
    \begin{center}
    \includegraphics[width = 0.6\textwidth]{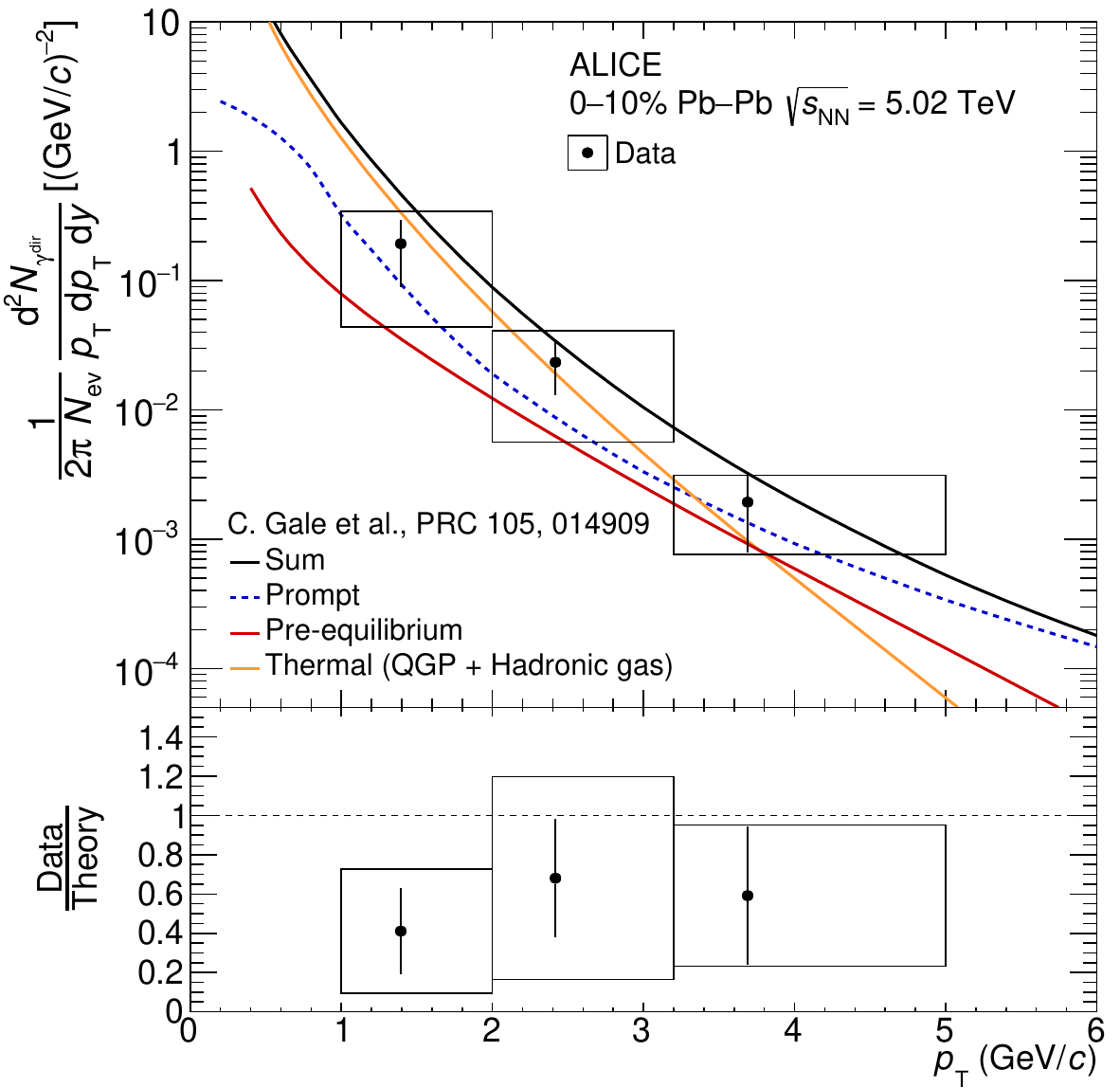}
    \end{center}
    \caption{Direct-photon invariant yield in the 10\% most central Pb--Pb collisions at \fivenn, compared with the predictions from a state-of-the-art model~\cite{Gale:2021emg}. The error bars and boxes represent the statistical and systematic uncertainties of the data, respectively.}
    \label{fig:dirphotonspectrum}
\end{figure}

The \pt-differential invariant yield of direct photons can be obtained from $r$ and the inclusive photon yield ($\frac{{\rm d}N_{\rm \gamma^{\rm inc}}}{{\rm d}p_{\rm T}}$). The latter is determined for each $p_{\rm T}$ interval from the yield of low mass \ee pairs in the range $\mee < 30$~MeV/$c^{\rm 2}$ ($\frac{{\rm d}N_{\rm ee}^{\rm data}}{{\rm d}p_{\rm T}}$) following the approach used by the PHENIX collaboration~\cite{PHENIX:2009gyd}. The relation between real photon production and the associated \ee production becomes process independent for very low \mee (within a few percent for $\mee < 30$~MeV/$c^{\rm 2}$). The ratio of the expected yield of real photons from known hadronic decays ($\frac{{\rm d}N_{\rm \gamma^{\rm cocktail}}}{{\rm d}p_{\rm T}}$) to the one of dielectrons from the same hadronic sources for $\mee < 30$~MeV/$c^{\rm 2}$ ($\frac{{\rm d}N_{\rm ee}^{\rm cocktail}}{{\rm d}p_{\rm T}}$) can be then used to estimate the inclusive real photon yield from the dielectron yield at low \mee, i.e. 
\begin{equation}
\frac{{\rm d}N_{\rm \gamma^{\rm inc}}}{{\rm d}p_{\rm T}} = \frac{{\rm d}N_{\rm ee}^{\rm data}}{{\rm d}p_{\rm T}} \times \left (\frac{{\rm d}N_{\rm \gamma^{\rm cocktail}}}{{\rm d}p_{\rm T}}\middle /\frac{{\rm d}N_{\rm ee}^{\rm cocktail}}{{\rm d}p_{\rm T}}\right ).
\label{ee}
\end{equation}
Most of the uncertainties related to the parameterized $\pt$ spectra used as input for the cocktail calculations cancel in the cocktail ratio. The resulting direct-photon \pt-differential spectrum in the 10\% most central \PbPb collisions at \fivenn is shown in Fig.~\ref{fig:dirphotonspectrum}. Within each \pt bin, the mean \pt is calculated with an iterative procedure using an exponential function to interpolate the \pt spectrum~\cite{LAFFERTY1995541}. The data are compared to a hybrid model that describes all stages of the heavy-ion collisions~\cite{Gale:2021emg}. The calculations include the contribution from prompt photons computed with next-to-leading-order perturbative QCD using INCNLO~\cite{Aurenche:1998gv}, nCTEQ15 parton distribution functions corrected for nuclear matter effects~\cite{Kovarik:2015cma}, and BFG-II fragmentation functions~\cite{Bourhis:1997yu}. The measured direct-photon invariant yield is consistent with the prompt photon contribution alone, but the central values of the data points are systematically higher. The model predicts additional contributions from the pre-equilibrium, as well as fluid-dynamical, phases. The very early nuclear medium is described using the IP-Glasma model~\cite{Schenke:2012wb,Schenke:2012hg} with a time-evolution determined by Yang--Mills equations. It is followed by an out-of-equilibrium phase where the energy-momentum tensor is evolved with non-equilibrium linear response functions~\cite{Kurkela:2018wud,Kurkela:2018vqr}. This pre-hydrodynamical K$\varnothing$MP$\varnothing$ST~\cite{Kurkela:2018vqr} phase feeds into a fluid-dynamical evolution, modelled with MUSIC~\cite{Schenke:2010nt}. The contribution of thermal photons is obtained by integrating photon emission rates over the entire space-time volume occupied by the fluid-dynamical phase. Thermal photons are the largest source of direct photons at $\pt < 3$~\GeVc, whereas the pre-equilibrium contribution never dominates, but exceeds the thermal contribution for $\pt > 3.5$~\GeVc. The predicted invariant yield of all direct-photon contributions together can describe the measurement, although it tends to overestimate the data by about $1 \sigma$. A better description of the data, within less than one sigma, is obtained if the thermal contribution in the model is neglected. While it is questionable why the dilepton emission rates would have to be considered only before the system reaches local equilibrium, it may point to a wrong equilibration time in the calculations.

\subsection{Charged-particle multiplicity dependence of direct-photon production}

\begin{figure}[tbh]
    \begin{center}
    \includegraphics[width = 0.6\textwidth]{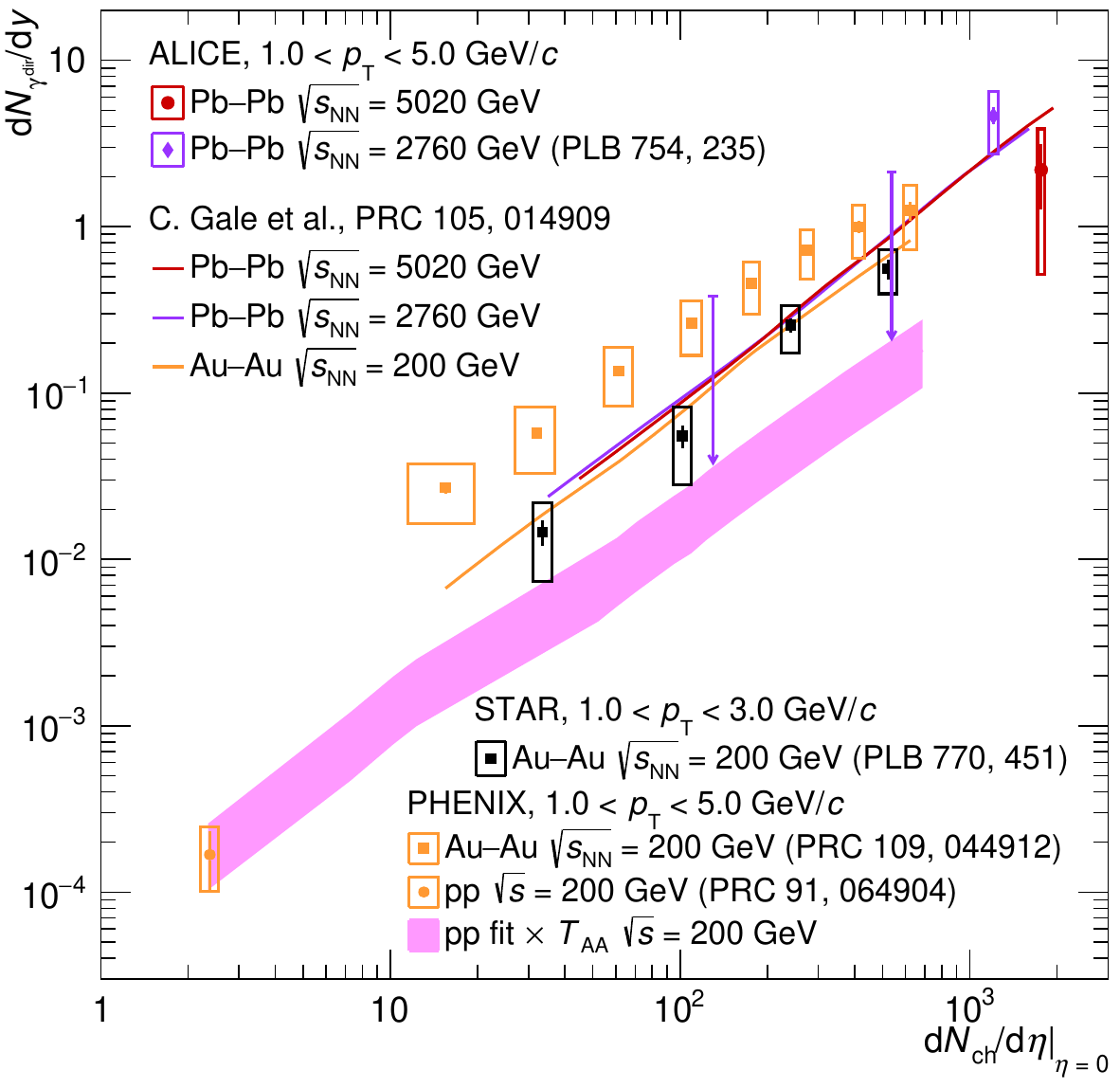}
    \end{center}
    \caption{Integrated direct-photon yield in the \pt range $1 < \pt < 5$~\GeVc (or $1 < \pt < 3$~\GeVc) in Pb--Pb collisions at \fivenn and \twosevensixnn~\cite{ALICE:2015xmh}, in \AuAu collisions at \twoHnn as measured by the PHENIX~\cite{PHENIX:2022rsx} and STAR~\cite{STAR:2016use} collaborations, and in \pp collisions at \twoH~\cite{PHENIX:2014nkk}. The data are compared to predictions from a state-of-the-art model~\cite{Gale:2021emg}. The error bars and boxes represent the statistical and systematic
uncertainties of the data along the y-axis, respectively. The horizontal width of the boxes shows the total uncertainty of the measured $\frac{{\rm d}N_{\rm ch}}{{\rm d}\eta}|_{\eta = 0}$~\cite{ALICE:2015juo}.
}
    \label{fig:dirphotondnde}
\end{figure}

The \pt-integrated direct-photon yields are shown in Fig.~\ref{fig:dirphotondnde} as a function of the measured charged-particle multiplicity at midrapidity~\cite{ALICE:2015juo} for \PbPb collisions at $\snn = 5.02$ (this analysis) and $2.76$~TeV~\cite{ALICE:2015xmh}, for \AuAu collisions at \twoHnn~\cite{PHENIX:2022rsx,STAR:2016use}, as well as for \pp collisions at \twoH~\cite{PHENIX:2014nkk}. The yields are integrated over the \pt range $1 < \pt < 5$~\GeVc, except for the STAR data where the range is $1 < \pt < 3$~\GeVc. The systematic uncertainties are assumed fully correlated as a function of \pt for the results at \fivenn. The magenta band gives the integrated direct-photon yield in \pp collisions at \twoH, scaled by the number of binary nucleon--nucleon collisions to the corresponding $\frac{{\rm d}N_{\rm ch}}{{\rm d}\eta}|_{\eta = 0}$ for \AuAu collisions at \twoHnn. The data are compared with predictions from the same model as the one used for the \pt-differential invariant yield discussed above~\cite{Gale:2021emg}. The calculations can reproduce the measured direct-photon yields at the LHC and in central \AuAu collisions at RHIC, although they tend to slightly overestimate the observed yield at the highest charged-particle multiplicity reported here. At lower $\frac{{\rm d}N_{\rm ch}}{{\rm d}\eta}|_{\eta = 0}$ ($\frac{{\rm d}N_{\rm ch}}{{\rm d}\eta}|_{\eta = 0} < 300$), the predicted direct-photon yields are smaller than the measured values by the PHENIX collaboration, with a difference increasing when moving to lower multiplicity. However, the comparison between the calculations and the data reported by the STAR collaboration for the same colliding system shows a better agreement. The discrepancy between the PHENIX and STAR results remains unresolved to this day.

The PHENIX collaboration reported an approximate power-law dependence on $\frac{{\rm d}N_{\rm ch}}{{\rm d}\eta}|_{\eta = 0}$ of the direct-photon yields at low \pt, with a power $\alpha$ independent of centrality or collision energy and similar to what would correspond to a scaling with the number of binary nucleon--nucleon collisions ($\alpha = 1.25 \pm 0.02$) shown with the magenta band in Fig.~\ref{fig:dirphotondnde}. However, the extracted value of $\alpha$ from the latest PHENIX \AuAu data presented here is smaller than 1.25, i.e. $\alpha = 1.11\pm 0.02\ (\rm stat.)^{+0.09}_{-0.08}\ (\rm syst.)$~\cite{PHENIX:2022rsx}. For comparison, the calculations predict $\alpha$ values of $1.25$, $1.32$ and $1.34$ for \AuAu collisions at \twoHnn and for \PbPb collisions at $\snn = 2.76$ and $5.02$~TeV, respectively. The results at LHC energies are not yet sensitive to the difference between the  different predicted or measured $\alpha$ values.

\section{Summary and outlook}
\label{summary}

The first measurement at the LHC of low-mass dielectrons ($0 < \mee < 3.5$\,\GeVmass) in the 10\% most central \PbPb collisions at \fivenn is presented. The \ee pair production is measured with the ALICE detector at midrapidity ($|\eta_{\rm e}| < 0.8$) and compared with a cocktail of expected contributions from decays of known hadrons. The data to cocktail ratio for $0.18 < \mee < 0.5$\,\GeVmass amounts to $1.40 \pm 0.11\  (\rm stat.) \pm 0.23\  (\rm syst.) \pm 0.16\  (\rm cocktail)$ for a cocktail where the background from open heavy-flavor hadron decays is taken from dielectron measurements in \pp collisions at the same center-of-mass energy per nucleon pair~\cite{ALICE:2020mfy} and scaled with $N_{\rm coll}$, and $1.42 \pm 0.11\  (\rm stat.) \pm 0.23\  (\rm syst.) ^{+0.24}_{-0.29}\  (\rm cocktail)$ for a cocktail including a modification of the heavy-flavor background based on the measured single electrons from heavy-flavor hadron decays~\cite{ALICE:2019nuy} and the EPS09 nPDF~\cite{Eskola:2009uj}. The latter approach suffers from large uncertainties. The excess spectra, obtained after subtracting the hadronic cocktails excluding the $\rho$ contribution, are consistent with predictions from two models that include thermal radiation from the partonic and hadronic phases~\cite{Rapp:2013nxa,Song:2018xca}, the latter being dominant in the \mee range considered ($0.2 < \mee < 0.7$~\GeVmass). The uncertainties of the data do not allow the extraction of a significant thermal signal. A different method, based on the pair transverse impact parameter of the \ee pairs, is used for the first time in \PbPb collisions to disentangle displaced dielectrons from heavy-flavor hadron decays from a prompt thermal contribution. The data are found to be consistent with a large suppression of \ee pairs from ${\rm c\overline{c}}$, larger than for \ee pairs from ${\rm b\overline{b}}$, and an additional prompt component in the intermediate-mass range ($1.2 < \mee < 2.6$~\GeVmass), where thermal radiation from the QGP is predicted~\cite{Rapp:2013nxa,Song:2018xca}. The small systematic uncertainties of the results, limited by the statistical precision of the data, show the potential of such an approach for future ALICE dielectron studies with the upgraded LHC and ALICE detector, for which a significant increase of the pointing detector resolution, as well as of the size of the data sample, are expected~\cite{ALICE:2023udb}. In the quasi-real virtual-photon region ($\mee < 0.34$~\GeVmass and $1 < \ptee < 5$~\GeVc), the fraction of direct photons over inclusive photons is obtained from a fit of the \mee spectra above the $\pi^{0}$ mass. Combined with the inclusive real-photon yield estimated from the inclusive dielectron yield at very low \mee, the first direct-photon measurement in the 10\% most central Pb--Pb collisions at \fivenn is reported. A model~\cite{Gale:2021emg} including prompt photons, as well as photons from the pre-equilibrium and fluid-dynamical (thermal photons) phases, can reproduce the results, while being at the upper edge of the data uncertainties.

\section*{Data availability}

This manuscript has associated data in a HEPData repository at~\cite{HEPData}.


\newenvironment{acknowledgement}{\relax}{\relax}
\begin{acknowledgement}
\section*{Acknowledgements}

The ALICE Collaboration would like to thank all its engineers and technicians for their invaluable contributions to the construction of the experiment and the CERN accelerator teams for the outstanding performance of the LHC complex.
The ALICE Collaboration gratefully acknowledges the resources and support provided by all Grid centres and the Worldwide LHC Computing Grid (WLCG) collaboration.
The ALICE Collaboration acknowledges the following funding agencies for their support in building and running the ALICE detector:
A. I. Alikhanyan National Science Laboratory (Yerevan Physics Institute) Foundation (ANSL), State Committee of Science and World Federation of Scientists (WFS), Armenia;
Austrian Academy of Sciences, Austrian Science Fund (FWF): [M 2467-N36] and Nationalstiftung f\"{u}r Forschung, Technologie und Entwicklung, Austria;
Ministry of Communications and High Technologies, National Nuclear Research Center, Azerbaijan;
Conselho Nacional de Desenvolvimento Cient\'{\i}fico e Tecnol\'{o}gico (CNPq), Financiadora de Estudos e Projetos (Finep), Funda\c{c}\~{a}o de Amparo \`{a} Pesquisa do Estado de S\~{a}o Paulo (FAPESP) and Universidade Federal do Rio Grande do Sul (UFRGS), Brazil;
Bulgarian Ministry of Education and Science, within the National Roadmap for Research Infrastructures 2020-2027 (object CERN), Bulgaria;
Ministry of Education of China (MOEC) , Ministry of Science \& Technology of China (MSTC) and National Natural Science Foundation of China (NSFC), China;
Ministry of Science and Education and Croatian Science Foundation, Croatia;
Centro de Aplicaciones Tecnol\'{o}gicas y Desarrollo Nuclear (CEADEN), Cubaenerg\'{\i}a, Cuba;
Ministry of Education, Youth and Sports of the Czech Republic, Czech Republic;
The Danish Council for Independent Research | Natural Sciences, the VILLUM FONDEN and Danish National Research Foundation (DNRF), Denmark;
Helsinki Institute of Physics (HIP), Finland;
Commissariat \`{a} l'Energie Atomique (CEA) and Institut National de Physique Nucl\'{e}aire et de Physique des Particules (IN2P3) and Centre National de la Recherche Scientifique (CNRS), France;
Bundesministerium f\"{u}r Bildung und Forschung (BMBF) and GSI Helmholtzzentrum f\"{u}r Schwerionenforschung GmbH, Germany;
General Secretariat for Research and Technology, Ministry of Education, Research and Religions, Greece;
National Research, Development and Innovation Office, Hungary;
Department of Atomic Energy Government of India (DAE), Department of Science and Technology, Government of India (DST), University Grants Commission, Government of India (UGC) and Council of Scientific and Industrial Research (CSIR), India;
National Research and Innovation Agency - BRIN, Indonesia;
Istituto Nazionale di Fisica Nucleare (INFN), Italy;
Japanese Ministry of Education, Culture, Sports, Science and Technology (MEXT) and Japan Society for the Promotion of Science (JSPS) KAKENHI, Japan;
Consejo Nacional de Ciencia (CONACYT) y Tecnolog\'{i}a, through Fondo de Cooperaci\'{o}n Internacional en Ciencia y Tecnolog\'{i}a (FONCICYT) and Direcci\'{o}n General de Asuntos del Personal Academico (DGAPA), Mexico;
Nederlandse Organisatie voor Wetenschappelijk Onderzoek (NWO), Netherlands;
The Research Council of Norway, Norway;
Commission on Science and Technology for Sustainable Development in the South (COMSATS), Pakistan;
Pontificia Universidad Cat\'{o}lica del Per\'{u}, Peru;
Ministry of Education and Science, National Science Centre and WUT ID-UB, Poland;
Korea Institute of Science and Technology Information and National Research Foundation of Korea (NRF), Republic of Korea;
Ministry of Education and Scientific Research, Institute of Atomic Physics, Ministry of Research and Innovation and Institute of Atomic Physics and Universitatea Nationala de Stiinta si Tehnologie Politehnica Bucuresti, Romania;
Ministry of Education, Science, Research and Sport of the Slovak Republic, Slovakia;
National Research Foundation of South Africa, South Africa;
Swedish Research Council (VR) and Knut \& Alice Wallenberg Foundation (KAW), Sweden;
European Organization for Nuclear Research, Switzerland;
Suranaree University of Technology (SUT), National Science and Technology Development Agency (NSTDA) and National Science, Research and Innovation Fund (NSRF via PMU-B B05F650021), Thailand;
Turkish Energy, Nuclear and Mineral Research Agency (TENMAK), Turkey;
National Academy of  Sciences of Ukraine, Ukraine;
Science and Technology Facilities Council (STFC), United Kingdom;
National Science Foundation of the United States of America (NSF) and United States Department of Energy, Office of Nuclear Physics (DOE NP), United States of America.
In addition, individual groups or members have received support from:
Czech Science Foundation (grant no. 23-07499S), Czech Republic;
European Research Council, Strong 2020 - Horizon 2020 (grant nos. 950692, 824093), European Union;
ICSC - Centro Nazionale di Ricerca in High Performance Computing, Big Data and Quantum Computing, European Union - NextGenerationEU;
Academy of Finland (Center of Excellence in Quark Matter) (grant nos. 346327, 346328), Finland.

\end{acknowledgement}

\bibliographystyle{utphys}   
\bibliography{bibliography}

\providecommand{\href}[2]{#2}\begingroup\raggedright\begin{thebibliography}{100}

\bibitem{ALICE:2022wpn}
{\bfseries ALICE} Collaboration, S.~Acharya {\em et~al.}, ``{The ALICE
  experiment: a journey through QCD}'',
  \href{https://doi.org/10.1140/epjc/s10052-024-12935-y}{{\em Eur. Phys. J. C}
  {\bfseries 84} (2024) 813},
  \href{https://arxiv.org/abs/2211.04384}{{\ttfamily arXiv:2211.04384
  [nucl-ex]}}.

\bibitem{Bazavov:2011nk}
A.~Bazavov {\em et~al.}, ``{The chiral and deconfinement aspects of the QCD
  transition}'', \href{https://doi.org/10.1103/PhysRevD.85.054503}{{\em Phys.
  Rev. D} {\bfseries 85} (2012) 054503},
  \href{https://arxiv.org/abs/1111.1710}{{\ttfamily arXiv:1111.1710
  [hep-lat]}}.

\bibitem{Borsanyi:2013bia}
S.~Borsanyi, Z.~Fodor, C.~Hoelbling, S.~D. Katz, S.~Krieg, and K.~K. Szabo,
  ``{Full result for the QCD equation of state with 2+1 flavors}'',
  \href{https://doi.org/10.1016/j.physletb.2014.01.007}{{\em Phys. Lett. B}
  {\bfseries 730} (2014) 99--104},
  \href{https://arxiv.org/abs/1309.5258}{{\ttfamily arXiv:1309.5258
  [hep-lat]}}.

\bibitem{HotQCD:2018pds}
{\bfseries HotQCD} Collaboration, A.~Bazavov {\em et~al.}, ``{Chiral crossover
  in QCD at zero and non-zero chemical potentials}'',
  \href{https://doi.org/10.1016/j.physletb.2019.05.013}{{\em Phys. Lett. B}
  {\bfseries 795} (2019) 15--21},
  \href{https://arxiv.org/abs/1812.08235}{{\ttfamily arXiv:1812.08235
  [hep-lat]}}.

\bibitem{Borsanyi:2020fev}
S.~Borsanyi, Z.~Fodor, J.~N. Guenther, R.~Kara, S.~D. Katz, P.~Parotto,
  A.~Pasztor, C.~Ratti, and K.~K. Szabo, ``{QCD Crossover at Finite Chemical
  Potential from Lattice Simulations}'',
  \href{https://doi.org/10.1103/PhysRevLett.125.052001}{{\em Phys. Rev. Lett.}
  {\bfseries 125} (2020) 052001},
  \href{https://arxiv.org/abs/2002.02821}{{\ttfamily arXiv:2002.02821
  [hep-lat]}}.

\bibitem{Bilic:1997sh}
N.~Bilic and H.~Nikolic, ``{Chiral symmetry restoration in the linear sigma
  model at nonzero temperature and baryon density}'',
  \href{https://doi.org/10.1007/s100529800923}{{\em Eur. Phys. J. C} {\bfseries
  6} (1999) 515--523}, \href{https://arxiv.org/abs/hep-ph/9711513}{{\ttfamily
  arXiv:hep-ph/9711513}}.

\bibitem{Dominguez:2012bs}
C.~A. Dominguez, M.~Loewe, and Y.~Zhang, ``{Chiral symmetry restoration and
  deconfinement in QCD at finite temperature}'',
  \href{https://doi.org/10.1103/PhysRevD.86.034030}{{\em Phys. Rev. D}
  {\bfseries 86} (2012) 034030},
  \href{https://arxiv.org/abs/1205.3361}{{\ttfamily arXiv:1205.3361 [hep-ph]}}.
  [Erratum: Phys.Rev.D 90, 039903 (2014)].

\bibitem{Gale:2009gc}
C.~Gale, ``{Photon Production in Hot and Dense Strongly Interacting Matter}'',
  \href{https://doi.org/10.1007/978-3-642-01539-7_15}{{\em Landolt-Bornstein}
  {\bfseries 23} (2010) 445}, \href{https://arxiv.org/abs/0904.2184}{{\ttfamily
  arXiv:0904.2184 [hep-ph]}}.

\bibitem{Turbide:2005fk}
S.~Turbide, C.~Gale, S.~Jeon, and G.~D. Moore, ``{Energy loss of leading
  hadrons and direct photon production in evolving quark-gluon plasma}'',
  \href{https://doi.org/10.1103/PhysRevC.72.014906}{{\em Phys. Rev. C}
  {\bfseries 72} (2005) 014906},
  \href{https://arxiv.org/abs/hep-ph/0502248}{{\ttfamily
  arXiv:hep-ph/0502248}}.

\bibitem{Qin:2009bk}
G.-Y. Qin, J.~Ruppert, C.~Gale, S.~Jeon, and G.~D. Moore, ``{Jet energy loss,
  photon production, and photon-hadron correlations at RHIC}'',
  \href{https://doi.org/10.1103/PhysRevC.80.054909}{{\em Phys. Rev. C}
  {\bfseries 80} (2009) 054909},
  \href{https://arxiv.org/abs/0906.3280}{{\ttfamily arXiv:0906.3280 [hep-ph]}}.

\bibitem{Kapusta:1991qp}
J.~I. Kapusta, P.~Lichard, and D.~Seibert, ``{High-energy photons from
  quark-gluon plasma versus hot hadronic gas}'',
  \href{https://doi.org/10.1103/PhysRevD.47.4171}{{\em Phys. Rev. D} {\bfseries
  44} (1991) 2774--2788}. [Erratum: Phys.Rev.D 47, 4171 (1993)].

\bibitem{Turbide:2003si}
S.~Turbide, R.~Rapp, and C.~Gale, ``{Hadronic production of thermal photons}'',
  \href{https://doi.org/10.1103/PhysRevC.69.014903}{{\em Phys. Rev. C}
  {\bfseries 69} (2004) 014903},
  \href{https://arxiv.org/abs/hep-ph/0308085}{{\ttfamily
  arXiv:hep-ph/0308085}}.

\bibitem{Gale:2021emg}
C.~Gale, J.-F. Paquet, B.~Schenke, and C.~Shen, ``{Multimessenger heavy-ion
  collision physics}'', \href{https://doi.org/10.1103/PhysRevC.105.014909}{{\em
  Phys. Rev. C} {\bfseries 105} (2022) 014909},
  \href{https://arxiv.org/abs/2106.11216}{{\ttfamily arXiv:2106.11216
  [nucl-th]}}.

\bibitem{WA98:2000vxl}
{\bfseries WA98} Collaboration, M.~M. Aggarwal {\em et~al.}, ``{Observation of
  direct photons in central 158-A-GeV Pb-208 + Pb-208 collisions}'',
  \href{https://doi.org/10.1103/PhysRevLett.85.3595}{{\em Phys. Rev. Lett.}
  {\bfseries 85} (2000) 3595--3599},
  \href{https://arxiv.org/abs/nucl-ex/0006008}{{\ttfamily
  arXiv:nucl-ex/0006008}}.

\bibitem{PHENIX:2005yls}
{\bfseries PHENIX} Collaboration, S.~S. Adler {\em et~al.}, ``{Centrality
  dependence of direct photon production in $\sqrt{s_{\rm NN}}=200$~GeV Au$+$Au
  collisions}'', \href{https://doi.org/10.1103/PhysRevLett.94.232301}{{\em
  Phys. Rev. Lett.} {\bfseries 94} (2005) 232301},
  \href{https://arxiv.org/abs/nucl-ex/0503003}{{\ttfamily
  arXiv:nucl-ex/0503003}}.

\bibitem{PHENIX:2008uif}
{\bfseries PHENIX} Collaboration, A.~Adare {\em et~al.}, ``{Enhanced production
  of direct photons in Au+Au collisions at $\sqrt{s_{_{\rm NN}}}=200$~GeV and
  implications for the initial temperature}'',
  \href{https://doi.org/10.1103/PhysRevLett.104.132301}{{\em Phys. Rev. Lett.}
  {\bfseries 104} (2010) 132301},
  \href{https://arxiv.org/abs/0804.4168}{{\ttfamily arXiv:0804.4168
  [nucl-ex]}}.

\bibitem{PHENIX:2011oxq}
{\bfseries PHENIX} Collaboration, A.~Adare {\em et~al.}, ``{Observation of
  direct-photon collective flow in $\sqrt{s_{\rm NN}}=200$~GeV Au$+$Au
  collisions}'', \href{https://doi.org/10.1103/PhysRevLett.109.122302}{{\em
  Phys. Rev. Lett.} {\bfseries 109} (2012) 122302},
  \href{https://arxiv.org/abs/1105.4126}{{\ttfamily arXiv:1105.4126
  [nucl-ex]}}.

\bibitem{PHENIX:2014nkk}
{\bfseries PHENIX} Collaboration, A.~Adare {\em et~al.}, ``{Centrality
  dependence of low-momentum direct-photon production in Au$+$Au collisions at
  $\sqrt{s_{_{\rm NN}}}=200$~GeV}'',
  \href{https://doi.org/10.1103/PhysRevC.91.064904}{{\em Phys. Rev. C}
  {\bfseries 91} (2015) 064904},
  \href{https://arxiv.org/abs/1405.3940}{{\ttfamily arXiv:1405.3940
  [nucl-ex]}}.

\bibitem{PHENIX:2022rsx}
{\bfseries PHENIX} Collaboration, N.~J. Abdulameer {\em et~al.}, ``{Nonprompt
  direct-photon production in Au+Au collisions at sNN=200 GeV}'',
  \href{https://doi.org/10.1103/PhysRevC.109.044912}{{\em Phys. Rev. C}
  {\bfseries 109} (2024) 044912},
  \href{https://arxiv.org/abs/2203.17187}{{\ttfamily arXiv:2203.17187
  [nucl-ex]}}.

\bibitem{PHENIX:2015igl}
{\bfseries PHENIX} Collaboration, A.~Adare {\em et~al.}, ``{Azimuthally
  anisotropic emission of low-momentum direct photons in Au$+$Au collisions at
  $\sqrt{s_{_{\rm NN}}}=200$~GeV}'',
  \href{https://doi.org/10.1103/PhysRevC.94.064901}{{\em Phys. Rev. C}
  {\bfseries 94} (2016) 064901},
  \href{https://arxiv.org/abs/1509.07758}{{\ttfamily arXiv:1509.07758
  [nucl-ex]}}.

\bibitem{PHENIX:2018for}
{\bfseries PHENIX} Collaboration, A.~Adare {\em et~al.}, ``{Beam Energy and
  Centrality Dependence of Direct-Photon Emission from Ultrarelativistic
  Heavy-Ion Collisions}'',
  \href{https://doi.org/10.1103/PhysRevLett.123.022301}{{\em Phys. Rev. Lett.}
  {\bfseries 123} (2019) 022301},
  \href{https://arxiv.org/abs/1805.04084}{{\ttfamily arXiv:1805.04084
  [hep-ex]}}.

\bibitem{PHENIX:2022qfp}
{\bfseries PHENIX} Collaboration, N.~J. Abdulameer {\em et~al.}, ``{Low-pT
  direct-photon production in Au+Au collisions at $\sqrt{s_{_{\rm NN}}}=39$ and
  62.4~GeV}'', \href{https://doi.org/10.1103/PhysRevC.107.024914}{{\em Phys.
  Rev. C} {\bfseries 107} (2023) 024914},
  \href{https://arxiv.org/abs/2203.12354}{{\ttfamily arXiv:2203.12354
  [nucl-ex]}}.

\bibitem{PHENIX:2006duh}
{\bfseries PHENIX} Collaboration, S.~S. Adler {\em et~al.}, ``{Measurement of
  direct photon production in $p+p$ collisions $\sqrt{s}=200$~GeV}'',
  \href{https://doi.org/10.1103/PhysRevLett.98.012002}{{\em Phys. Rev. Lett.}
  {\bfseries 98} (2007) 012002},
  \href{https://arxiv.org/abs/hep-ex/0609031}{{\ttfamily
  arXiv:hep-ex/0609031}}.

\bibitem{PHENIX:2012jgx}
{\bfseries PHENIX} Collaboration, A.~Adare {\em et~al.}, ``{Direct-Photon
  Production in $p+p$ Collisions at $\sqrt{s}=200$~GeV at Midrapidity}'',
  \href{https://doi.org/10.1103/PhysRevD.86.072008}{{\em Phys. Rev. D}
  {\bfseries 86} (2012) 072008},
  \href{https://arxiv.org/abs/1205.5533}{{\ttfamily arXiv:1205.5533 [hep-ex]}}.

\bibitem{PHENIX:2012krx}
{\bfseries PHENIX} Collaboration, A.~Adare {\em et~al.}, ``{Direct photon
  production in d$+$Au collisions at $\sqrt{s_{\rm NN}}=200$~GeV}'',
  \href{https://doi.org/10.1103/PhysRevC.87.054907}{{\em Phys. Rev. C}
  {\bfseries 87} (2013) 054907},
  \href{https://arxiv.org/abs/1208.1234}{{\ttfamily arXiv:1208.1234
  [nucl-ex]}}.

\bibitem{Paquet:2015lta}
J.-F. Paquet, C.~Shen, G.~S. Denicol, M.~Luzum, B.~Schenke, S.~Jeon, and
  C.~Gale, ``{Production of photons in relativistic heavy-ion collisions}'',
  \href{https://doi.org/10.1103/PhysRevC.93.044906}{{\em Phys. Rev. C}
  {\bfseries 93} (2016) 044906},
  \href{https://arxiv.org/abs/1509.06738}{{\ttfamily arXiv:1509.06738
  [hep-ph]}}.

\bibitem{STAR:2016use}
{\bfseries STAR} Collaboration, L.~Adamczyk {\em et~al.}, ``{Direct virtual
  photon production in Au$+$Au collisions at $\sqrt{s_{\rm NN}}= 200$~GeV}'',
  \href{https://doi.org/10.1016/j.physletb.2017.04.050}{{\em Phys. Lett. B}
  {\bfseries 770} (2017) 451--458},
  \href{https://arxiv.org/abs/1607.01447}{{\ttfamily arXiv:1607.01447
  [nucl-ex]}}.

\bibitem{Chatterjee:2013naa}
R.~Chatterjee, H.~Holopainen, I.~Helenius, T.~Renk, and K.~J. Eskola,
  ``{Elliptic flow of thermal photons from event-by-event hydrodynamic
  model}'', \href{https://doi.org/10.1103/PhysRevC.88.034901}{{\em Phys. Rev.
  C} {\bfseries 88} (2013) 034901},
  \href{https://arxiv.org/abs/1305.6443}{{\ttfamily arXiv:1305.6443 [hep-ph]}}.

\bibitem{Gale:2014dfa}
C.~Gale, Y.~Hidaka, S.~Jeon, S.~Lin, J.-F. Paquet, R.~D. Pisarski, D.~Satow,
  V.~V. Skokov, and G.~Vujanovic, ``{Production and Elliptic Flow of Dileptons
  and Photons in a Matrix Model of the Quark-Gluon Plasma}'',
  \href{https://doi.org/10.1103/PhysRevLett.114.072301}{{\em Phys. Rev. Lett.}
  {\bfseries 114} (2015) 072301},
  \href{https://arxiv.org/abs/1409.4778}{{\ttfamily arXiv:1409.4778 [hep-ph]}}.

\bibitem{Chatterjee:2017akg}
R.~Chatterjee, P.~Dasgupta, and D.~K. Srivastava, ``{Anisotropic flow of
  thermal photons at energies available at the BNL Relativistic Heavy Ion
  Collider and at the CERN Large Hadron Collider}'',
  \href{https://doi.org/10.1103/PhysRevC.96.014911}{{\em Phys. Rev. C}
  {\bfseries 96} (2017) 014911},
  \href{https://arxiv.org/abs/1702.02378}{{\ttfamily arXiv:1702.02378
  [nucl-th]}}.

\bibitem{Linnyk:2015tha}
O.~Linnyk, V.~Konchakovski, T.~Steinert, W.~Cassing, and E.~L. Bratkovskaya,
  ``{Hadronic and partonic sources of direct photons in relativistic heavy-ion
  collisions}'', \href{https://doi.org/10.1103/PhysRevC.92.054914}{{\em Phys.
  Rev. C} {\bfseries 92} (2015) 054914},
  \href{https://arxiv.org/abs/1504.05699}{{\ttfamily arXiv:1504.05699
  [nucl-th]}}.

\bibitem{ALICE:2015xmh}
{\bfseries ALICE} Collaboration, J.~Adam {\em et~al.}, ``{Direct photon
  production in Pb--Pb collisions at $\sqrt{s_{\rm NN}} =$ 2.76~TeV}'',
  \href{https://doi.org/10.1016/j.physletb.2016.01.020}{{\em Phys. Lett. B}
  {\bfseries 754} (2016) 235--248},
  \href{https://arxiv.org/abs/1509.07324}{{\ttfamily arXiv:1509.07324
  [nucl-ex]}}.

\bibitem{ALICE:2018dti}
{\bfseries ALICE} Collaboration, S.~Acharya {\em et~al.}, ``{Direct photon
  elliptic flow in Pb--Pb collisions at $\sqrt{s_{\rm NN}}$ = 2.76~TeV}'',
  \href{https://doi.org/10.1016/j.physletb.2018.11.039}{{\em Phys. Lett. B}
  {\bfseries 789} (2019) 308--322},
  \href{https://arxiv.org/abs/1805.04403}{{\ttfamily arXiv:1805.04403
  [nucl-ex]}}.

\bibitem{Shen:2013vja}
C.~Shen, U.~W. Heinz, J.-F. Paquet, and C.~Gale, ``{Thermal photons as a
  quark-gluon plasma thermometer reexamined}'',
  \href{https://doi.org/10.1103/PhysRevC.89.044910}{{\em Phys. Rev. C}
  {\bfseries 89} (2014) 044910},
  \href{https://arxiv.org/abs/1308.2440}{{\ttfamily arXiv:1308.2440
  [nucl-th]}}.

\bibitem{Rapp:2014hha}
R.~Rapp and H.~van Hees, ``{Thermal Dileptons as Fireball Thermometer and
  Chronometer}'', \href{https://doi.org/10.1016/j.physletb.2015.12.065}{{\em
  Phys. Lett. B} {\bfseries 753} (2016) 586--590},
  \href{https://arxiv.org/abs/1411.4612}{{\ttfamily arXiv:1411.4612 [hep-ph]}}.

\bibitem{Rapp:2013nxa}
R.~Rapp, ``{Dilepton Spectroscopy of QCD Matter at Collider Energies}'',
  \href{https://doi.org/10.1155/2013/148253}{{\em Adv. High Energy Phys.}
  {\bfseries 2013} (2013) 148253},
  \href{https://arxiv.org/abs/1304.2309}{{\ttfamily arXiv:1304.2309 [hep-ph]}}.

\bibitem{Coquet:2021lca}
M.~Coquet, X.~Du, J.-Y. Ollitrault, S.~Schlichting, and M.~Winn,
  ``{Intermediate mass dileptons as pre-equilibrium probes in heavy ion
  collisions}'', \href{https://doi.org/10.1016/j.physletb.2021.136626}{{\em
  Phys. Lett. B} {\bfseries 821} (2021) 136626},
  \href{https://arxiv.org/abs/2104.07622}{{\ttfamily arXiv:2104.07622
  [nucl-th]}}.

\bibitem{Coquet:2021gms}
M.~Coquet, X.~Du, J.-Y. Ollitrault, S.~Schlichting, and M.~Winn, ``{Transverse
  mass scaling of dilepton radiation off a quark-gluon plasma}'',
  \href{https://doi.org/10.1016/j.nuclphysa.2022.122579}{{\em Nucl. Phys. A}
  {\bfseries 1030} (2023) 122579},
  \href{https://arxiv.org/abs/2112.13876}{{\ttfamily arXiv:2112.13876
  [nucl-th]}}.

\bibitem{Pisarski:1995xu}
R.~D. Pisarski, ``{Where does the $\rho$ go? Chirally symmetric vector mesons
  in the quark-gluon plasma}'',
  \href{https://doi.org/10.1103/PhysRevD.52.R3773}{{\em Phys. Rev. D}
  {\bfseries 52} (1995) R3773--R3776},
  \href{https://arxiv.org/abs/hep-ph/9503328}{{\ttfamily
  arXiv:hep-ph/9503328}}.

\bibitem{Hohler:2013eba}
P.~M. Hohler and R.~Rapp, ``{Is $\rho$-Meson Melting Compatible with Chiral
  Restoration?}'', \href{https://doi.org/10.1016/j.physletb.2014.02.021}{{\em
  Phys. Lett. B} {\bfseries 731} (2014) 103--109},
  \href{https://arxiv.org/abs/1311.2921}{{\ttfamily arXiv:1311.2921 [hep-ph]}}.

\bibitem{PHENIX:2009gyd}
{\bfseries PHENIX} Collaboration, A.~Adare {\em et~al.}, ``{Detailed
  measurement of the $e^+ e^-$ pair continuum in $p+p$ and Au+Au collisions at
  $\sqrt{s_{NN}} = 200$ GeV and implications for direct photon production}'',
  \href{https://doi.org/10.1103/PhysRevC.81.034911}{{\em Phys. Rev. C}
  {\bfseries 81} (2010) 034911},
  \href{https://arxiv.org/abs/0912.0244}{{\ttfamily arXiv:0912.0244
  [nucl-ex]}}.

\bibitem{CERESNA45:1997tgc}
{\bfseries CERES/NA45} Collaboration, G.~Agakichiev {\em et~al.}, ``{Low mass
  $e^{+} e^{-}$ pair production in 158/A-GeV Pb-Au collisions at the CERN SPS,
  its dependence on multiplicity and transverse momentum}'',
  \href{https://doi.org/10.1016/S0370-2693(98)00083-5}{{\em Phys. Lett. B}
  {\bfseries 422} (1998) 405--412},
  \href{https://arxiv.org/abs/nucl-ex/9712008}{{\ttfamily
  arXiv:nucl-ex/9712008}}.

\bibitem{CERESNA45:2002gnc}
{\bfseries CERES/NA45} Collaboration, D.~Adamova {\em et~al.}, ``{Enhanced
  production of low mass electron pairs in 40-AGeV Pb-Au collisions at the CERN
  SPS}'', \href{https://doi.org/10.1103/PhysRevLett.91.042301}{{\em Phys. Rev.
  Lett.} {\bfseries 91} (2003) 042301},
  \href{https://arxiv.org/abs/nucl-ex/0209024}{{\ttfamily
  arXiv:nucl-ex/0209024}}.

\bibitem{CERES:2006wcq}
{\bfseries CERES} Collaboration, D.~Adamova {\em et~al.}, ``{Modification of
  the $\rho$-meson detected by low-mass electron-positron pairs in central
  Pb-Au collisions at 158-A-GeV/c}'',
  \href{https://doi.org/10.1016/j.physletb.2008.07.104}{{\em Phys. Lett. B}
  {\bfseries 666} (2008) 425--429},
  \href{https://arxiv.org/abs/nucl-ex/0611022}{{\ttfamily
  arXiv:nucl-ex/0611022}}.

\bibitem{NA60:2008dcb}
{\bfseries NA60} Collaboration, R.~Arnaldi {\em et~al.}, ``{Evidence for the
  production of thermal-like muon pairs with masses above 1~GeV/$c^{\rm 2}$ in
  158-A-GeV Indium-Indium Collisions}'',
  \href{https://doi.org/10.1140/epjc/s10052-008-0857-2}{{\em Eur. Phys. J. C}
  {\bfseries 59} (2009) 607--623},
  \href{https://arxiv.org/abs/0810.3204}{{\ttfamily arXiv:0810.3204
  [nucl-ex]}}.

\bibitem{NA60:2006ymb}
{\bfseries NA60} Collaboration, R.~Arnaldi {\em et~al.}, ``{First measurement
  of the $\rho$ spectral function in high-energy nuclear collisions}'',
  \href{https://doi.org/10.1103/PhysRevLett.96.162302}{{\em Phys. Rev. Lett.}
  {\bfseries 96} (2006) 162302},
  \href{https://arxiv.org/abs/nucl-ex/0605007}{{\ttfamily
  arXiv:nucl-ex/0605007}}.

\bibitem{NA60:2008ctj}
{\bfseries NA60} Collaboration, R.~Arnaldi {\em et~al.}, ``{NA60 results on
  thermal dimuons}'',
  \href{https://doi.org/10.1140/epjc/s10052-009-0878-5}{{\em Eur. Phys. J. C}
  {\bfseries 61} (2009) 711--720},
  \href{https://arxiv.org/abs/0812.3053}{{\ttfamily arXiv:0812.3053
  [nucl-ex]}}.

\bibitem{Specht:2010xu}
{\bfseries NA60} Collaboration, H.~J. Specht, ``{Thermal Dileptons from Hot and
  Dense Strongly Interacting Matter}'',
  \href{https://doi.org/10.1063/1.3541982}{{\em AIP Conf. Proc.} {\bfseries
  1322} (2010) 1--10}, \href{https://arxiv.org/abs/1011.0615}{{\ttfamily
  arXiv:1011.0615 [nucl-ex]}}.

\bibitem{Renk:2006qr}
T.~Renk and J.~Ruppert, ``{Dimuon transverse momentum spectra as a tool to
  characterize the emission region in heavy-ion collisions}'',
  \href{https://doi.org/10.1103/PhysRevC.77.024907}{{\em Phys. Rev. C}
  {\bfseries 77} (2008) 024907},
  \href{https://arxiv.org/abs/hep-ph/0612113}{{\ttfamily
  arXiv:hep-ph/0612113}}.

\bibitem{Ruppert:2007cr}
J.~Ruppert, C.~Gale, T.~Renk, P.~Lichard, and J.~I. Kapusta, ``{Low mass
  dimuons produced in relativistic nuclear collisions}'',
  \href{https://doi.org/10.1103/PhysRevLett.100.162301}{{\em Phys. Rev. Lett.}
  {\bfseries 100} (2008) 162301},
  \href{https://arxiv.org/abs/0706.1934}{{\ttfamily arXiv:0706.1934 [hep-ph]}}.

\bibitem{PHENIX:2015vek}
{\bfseries PHENIX} Collaboration, A.~Adare {\em et~al.}, ``{Dielectron
  production in Au$+$Au collisions at $\sqrt{s_{NN}}$=200 GeV}'',
  \href{https://doi.org/10.1103/PhysRevC.93.014904}{{\em Phys. Rev. C}
  {\bfseries 93} (2016) 014904},
  \href{https://arxiv.org/abs/1509.04667}{{\ttfamily arXiv:1509.04667
  [nucl-ex]}}.

\bibitem{STAR:2013pwb}
{\bfseries STAR} Collaboration, L.~Adamczyk {\em et~al.}, ``{Dielectron Mass
  Spectra from Au+Au Collisions at $\sqrt{s_{\rm NN}}$ = 200 GeV}'',
  \href{https://doi.org/10.1103/PhysRevLett.113.022301}{{\em Phys. Rev. Lett.}
  {\bfseries 113} (2014) 022301},
  \href{https://arxiv.org/abs/1312.7397}{{\ttfamily arXiv:1312.7397 [hep-ex]}}.
  [Addendum: Phys.Rev.Lett. 113, 049903 (2014)].

\bibitem{ALICE:2018ael}
{\bfseries ALICE} Collaboration, S.~Acharya {\em et~al.}, ``{Measurement of
  dielectron production in central Pb--Pb collisions at
  $\sqrt{{\textit{s}}_{\mathrm{NN}}}$ = 2.76 TeV}'',
  \href{https://doi.org/10.1103/PhysRevC.99.024002}{{\em Phys. Rev. C}
  {\bfseries 99} (2019) 024002},
  \href{https://arxiv.org/abs/1807.00923}{{\ttfamily arXiv:1807.00923
  [nucl-ex]}}.

\bibitem{Rapp:1999us}
R.~Rapp and J.~Wambach, ``{Low mass dileptons at the CERN SPS: Evidence for
  chiral restoration?}'', \href{https://doi.org/10.1007/s100500050364}{{\em
  Eur. Phys. J. A} {\bfseries 6} (1999) 415--420},
  \href{https://arxiv.org/abs/hep-ph/9907502}{{\ttfamily
  arXiv:hep-ph/9907502}}.

\bibitem{vanHees:2007th}
H.~van Hees and R.~Rapp, ``{Dilepton Radiation at the CERN Super Proton
  Synchrotron}'', \href{https://doi.org/10.1016/j.nuclphysa.2008.03.009}{{\em
  Nucl. Phys. A} {\bfseries 806} (2008) 339--387},
  \href{https://arxiv.org/abs/0711.3444}{{\ttfamily arXiv:0711.3444 [hep-ph]}}.

\bibitem{Rapp:2000pe}
R.~Rapp, ``{Signatures of thermal dilepton radiation at RHIC}'',
  \href{https://doi.org/10.1103/PhysRevC.63.054907}{{\em Phys. Rev. C}
  {\bfseries 63} (2001) 054907},
  \href{https://arxiv.org/abs/hep-ph/0010101}{{\ttfamily
  arXiv:hep-ph/0010101}}.

\bibitem{Linnyk:2011vx}
O.~Linnyk, W.~Cassing, J.~Manninen, E.~L. Bratkovskaya, and C.~M. Ko,
  ``{Analysis of dilepton production in Au+Au collisions at $\sqrt{s_{NN}}=200$
  GeV within the Parton-Hadron-String Dynamics (PHSD) transport approach}'',
  \href{https://doi.org/10.1103/PhysRevC.85.024910}{{\em Phys. Rev. C}
  {\bfseries 85} (2012) 024910},
  \href{https://arxiv.org/abs/1111.2975}{{\ttfamily arXiv:1111.2975
  [nucl-th]}}.

\bibitem{ALICE:2008ngc}
{\bfseries ALICE} Collaboration, K.~Aamodt {\em et~al.}, ``{The ALICE
  experiment at the CERN LHC}'',
  \href{https://doi.org/10.1088/1748-0221/3/08/S08002}{{\em JINST} {\bfseries
  3} (2008) S08002}.

\bibitem{ALICE:2014sbx}
{\bfseries ALICE} Collaboration, B.~Abelev {\em et~al.}, ``{Performance of the
  ALICE Experiment at the CERN LHC}'',
  \href{https://doi.org/10.1142/S0217751X14300440}{{\em Int. J. Mod. Phys. A}
  {\bfseries 29} (2014) 1430044},
  \href{https://arxiv.org/abs/1402.4476}{{\ttfamily arXiv:1402.4476
  [nucl-ex]}}.

\bibitem{ALICE:2010tia}
{\bfseries ALICE} Collaboration, K.~Aamodt {\em et~al.}, ``{Alignment of the
  ALICE Inner Tracking System with cosmic-ray tracks}'',
  \href{https://doi.org/10.1088/1748-0221/5/03/P03003}{{\em JINST} {\bfseries
  5} (2010) P03003}, \href{https://arxiv.org/abs/1001.0502}{{\ttfamily
  arXiv:1001.0502 [physics.ins-det]}}.

\bibitem{Alme:2010ke}
J.~Alme {\em et~al.}, ``{The ALICE TPC, a large 3-dimensional tracking device
  with fast readout for ultra-high multiplicity events}'',
  \href{https://doi.org/10.1016/j.nima.2010.04.042}{{\em Nucl. Instrum. Meth.
  A} {\bfseries 622} (2010) 316--367},
  \href{https://arxiv.org/abs/1001.1950}{{\ttfamily arXiv:1001.1950
  [physics.ins-det]}}.

\bibitem{Akindinov:2013tea}
A.~Akindinov {\em et~al.}, ``{Performance of the ALICE Time-Of-Flight detector
  at the LHC}'', \href{https://doi.org/10.1140/epjp/i2013-13044-x}{{\em Eur.
  Phys. J. Plus} {\bfseries 128} (2013) 44}.

\bibitem{ALICE:2013axi}
{\bfseries ALICE} Collaboration, E.~Abbas {\em et~al.}, ``{Performance of the
  ALICE VZERO system}'',
  \href{https://doi.org/10.1088/1748-0221/8/10/P10016}{{\em JINST} {\bfseries
  8} (2013) P10016}, \href{https://arxiv.org/abs/1306.3130}{{\ttfamily
  arXiv:1306.3130 [nucl-ex]}}.

\bibitem{Arnaldi:1999zz}
R.~Arnaldi {\em et~al.}, ``{The Zero degree calorimeters for the ALICE
  experiment}'', \href{https://doi.org/10.1016/j.nima.2008.04.009}{{\em Nucl.
  Instrum. Meth. A} {\bfseries 581} (2007) 397--401}. [Erratum:
  Nucl.Instrum.Meth.A 604, 765 (2009)].

\bibitem{ALICE:2018tvk}
{\bfseries ALICE} Collaboration, S.~Acharya {\em et~al.}, ``{Centrality
  determination in heavy ion collisions}'', {\em ALICE-PUBLIC-2018-011} (2018)
  . \url{http://cds.cern.ch/record/2636623}.

\bibitem{ALICE:2022xir}
{\bfseries ALICE} Collaboration, S.~Acharya {\em et~al.}, ``{ALICE luminosity
  determination for Pb$-$Pb collisions at $\sqrt{s_{\mathrm{NN}}} = 5.02$
  TeV}'', \href{https://doi.org/10.1088/1748-0221/19/02/P02039}{{\em JINST}
  {\bfseries 19} (2024) P02039},
  \href{https://arxiv.org/abs/2204.10148}{{\ttfamily arXiv:2204.10148
  [nucl-ex]}}.

\bibitem{ALICE:2023kzv}
{\bfseries ALICE} Collaboration, S.~Acharya {\em et~al.}, ``{Data-driven
  precision determination of the material budget in ALICE}'',
  \href{https://arxiv.org/abs/2303.15317}{{\ttfamily arXiv:2303.15317
  [physics.ins-det]}}.

\bibitem{ALICE:2022hvk}
{\bfseries ALICE} Collaboration, S.~Acharya {\em et~al.}, ``{Dielectron
  production at midrapidity at low transverse momentum in peripheral and
  semi-peripheral Pb\textendash{}Pb collisions at $ {\sqrt{s}}_{\textrm{NN}} $
  = 5.02 TeV}'', \href{https://doi.org/10.1007/JHEP06(2023)024}{{\em JHEP}
  {\bfseries 06} (2023) 024},
  \href{https://arxiv.org/abs/2204.11732}{{\ttfamily arXiv:2204.11732
  [nucl-ex]}}.

\bibitem{Shahoyan:2009zz}
{\bfseries NA60} Collaboration, R.~Shahoyan, ``{Evidence for the production of
  thermal dimuons with masses above 1-GeV/c**2 in 158A-GeV In-In collisions}'',
  \href{https://doi.org/10.1016/j.nuclphysa.2009.05.077}{{\em Nucl. Phys. A}
  {\bfseries 827} (2009) 353C--355C}.

\bibitem{ALICE:2018fvj}
{\bfseries ALICE} Collaboration, S.~Acharya {\em et~al.}, ``{Dielectron
  production in proton--proton collisions at $ \sqrt{s}=7 $ TeV}'',
  \href{https://doi.org/10.1007/JHEP09(2018)064}{{\em JHEP} {\bfseries 09}
  (2018) 064}, \href{https://arxiv.org/abs/1805.04391}{{\ttfamily
  arXiv:1805.04391 [hep-ex]}}.

\bibitem{STAR:2015tnn}
{\bfseries STAR} Collaboration, L.~Adamczyk {\em et~al.}, ``{Measurements of
  Dielectron Production in Au$+$Au Collisions at $\sqrt{s_{\rm NN}}$ = 200 GeV
  from the STAR Experiment}'',
  \href{https://doi.org/10.1103/PhysRevC.92.024912}{{\em Phys. Rev. C}
  {\bfseries 92} (2015) 024912},
  \href{https://arxiv.org/abs/1504.01317}{{\ttfamily arXiv:1504.01317
  [hep-ex]}}.

\bibitem{Brun:1119728}
R.~Brun, F.~Bruyant, M.~Maire, A.~C. McPherson, and P.~Zanarini, {\em {GEANT 3:
  user's guide Geant 3.10, Geant 3.11; rev. version}}.
\newblock CERN, Geneva, 1987.
\newblock \url{https://cds.cern.ch/record/1119728}.

\bibitem{Wang:1991hta}
X.-N. Wang and M.~Gyulassy, ``{HIJING: A Monte Carlo model for multiple jet
  production in p p, p A and A A collisions}'',
  \href{https://doi.org/10.1103/PhysRevD.44.3501}{{\em Phys. Rev. D} {\bfseries
  44} (1991) 3501--3516}.

\bibitem{Golonka:2005pn}
P.~Golonka and Z.~Was, ``{PHOTOS Monte Carlo: A Precision tool for QED
  corrections in $Z$ and $W$ decays}'',
  \href{https://doi.org/10.1140/epjc/s2005-02396-4}{{\em Eur. Phys. J. C}
  {\bfseries 45} (2006) 97--107},
  \href{https://arxiv.org/abs/hep-ph/0506026}{{\ttfamily
  arXiv:hep-ph/0506026}}.

\bibitem{Skands:2010ak}
P.~Z. Skands, ``{Tuning Monte Carlo Generators: The Perugia Tunes}'',
  \href{https://doi.org/10.1103/PhysRevD.82.074018}{{\em Phys. Rev. D}
  {\bfseries 82} (2010) 074018},
  \href{https://arxiv.org/abs/1005.3457}{{\ttfamily arXiv:1005.3457 [hep-ph]}}.

\bibitem{Sjostrand:2006za}
T.~Sjostrand, S.~Mrenna, and P.~Z. Skands, ``{PYTHIA 6.4 Physics and Manual}'',
  \href{https://doi.org/10.1088/1126-6708/2006/05/026}{{\em JHEP} {\bfseries
  05} (2006) 026}, \href{https://arxiv.org/abs/hep-ph/0603175}{{\ttfamily
  arXiv:hep-ph/0603175}}.

\bibitem{ALICE:2017tau}
{\bfseries ALICE} Collaboration, S.~Acharya {\em et~al.}, ``{Momentum
  transformation matrix for dielectron simulations in Pb--Pb collisions at
  $\sqrt{s_{\mathrm{NN}}}=2.76\ \mathrm{TeV}$}'', {\em ALICE-PUBLIC-2017-011}
  (2017) . \url{http://cds.cern.ch/record/2289779}.

\bibitem{ALICE:2019hno}
{\bfseries ALICE} Collaboration, S.~Acharya {\em et~al.}, ``{Production of
  charged pions, kaons, and (anti-)protons in Pb--Pb and inelastic $pp$
  collisions at $\sqrt {s_{NN}}$ = 5.02 TeV}'',
  \href{https://doi.org/10.1103/PhysRevC.101.044907}{{\em Phys. Rev. C}
  {\bfseries 101} (2020) 044907},
  \href{https://arxiv.org/abs/1910.07678}{{\ttfamily arXiv:1910.07678
  [nucl-ex]}}.

\bibitem{ALICE:2020umb}
{\bfseries ALICE} Collaboration, S.~Acharya {\em et~al.}, ``{Soft-Dielectron
  Excess in Proton--Proton Collisions at $\sqrt{s}$ = 13 TeV}'',
  \href{https://doi.org/10.1103/PhysRevLett.127.042302}{{\em Phys. Rev. Lett.}
  {\bfseries 127} (2021) 042302},
  \href{https://arxiv.org/abs/2005.14522}{{\ttfamily arXiv:2005.14522
  [nucl-ex]}}.

\bibitem{Ren:2021pzi}
Y.~Ren and A.~Drees, ``{Study of the $\eta$ to $\pi^0$ Ratio in Heavy-Ion
  Collisions}'', \href{https://doi.org/10.1103/PhysRevC.104.054902}{{\em Phys.
  Rev. C} {\bfseries 104} (2021) 054902},
  \href{https://arxiv.org/abs/2102.05220}{{\ttfamily arXiv:2102.05220
  [nucl-ex]}}.

\bibitem{ALICE:2012wos}
{\bfseries ALICE} Collaboration, B.~Abelev {\em et~al.}, ``{Neutral pion and
  $\eta$ meson production in proton--proton collisions at $\sqrt{s}=0.9$ TeV
  and $\sqrt{s}=7$ TeV}'',
  \href{https://doi.org/10.1016/j.physletb.2012.09.015}{{\em Phys. Lett. B}
  {\bfseries 717} (2012) 162--172},
  \href{https://arxiv.org/abs/1205.5724}{{\ttfamily arXiv:1205.5724 [hep-ex]}}.

\bibitem{ALICE:2017ryd}
{\bfseries ALICE} Collaboration, S.~Acharya {\em et~al.}, ``{$\pi ^{0}$ and
  $\eta $ meson production in proton--proton collisions at $\sqrt{s}=8$ TeV}'',
  \href{https://doi.org/10.1140/epjc/s10052-018-5612-8}{{\em Eur. Phys. J. C}
  {\bfseries 78} (2018) 263},
  \href{https://arxiv.org/abs/1708.08745}{{\ttfamily arXiv:1708.08745
  [hep-ex]}}.

\bibitem{ALICE:2018vhm}
{\bfseries ALICE} Collaboration, S.~Acharya {\em et~al.}, ``{Neutral pion and
  $\eta$ meson production in p--Pb collisions at $\sqrt{s_\mathrm{NN}} = 5.02$
  TeV}'', \href{https://doi.org/10.1140/epjc/s10052-018-6013-8}{{\em Eur. Phys.
  J. C} {\bfseries 78} (2018) 624},
  \href{https://arxiv.org/abs/1801.07051}{{\ttfamily arXiv:1801.07051
  [nucl-ex]}}.

\bibitem{ALICE:2017nce}
{\bfseries ALICE} Collaboration, S.~Acharya {\em et~al.}, ``{Production of
  ${\pi ^0}$ and $\eta $ mesons up to high transverse momentum in pp collisions
  at $\sqrt{s}=2.76$ TeV}'',
  \href{https://doi.org/10.1140/epjc/s10052-017-4890-x}{{\em Eur. Phys. J. C}
  {\bfseries 77} (2017) 339},
  \href{https://arxiv.org/abs/1702.00917}{{\ttfamily arXiv:1702.00917
  [hep-ex]}}.

\bibitem{Agakichiev:1998ign}
G.~Agakichiev {\em et~al.}, ``{Neutral meson production in p Be and p Au
  collisions at 450-GeV beam energy}'',
  \href{https://doi.org/10.1007/s100529800804}{{\em Eur. Phys. J. C} {\bfseries
  4} (1998) 249--257}.

\bibitem{ALICE:2018mdl}
{\bfseries ALICE} Collaboration, S.~Acharya {\em et~al.}, ``{Neutral pion and
  $\eta$ meson production at midrapidity in Pb--Pb collisions at
  $\sqrt{s_\mathrm{NN}}$ = 2.76 TeV}'',
  \href{https://doi.org/10.1103/PhysRevC.98.044901}{{\em Phys. Rev. C}
  {\bfseries 98} (2018) 044901},
  \href{https://arxiv.org/abs/1803.05490}{{\ttfamily arXiv:1803.05490
  [nucl-ex]}}.

\bibitem{ALICE:2019xyr}
{\bfseries ALICE} Collaboration, S.~Acharya {\em et~al.}, ``{Evidence of
  rescattering effect in Pb-Pb collisions at the LHC through production of
  $\rm{K}^{*}(892)^{0}$ and $\phi(1020)$ mesons}'',
  \href{https://doi.org/10.1016/j.physletb.2020.135225}{{\em Phys. Lett. B}
  {\bfseries 802} (2020) 135225},
  \href{https://arxiv.org/abs/1910.14419}{{\ttfamily arXiv:1910.14419
  [nucl-ex]}}.

\bibitem{ALICE:2023hou}
{\bfseries ALICE} Collaboration, S.~Acharya {\em et~al.}, ``{Prompt and
  non-prompt J$/\psi$ production at midrapidity in Pb$-$Pb collisions at
  $\sqrt{s_{\mathrm{NN}}}$ = 5.02 TeV}'',
  \href{https://arxiv.org/abs/2308.16125}{{\ttfamily arXiv:2308.16125
  [nucl-ex]}}.

\bibitem{ALICE:2019nuy}
{\bfseries ALICE} Collaboration, S.~Acharya {\em et~al.}, ``{Measurement of
  electrons from semileptonic heavy-flavour hadron decays at midrapidity in pp
  and Pb--Pb collisions at $\sqrt{s_{\rm{NN}}}$ = 5.02 TeV}'',
  \href{https://doi.org/10.1016/j.physletb.2020.135377}{{\em Phys. Lett. B}
  {\bfseries 804} (2020) 135377},
  \href{https://arxiv.org/abs/1910.09110}{{\ttfamily arXiv:1910.09110
  [nucl-ex]}}.

\bibitem{Eskola:2009uj}
K.~J. Eskola, H.~Paukkunen, and C.~A. Salgado, ``{EPS09: A New Generation of
  NLO and LO Nuclear Parton Distribution Functions}'',
  \href{https://doi.org/10.1088/1126-6708/2009/04/065}{{\em JHEP} {\bfseries
  04} (2009) 065}, \href{https://arxiv.org/abs/0902.4154}{{\ttfamily
  arXiv:0902.4154 [hep-ph]}}.

\bibitem{ALICE:2020mfy}
{\bfseries ALICE} Collaboration, S.~Acharya {\em et~al.}, ``{Dielectron
  production in proton--proton and proton--lead collisions at
  $\sqrt{s_\mathrm{NN}}=$ 5.02 TeV}'',
  \href{https://doi.org/10.1103/PhysRevC.102.055204}{{\em Phys. Rev. C}
  {\bfseries 102} (2020) 055204},
  \href{https://arxiv.org/abs/2005.11995}{{\ttfamily arXiv:2005.11995
  [nucl-ex]}}.

\bibitem{powheg1}
P.~Nason, ``{A New method for combining NLO QCD with shower Monte Carlo
  algorithms}'', \href{https://doi.org/10.1088/1126-6708/2004/11/040}{{\em
  JHEP} {\bfseries 11} (2004) 040},
\href{https://arxiv.org/abs/hep-ph/0409146}{{\ttfamily arXiv:hep-ph/0409146
  [hep-ph]}}.

\bibitem{powheg2}
S.~Frixione, P.~Nason, and G.~Ridolfi, ``{A Positive-weight
  next-to-leading-order Monte Carlo for heavy flavour hadroproduction}'',
  \href{https://doi.org/10.1088/1126-6708/2007/09/126}{{\em JHEP} {\bfseries
  09} (2007) 126}.

\bibitem{powheg3}
S.~Frixione, P.~Nason, and C.~Oleari, ``{Matching NLO QCD computations with
  Parton Shower simulations: the POWHEG method}'',
  \href{https://doi.org/10.1088/1126-6708/2007/11/070}{{\em JHEP} {\bfseries
  11} (2007) 070}.

\bibitem{powheg4}
S.~Alioli, P.~Nason, C.~Oleari, and E.~Re, ``{A general framework for
  implementing NLO calculations in shower Monte Carlo programs: the POWHEG
  BOX}'', \href{https://doi.org/10.1007/JHEP06(2010)043}{{\em JHEP} {\bfseries
  06} (2010) 043}, \href{https://arxiv.org/abs/1002.2581}{{\ttfamily
  arXiv:1002.2581 [hep-ph]}}.

\bibitem{ParticleDataGroup:2022pth}
{\bfseries Particle Data Group} Collaboration, R.~L. Workman {\em et~al.},
  ``{Review of Particle Physics}'',
  \href{https://doi.org/10.1093/ptep/ptac097}{{\em PTEP} {\bfseries 2022}
  (2022) 083C01}.

\bibitem{ALICE:2021dhb}
{\bfseries ALICE} Collaboration, S.~Acharya {\em et~al.}, ``{Charm-quark
  fragmentation fractions and production cross section at midrapidity in pp
  collisions at the LHC}'',
  \href{https://doi.org/10.1103/PhysRevD.105.L011103}{{\em Phys. Rev. D}
  {\bfseries 105} (2022) L011103},
  \href{https://arxiv.org/abs/2105.06335}{{\ttfamily arXiv:2105.06335
  [nucl-ex]}}.

\bibitem{ALICE:2021rxa}
{\bfseries ALICE} Collaboration, S.~Acharya {\em et~al.}, ``{Prompt D$^{0}$,
  D$^{+}$, and D$^{*+}$ production in Pb\textendash{}Pb collisions at $
  \sqrt{s_{\mathrm{NN}}} $ = 5.02 TeV}'',
  \href{https://doi.org/10.1007/JHEP01(2022)174}{{\em JHEP} {\bfseries 01}
  (2022) 174}, \href{https://arxiv.org/abs/2110.09420}{{\ttfamily
  arXiv:2110.09420 [nucl-ex]}}.

\bibitem{ALICE:2021bib}
{\bfseries ALICE} Collaboration, S.~Acharya {\em et~al.}, ``{Constraining
  hadronization mechanisms with \ensuremath{\Lambda}c+/D0 production ratios in
  Pb\textendash{}Pb collisions at sNN=5.02 TeV}'',
  \href{https://doi.org/10.1016/j.physletb.2023.137796}{{\em Phys. Lett. B}
  {\bfseries 839} (2023) 137796},
  \href{https://arxiv.org/abs/2112.08156}{{\ttfamily arXiv:2112.08156
  [nucl-ex]}}.

\bibitem{Song:2018xca}
T.~Song, W.~Cassing, P.~Moreau, and E.~Bratkovskaya, ``{Open charm and
  dileptons from relativistic heavy-ion collisions}'',
  \href{https://doi.org/10.1103/PhysRevC.97.064907}{{\em Phys. Rev. C}
  {\bfseries 97} (2018) 064907},
  \href{https://arxiv.org/abs/1803.02698}{{\ttfamily arXiv:1803.02698
  [nucl-th]}}.

\bibitem{Feldman:1997qc}
G.~J. Feldman and R.~D. Cousins, ``{A Unified approach to the classical
  statistical analysis of small signals}'',
  \href{https://doi.org/10.1103/PhysRevD.57.3873}{{\em Phys. Rev. D} {\bfseries
  57} (1998) 3873--3889},
  \href{https://arxiv.org/abs/physics/9711021}{{\ttfamily
  arXiv:physics/9711021}}.

\bibitem{Linnyk:2011ee}
O.~Linnyk, E.~L. Bratkovskaya, J.~Manninen, and W.~Cassing, ``{Dilepton
  production from parton interactions in the early stage of relativistic
  heavy-ion collisions}'',
  \href{https://doi.org/10.1088/1742-6596/312/1/012010}{{\em J. Phys. Conf.
  Ser.} {\bfseries 312} (2011) 012010},
  \href{https://arxiv.org/abs/1102.3624}{{\ttfamily arXiv:1102.3624
  [nucl-th]}}.

\bibitem{CERN-LHCC-2013-020}
{\bfseries {ALICE}} Collaboration, B.~Abelev {\em et~al.}, ``{Upgrade of the
  ALICE Time Projection Chamber}'', {\em CERN-LHCC-2013-020} (2013) .
  \url{https://cds.cern.ch/record/1622286}.

\bibitem{CERN-LHCC-2015-002}
{\bfseries ALICE} Collaboration, J.~Adam {\em et~al.}, ``{Addendum to the
  Technical Design Report for the Upgrade of the ALICE Time Projection
  Chamber}'', {\em CERN-LHCC-2015-002} (2015) .
  \url{https://cds.cern.ch/record/1984329}.

\bibitem{ALICETPC:2020ann}
{\bfseries ALICE TPC} Collaboration, J.~Adolfsson {\em et~al.}, ``{The upgrade
  of the ALICE TPC with GEMs and continuous readout}'',
  \href{https://doi.org/10.1088/1748-0221/16/03/P03022}{{\em JINST} {\bfseries
  16} (2021) P03022}, \href{https://arxiv.org/abs/2012.09518}{{\ttfamily
  arXiv:2012.09518 [physics.ins-det]}}.

\bibitem{newReadoutALICE}
{\bfseries ALICE} Collaboration, P.~Antonioli, A.~Kluge, and W.~Riegler,
  ``{Upgrade of the ALICE Readout and Trigger System}'', {\em
  CERN-LHCC-2013-019} (2013) . \url{https://cds.cern.ch/record/1603472}.

\bibitem{newCompFrameworkALICE}
{\bfseries ALICE} Collaboration, P.~Buncic, M.~Krzewicki, and P.~Vande~Vyvre,
  ``{Technical Design Report for the Upgrade of the Online-Offline Computing
  System}'', {\em CERN-LHCC-2015-006} (2015) .
  \url{https://cds.cern.ch/record/2011297}.

\bibitem{Musa:1475244}
{\bfseries {ALICE}} Collaboration, ``{Upgrade of the Inner Tracking System
  Conceptual Design Report. Conceptual Design Report for the Upgrade of the
  ALICE ITS}'', {\em CERN-LHCC-2012-013} (2012) .
  \url{https://cds.cern.ch/record/1475244}.

\bibitem{ALICE:2018fuj}
{\bfseries ALICE} Collaboration, ``{Expression of Interest for an ALICE ITS
  Upgrade in LS3}'', {\em ALICE-PUBLIC-2018-013} (2018) .
  \url{http://cds.cern.ch/record/2644611}.

\bibitem{Citron:2018lsq}
Z.~Citron {\em et~al.}, ``{Report from Working Group 5}: {Future physics
  opportunities for high-density QCD at the LHC with heavy-ion and proton
  beams}'', \href{https://doi.org/10.23731/CYRM-2019-007.1159}{{\em CERN Yellow
  Rep. Monogr.} {\bfseries 7} (2019) 1159--1410},
  \href{https://arxiv.org/abs/1812.06772}{{\ttfamily arXiv:1812.06772
  [hep-ph]}}.

\bibitem{ALICE:2023udb}
{\bfseries ALICE} Collaboration, S.~Acharya {\em et~al.}, ``{ALICE upgrades
  during the LHC Long Shutdown 2}'',
  \href{https://doi.org/10.1088/1748-0221/19/05/P05062}{{\em JINST} {\bfseries
  19} (2024) P05062}, \href{https://arxiv.org/abs/2302.01238}{{\ttfamily
  arXiv:2302.01238 [physics.ins-det]}}.

\bibitem{Kroll:1955zu}
N.~M. Kroll and W.~Wada, ``{Internal pair production associated with the
  emission of high-energy gamma rays}'',
  \href{https://doi.org/10.1103/PhysRev.98.1355}{{\em Phys. Rev.} {\bfseries
  98} (1955) 1355--1359}.

\bibitem{ALICE:2018gev}
{\bfseries ALICE} Collaboration, S.~Acharya {\em et~al.}, ``{Dielectron and
  heavy-quark production in inelastic and high-multiplicity
  proton\textendash{}proton collisions at $\sqrt {s_{NN}}=$ 13TeV}'',
  \href{https://doi.org/10.1016/j.physletb.2018.11.009}{{\em Phys. Lett. B}
  {\bfseries 788} (2019) 505--518},
  \href{https://arxiv.org/abs/1805.04407}{{\ttfamily arXiv:1805.04407
  [hep-ex]}}.

\bibitem{LAFFERTY1995541}
G.~Lafferty and T.~Wyatt, ``Where to stick your data points: The treatment of
  measurements within wide bins'',
  \href{https://doi.org/https://doi.org/10.1016/0168-9002(94)01112-5}{{\em
  Nuclear Instruments and Methods in Physics Research Section A: Accelerators,
  Spectrometers, Detectors and Associated Equipment} {\bfseries 355} (1995)
  541--547}.
  \url{https://www.sciencedirect.com/science/article/pii/0168900294011125}.

\bibitem{Aurenche:1998gv}
P.~Aurenche, M.~Fontannaz, J.~P. Guillet, B.~A. Kniehl, E.~Pilon, and
  M.~Werlen, ``{A Critical phenomenological study of inclusive photon
  production in hadronic collisions}'',
  \href{https://doi.org/10.1007/s100529900018}{{\em Eur. Phys. J. C} {\bfseries
  9} (1999) 107--119}, \href{https://arxiv.org/abs/hep-ph/9811382}{{\ttfamily
  arXiv:hep-ph/9811382}}.

\bibitem{Kovarik:2015cma}
K.~Kovarik {\em et~al.}, ``{nCTEQ15 - Global analysis of nuclear parton
  distributions with uncertainties in the CTEQ framework}'',
  \href{https://doi.org/10.1103/PhysRevD.93.085037}{{\em Phys. Rev. D}
  {\bfseries 93} (2016) 085037},
  \href{https://arxiv.org/abs/1509.00792}{{\ttfamily arXiv:1509.00792
  [hep-ph]}}.

\bibitem{Bourhis:1997yu}
L.~Bourhis, M.~Fontannaz, and J.~P. Guillet, ``{Quarks and gluon fragmentation
  functions into photons}'', \href{https://doi.org/10.1007/s100520050158}{{\em
  Eur. Phys. J. C} {\bfseries 2} (1998) 529--537},
  \href{https://arxiv.org/abs/hep-ph/9704447}{{\ttfamily
  arXiv:hep-ph/9704447}}.

\bibitem{Schenke:2012wb}
B.~Schenke, P.~Tribedy, and R.~Venugopalan, ``{Fluctuating Glasma initial
  conditions and flow in heavy ion collisions}'',
  \href{https://doi.org/10.1103/PhysRevLett.108.252301}{{\em Phys. Rev. Lett.}
  {\bfseries 108} (2012) 252301},
  \href{https://arxiv.org/abs/1202.6646}{{\ttfamily arXiv:1202.6646
  [nucl-th]}}.

\bibitem{Schenke:2012hg}
B.~Schenke, P.~Tribedy, and R.~Venugopalan, ``{Event-by-event gluon
  multiplicity, energy density, and eccentricities in ultrarelativistic
  heavy-ion collisions}'',
  \href{https://doi.org/10.1103/PhysRevC.86.034908}{{\em Phys. Rev. C}
  {\bfseries 86} (2012) 034908},
  \href{https://arxiv.org/abs/1206.6805}{{\ttfamily arXiv:1206.6805 [hep-ph]}}.

\bibitem{Kurkela:2018wud}
A.~Kurkela, A.~Mazeliauskas, J.-F. Paquet, S.~Schlichting, and D.~Teaney,
  ``{Matching the Nonequilibrium Initial Stage of Heavy Ion Collisions to
  Hydrodynamics with QCD Kinetic Theory}'',
  \href{https://doi.org/10.1103/PhysRevLett.122.122302}{{\em Phys. Rev. Lett.}
  {\bfseries 122} (2019) 122302},
  \href{https://arxiv.org/abs/1805.01604}{{\ttfamily arXiv:1805.01604
  [hep-ph]}}.

\bibitem{Kurkela:2018vqr}
A.~Kurkela, A.~Mazeliauskas, J.-F. Paquet, S.~Schlichting, and D.~Teaney,
  ``{Effective kinetic description of event-by-event pre-equilibrium dynamics
  in high-energy heavy-ion collisions}'',
  \href{https://doi.org/10.1103/PhysRevC.99.034910}{{\em Phys. Rev. C}
  {\bfseries 99} (2019) 034910},
  \href{https://arxiv.org/abs/1805.00961}{{\ttfamily arXiv:1805.00961
  [hep-ph]}}.

\bibitem{Schenke:2010nt}
B.~Schenke, S.~Jeon, and C.~Gale, ``{(3+1)D hydrodynamic simulation of
  relativistic heavy-ion collisions}'',
  \href{https://doi.org/10.1103/PhysRevC.82.014903}{{\em Phys. Rev. C}
  {\bfseries 82} (2010) 014903},
  \href{https://arxiv.org/abs/1004.1408}{{\ttfamily arXiv:1004.1408 [hep-ph]}}.

\bibitem{ALICE:2015juo}
{\bfseries ALICE} Collaboration, J.~Adam {\em et~al.}, ``{Centrality Dependence
  of the Charged-Particle Multiplicity Density at Midrapidity in Pb-Pb
  Collisions at $\sqrt{s_{\rm NN}}$ = 5.02 TeV}'',
  \href{https://doi.org/10.1103/PhysRevLett.116.222302}{{\em Phys. Rev. Lett.}
  {\bfseries 116} (2016) 222302},
  \href{https://arxiv.org/abs/1512.06104}{{\ttfamily arXiv:1512.06104
  [nucl-ex]}}.

\bibitem{HEPData}
{\bfseries ALICE} Collaboration,
  \href{https://www.hepdata.net/record/ins2693299}{https://www.hepdata.net/record/ins2693299},
  2025.

\end{thebibliography}\endgroup

\newpage
\appendix

%
%

\section{The ALICE Collaboration}
\label{app:collab}
\begin{flushleft} 
\small

S.~Acharya\,\orcidlink{0000-0002-9213-5329}\,$^{\rm 128}$, 
D.~Adamov\'{a}\,\orcidlink{0000-0002-0504-7428}\,$^{\rm 87}$, 
G.~Aglieri Rinella\,\orcidlink{0000-0002-9611-3696}\,$^{\rm 33}$, 
M.~Agnello\,\orcidlink{0000-0002-0760-5075}\,$^{\rm 30}$, 
N.~Agrawal\,\orcidlink{0000-0003-0348-9836}\,$^{\rm 52}$, 
Z.~Ahammed\,\orcidlink{0000-0001-5241-7412}\,$^{\rm 136}$, 
S.~Ahmad\,\orcidlink{0000-0003-0497-5705}\,$^{\rm 16}$, 
S.U.~Ahn\,\orcidlink{0000-0001-8847-489X}\,$^{\rm 72}$, 
I.~Ahuja\,\orcidlink{0000-0002-4417-1392}\,$^{\rm 38}$, 
A.~Akindinov\,\orcidlink{0000-0002-7388-3022}\,$^{\rm 142}$
% ...

\section*{Affiliation Notes}

$^{\rm I}$ Deceased\\
$^{\rm II}$ Also at: Max-Planck-Institut fur Physik, Munich, Germany\\
$^{\rm III}$ Also at: Italian National Agency for New Technologies, Energy and Sustainable Economic Development (ENEA), Bologna, Italy\\
$^{\rm IV}$ Also at: Dipartimento DET del Politecnico di Torino, Turin, Italy\\
$^{\rm V}$ Also at: Yildiz Technical University, Istanbul, T\"{u}rkiye\\
$^{\rm VI}$ Also at: Department of Applied Physics, Aligarh Muslim University, Aligarh, India\\
$^{\rm VII}$ Also at: Institute of Theoretical Physics, University of Wroclaw, Poland\\
$^{\rm VIII}$ Also at: An institution covered by a cooperation agreement with CERN\\

\section*{Collaboration Institutes}

$^{1}$ A.I. Alikhanyan National Science Laboratory (Yerevan Physics Institute) Foundation, Yerevan, Armenia\\
$^{2}$ AGH University of Krakow, Cracow, Poland\\
$^{3}$ Bogolyubov Institute for Theoretical Physics, National Academy of Sciences of Ukraine, Kiev, Ukraine\\
$^{4}$ Bose Institute, Department of Physics  and Centre for Astroparticle Physics and Space Science (CAPSS), Kolkata, India\\
$^{5}$ California Polytechnic State University, San Luis Obispo, California, United States\\
$^{6}$ Central China Normal University, Wuhan, China\\
$^{7}$ Centro de Aplicaciones Tecnol\'{o}gicas y Desarrollo Nuclear (CEADEN), Havana, Cuba\\
$^{8}$ Centro de Investigaci\'{o}n y de Estudios Avanzados (CINVESTAV), Mexico City and M\'{e}rida, Mexico\\
$^{9}$ Chicago State University, Chicago, Illinois, United States\\
$^{10}$ China Institute of Atomic Energy, Beijing, China\\
$^{11}$ China University of Geosciences, Wuhan, China\\
$^{12}$ Chungbuk National University, Cheongju, Republic of Korea\\
$^{13}$ Comenius University Bratislava, Faculty of Mathematics, Physics and Informatics, Bratislava, Slovak Republic\\
$^{14}$ COMSATS University Islamabad, Islamabad, Pakistan\\
$^{15}$ Creighton University, Omaha, Nebraska, United States\\
$^{16}$ Department of Physics, Aligarh Muslim University, Aligarh, India\\
$^{17}$ Department of Physics, Pusan National University, Pusan, Republic of Korea\\
$^{18}$ Department of Physics, Sejong University, Seoul, Republic of Korea\\
$^{19}$ Department of Physics, University of California, Berkeley, California, United States\\
$^{20}$ Department of Physics, University of Oslo, Oslo, Norway\\
$^{21}$ Department of Physics and Technology, University of Bergen, Bergen, Norway\\
$^{22}$ Dipartimento di Fisica, Universit\`{a} di Pavia, Pavia, Italy\\
$^{23}$ Dipartimento di Fisica dell'Universit\`{a} and Sezione INFN, Cagliari, Italy\\
$^{24}$ Dipartimento di Fisica dell'Universit\`{a} and Sezione INFN, Trieste, Italy\\
$^{25}$ Dipartimento di Fisica dell'Universit\`{a} and Sezione INFN, Turin, Italy\\
$^{26}$ Dipartimento di Fisica e Astronomia dell'Universit\`{a} and Sezione INFN, Bologna, Italy\\
$^{27}$ Dipartimento di Fisica e Astronomia dell'Universit\`{a} and Sezione INFN, Catania, Italy\\
$^{28}$ Dipartimento di Fisica e Astronomia dell'Universit\`{a} and Sezione INFN, Padova, Italy\\
$^{29}$ Dipartimento di Fisica `E.R.~Caianiello' dell'Universit\`{a} and Gruppo Collegato INFN, Salerno, Italy\\
$^{30}$ Dipartimento DISAT del Politecnico and Sezione INFN, Turin, Italy\\
$^{31}$ Dipartimento di Scienze MIFT, Universit\`{a} di Messina, Messina, Italy\\
$^{32}$ Dipartimento Interateneo di Fisica `M.~Merlin' and Sezione INFN, Bari, Italy\\
$^{33}$ European Organization for Nuclear Research (CERN), Geneva, Switzerland\\
$^{34}$ Faculty of Electrical Engineering, Mechanical Engineering and Naval Architecture, University of Split, Split, Croatia\\
$^{35}$ Faculty of Engineering and Science, Western Norway University of Applied Sciences, Bergen, Norway\\
$^{36}$ Faculty of Nuclear Sciences and Physical Engineering, Czech Technical University in Prague, Prague, Czech Republic\\
$^{37}$ Faculty of Physics, Sofia University, Sofia, Bulgaria\\
$^{38}$ Faculty of Science, P.J.~\v{S}af\'{a}rik University, Ko\v{s}ice, Slovak Republic\\
$^{39}$ Frankfurt Institute for Advanced Studies, Johann Wolfgang Goethe-Universit\"{a}t Frankfurt, Frankfurt, Germany\\
$^{40}$ Fudan University, Shanghai, China\\
$^{41}$ Gangneung-Wonju National University, Gangneung, Republic of Korea\\
$^{42}$ Gauhati University, Department of Physics, Guwahati, India\\
$^{43}$ Helmholtz-Institut f\"{u}r Strahlen- und Kernphysik, Rheinische Friedrich-Wilhelms-Universit\"{a}t Bonn, Bonn, Germany\\
$^{44}$ Helsinki Institute of Physics (HIP), Helsinki, Finland\\
$^{45}$ High Energy Physics Group,  Universidad Aut\'{o}noma de Puebla, Puebla, Mexico\\
$^{46}$ Horia Hulubei National Institute of Physics and Nuclear Engineering, Bucharest, Romania\\
$^{47}$ HUN-REN Wigner Research Centre for Physics, Budapest, Hungary\\
$^{48}$ Indian Institute of Technology Bombay (IIT), Mumbai, India\\
$^{49}$ Indian Institute of Technology Indore, Indore, India\\
$^{50}$ INFN, Laboratori Nazionali di Frascati, Frascati, Italy\\
$^{51}$ INFN, Sezione di Bari, Bari, Italy\\
$^{52}$ INFN, Sezione di Bologna, Bologna, Italy\\
$^{53}$ INFN, Sezione di Cagliari, Cagliari, Italy\\
$^{54}$ INFN, Sezione di Catania, Catania, Italy\\
$^{55}$ INFN, Sezione di Padova, Padova, Italy\\
$^{56}$ INFN, Sezione di Pavia, Pavia, Italy\\
$^{57}$ INFN, Sezione di Torino, Turin, Italy\\
$^{58}$ INFN, Sezione di Trieste, Trieste, Italy\\
$^{59}$ Inha University, Incheon, Republic of Korea\\
$^{60}$ Institute for Gravitational and Subatomic Physics (GRASP), Utrecht University/Nikhef, Utrecht, Netherlands\\
$^{61}$ Institute of Experimental Physics, Slovak Academy of Sciences, Ko\v{s}ice, Slovak Republic\\
$^{62}$ Institute of Physics, Homi Bhabha National Institute, Bhubaneswar, India\\
$^{63}$ Institute of Physics of the Czech Academy of Sciences, Prague, Czech Republic\\
$^{64}$ Institute of Space Science (ISS), Bucharest, Romania\\
$^{65}$ Institut f\"{u}r Kernphysik, Johann Wolfgang Goethe-Universit\"{a}t Frankfurt, Frankfurt, Germany\\
$^{66}$ Instituto de Ciencias Nucleares, Universidad Nacional Aut\'{o}noma de M\'{e}xico, Mexico City, Mexico\\
$^{67}$ Instituto de F\'{i}sica, Universidade Federal do Rio Grande do Sul (UFRGS), Porto Alegre, Brazil\\
$^{68}$ Instituto de F\'{\i}sica, Universidad Nacional Aut\'{o}noma de M\'{e}xico, Mexico City, Mexico\\
$^{69}$ iThemba LABS, National Research Foundation, Somerset West, South Africa\\
$^{70}$ Jeonbuk National University, Jeonju, Republic of Korea\\
$^{71}$ Johann-Wolfgang-Goethe Universit\"{a}t Frankfurt Institut f\"{u}r Informatik, Fachbereich Informatik und Mathematik, Frankfurt, Germany\\
$^{72}$ Korea Institute of Science and Technology Information, Daejeon, Republic of Korea\\
$^{73}$ KTO Karatay University, Konya, Turkey\\
$^{74}$ Laboratoire de Physique Subatomique et de Cosmologie, Universit\'{e} Grenoble-Alpes, CNRS-IN2P3, Grenoble, France\\
$^{75}$ Lawrence Berkeley National Laboratory, Berkeley, California, United States\\
$^{76}$ Lund University Department of Physics, Division of Particle Physics, Lund, Sweden\\
$^{77}$ Nagasaki Institute of Applied Science, Nagasaki, Japan\\
$^{78}$ Nara Women{'}s University (NWU), Nara, Japan\\
$^{79}$ National and Kapodistrian University of Athens, School of Science, Department of Physics , Athens, Greece\\
$^{80}$ National Centre for Nuclear Research, Warsaw, Poland\\
$^{81}$ National Institute of Science Education and Research, Homi Bhabha National Institute, Jatni, India\\
$^{82}$ National Nuclear Research Center, Baku, Azerbaijan\\
$^{83}$ National Research and Innovation Agency - BRIN, Jakarta, Indonesia\\
$^{84}$ Niels Bohr Institute, University of Copenhagen, Copenhagen, Denmark\\
$^{85}$ Nikhef, National institute for subatomic physics, Amsterdam, Netherlands\\
$^{86}$ Nuclear Physics Group, STFC Daresbury Laboratory, Daresbury, United Kingdom\\
$^{87}$ Nuclear Physics Institute of the Czech Academy of Sciences, Husinec-\v{R}e\v{z}, Czech Republic\\
$^{88}$ Oak Ridge National Laboratory, Oak Ridge, Tennessee, United States\\
$^{89}$ Ohio State University, Columbus, Ohio, United States\\
$^{90}$ Physics department, Faculty of science, University of Zagreb, Zagreb, Croatia\\
$^{91}$ Physics Department, Panjab University, Chandigarh, India\\
$^{92}$ Physics Department, University of Jammu, Jammu, India\\
$^{93}$ Physics Program and International Institute for Sustainability with Knotted Chiral Meta Matter (SKCM2), Hiroshima University, Hiroshima, Japan\\
$^{94}$ Physikalisches Institut, Eberhard-Karls-Universit\"{a}t T\"{u}bingen, T\"{u}bingen, Germany\\
$^{95}$ Physikalisches Institut, Ruprecht-Karls-Universit\"{a}t Heidelberg, Heidelberg, Germany\\
$^{96}$ Physik Department, Technische Universit\"{a}t M\"{u}nchen, Munich, Germany\\
$^{97}$ Politecnico di Bari and Sezione INFN, Bari, Italy\\
$^{98}$ Research Division and ExtreMe Matter Institute EMMI, GSI Helmholtzzentrum f\"ur Schwerionenforschung GmbH, Darmstadt, Germany\\
$^{99}$ Saga University, Saga, Japan\\
$^{100}$ Saha Institute of Nuclear Physics, Homi Bhabha National Institute, Kolkata, India\\
$^{101}$ School of Physics and Astronomy, University of Birmingham, Birmingham, United Kingdom\\
$^{102}$ Secci\'{o}n F\'{\i}sica, Departamento de Ciencias, Pontificia Universidad Cat\'{o}lica del Per\'{u}, Lima, Peru\\
$^{103}$ Stefan Meyer Institut f\"{u}r Subatomare Physik (SMI), Vienna, Austria\\
$^{104}$ SUBATECH, IMT Atlantique, Nantes Universit\'{e}, CNRS-IN2P3, Nantes, France\\
$^{105}$ Sungkyunkwan University, Suwon City, Republic of Korea\\
$^{106}$ Suranaree University of Technology, Nakhon Ratchasima, Thailand\\
$^{107}$ Technical University of Ko\v{s}ice, Ko\v{s}ice, Slovak Republic\\
$^{108}$ The Henryk Niewodniczanski Institute of Nuclear Physics, Polish Academy of Sciences, Cracow, Poland\\
$^{109}$ The University of Texas at Austin, Austin, Texas, United States\\
$^{110}$ Universidad Aut\'{o}noma de Sinaloa, Culiac\'{a}n, Mexico\\
$^{111}$ Universidade de S\~{a}o Paulo (USP), S\~{a}o Paulo, Brazil\\
$^{112}$ Universidade Estadual de Campinas (UNICAMP), Campinas, Brazil\\
$^{113}$ Universidade Federal do ABC, Santo Andre, Brazil\\
$^{114}$ Universitatea Nationala de Stiinta si Tehnologie Politehnica Bucuresti, Bucharest, Romania\\
$^{115}$ University of Cape Town, Cape Town, South Africa\\
$^{116}$ University of Derby, Derby, United Kingdom\\
$^{117}$ University of Houston, Houston, Texas, United States\\
$^{118}$ University of Jyv\"{a}skyl\"{a}, Jyv\"{a}skyl\"{a}, Finland\\
$^{119}$ University of Kansas, Lawrence, Kansas, United States\\
$^{120}$ University of Liverpool, Liverpool, United Kingdom\\
$^{121}$ University of Science and Technology of China, Hefei, China\\
$^{122}$ University of South-Eastern Norway, Kongsberg, Norway\\
$^{123}$ University of Tennessee, Knoxville, Tennessee, United States\\
$^{124}$ University of the Witwatersrand, Johannesburg, South Africa\\
$^{125}$ University of Tokyo, Tokyo, Japan\\
$^{126}$ University of Tsukuba, Tsukuba, Japan\\
$^{127}$ Universit\"{a}t M\"{u}nster, Institut f\"{u}r Kernphysik, M\"{u}nster, Germany\\
$^{128}$ Universit\'{e} Clermont Auvergne, CNRS/IN2P3, LPC, Clermont-Ferrand, France\\
$^{129}$ Universit\'{e} de Lyon, CNRS/IN2P3, Institut de Physique des 2 Infinis de Lyon, Lyon, France\\
$^{130}$ Universit\'{e} de Strasbourg, CNRS, IPHC UMR 7178, F-67000 Strasbourg, France, Strasbourg, France\\
$^{131}$ Universit\'{e} Paris-Saclay, Centre d'Etudes de Saclay (CEA), IRFU, D\'{e}partment de Physique Nucl\'{e}aire (DPhN), Saclay, France\\
$^{132}$ Universit\'{e}  Paris-Saclay, CNRS/IN2P3, IJCLab, Orsay, France\\
$^{133}$ Universit\`{a} degli Studi di Foggia, Foggia, Italy\\
$^{134}$ Universit\`{a} del Piemonte Orientale, Vercelli, Italy\\
$^{135}$ Universit\`{a} di Brescia, Brescia, Italy\\
$^{136}$ Variable Energy Cyclotron Centre, Homi Bhabha National Institute, Kolkata, India\\
$^{137}$ Warsaw University of Technology, Warsaw, Poland\\
$^{138}$ Wayne State University, Detroit, Michigan, United States\\
$^{139}$ Yale University, New Haven, Connecticut, United States\\
$^{140}$ Yonsei University, Seoul, Republic of Korea\\
$^{141}$  Zentrum  f\"{u}r Technologie und Transfer (ZTT), Worms, Germany\\
$^{142}$ Affiliated with an institute covered by a cooperation agreement with CERN\\
$^{143}$ Affiliated with an international laboratory covered by a cooperation agreement with CERN.\\

\end{flushleft} 

\end{document}